\documentclass[aps,prd,showpacs,floats,onecolumn,nofootinbib]{revtex4}
\pdfoutput=1  % togliere il commento se le figure sono pdf, lasciarlo se sono eps
\usepackage{graphicx,color}
\usepackage{appendix}
\usepackage{subfigure}
\usepackage{latexsym,amsmath,amssymb,graphicx,booktabs}
\usepackage{hyperref}
\definecolor{MyBlue}{rgb}{0.15,0.15,0.70}

\hypersetup{
colorlinks=true,
citecolor=MyBlue,
linkcolor=MyBlue,
urlcolor=MyBlue
}

\newcommand{\bk}{{\mathbf k}}
\newcommand{\bfe}{{\mathbf e}}

\newcommand{\EE}{{\cal E}}
\newcommand{\HH}{{\cal H}}

\newcommand{\MM}{{\cal M}}
\newcommand{\OO}{{\cal O}}

\newcommand{\al}{\alpha}

\newcommand{\de}{\delta}

\newcommand{\si}{\sigma}

\newcommand{\tr}{\mbox{tr}}
\newcommand{\La}{\Lambda}
\newcommand{\la}{\lambda}
\newcommand{\gsim}{\stackrel{>}{\sim}}
\newcommand{\lsim}{\stackrel{<}{\sim}}
\newcommand{\be}{\begin{equation}}
\newcommand{\ee}{\end{equation}}
\newcommand{\bea}{\begin{eqnarray}}
\newcommand{\eea}{\end{eqnarray}}
\newcommand{\bean}{\begin{eqnarray*}}
\newcommand{\eean}{\end{eqnarray*}}
\newcommand{\dd}{\partial}

\def\id{{\rm 1\kern -2.5pt I}}

\usepackage{amssymb}
\usepackage{amsmath}
\usepackage{amsfonts}
\usepackage{upgreek}
\usepackage{latexsym}
\usepackage{placeins}

\def\id{{\rm 1\kern -2.5pt I}}

\graphicspath{ {./Graphs/} }

\newcommand{\nn}{\nonumber}
\renewcommand\({\left(}
\renewcommand\){\right)}
\renewcommand\[{\left[}
\renewcommand\]{\right]}

\newcommand{\ra}{\rightarrow}

\newcommand{\lc}{\ell}
\newcommand{\fc}{f}
\newcommand{\gc}{g}

\begin{document}

\vspace*{2cm}
%Here a list of possible titles 
%\title{Bigravity on general backgrounds}
%\title{Massive bi-graviton on general backgrounds}
%\title{General formalism for bigravity perturbations}
\title{A general mass term for bigravity}
\author{Giulia Cusin, Ruth Durrer, Pietro Guarato and  Mariele Motta}
\affiliation{D\'epartement de Physique Th\'eorique and Center for Astroparticle Physics, Universit\'e de Gen\`eve, 24 quai Ansermet, CH--1211 Gen\`eve 4, Switzerland}
\email{giulia.cusin@unige.ch, ruth.durrer@unige.ch, pietro.guarato@unige.ch, mariele.motta@unige.ch}

\date{\today}

\begin{abstract}

We introduce a new formalism to study perturbations of Hassan-Rosen bigravity theory, around general backgrounds for the two dynamical metrics. In particular, we derive the general expression for the mass term of the perturbations and we explicitly compute it for  cosmological settings. We study tensor perturbations in a specific branch of bigravity %\textcolor{red}{of ref.~\cite{Pilo0}} 
using this formalism. We show that the tensor sector is affected by a late-time instability, which sets in when the mass matrix is no longer positive definite. 

\end{abstract}


\maketitle

\section{Introduction}

The discovery of the accelerated expansion of the Universe~\cite{Riess:1998cb,Perlmutter:1998np,Ade:2015xua} has reactivated  attempts to modify gravity, i.e., General Relativity. Maybe the observed acceleration is not due to a highly fine tuned cosmological constant or some other form of dark energy, but to the fact that gravity becomes weaker on very large scales, see~\cite{Durrer:2007re} for a review of the dark energy/modified gravity problem. This 'degravitation'~\cite{Dvali:2007kt,deRham:2007rw} can be achieved by different means, one of them is to give the graviton a mass of the order of the present Hubble parameter, $H_0$, so that gravity has finite range.

The idea of massive gravity goes back to Fierz and Pauli in 1939~\cite{Fierz:1939ix,Pauli:1939xp} who found the unique quadratic expression in the deviation of the metric from a flat background which is free of ghosts. Later, in the 70s, Boulware and Deser have shown that higher order terms in the 'metric potential' generally re-introduce 
a ghost~\cite{Boulware:1973my}. Only a few years ago, de~Rham, Gabadadze and Tolley have successfully derived the first metric potential which is ghost-free at least up to fourth order in perturbation theory around flat space and in the decoupling limit~\cite{deRham:2010kj}. This potential has been proved to be the most general metric potential in~\cite{Hassan:2011vm} and the proof of the absence of the Boulware-Deser ghost in the full non-perturbative theory away from the decoupling limit has been worked out in~\cite{Hassan:2011hr}, see also~\cite{Hassan:2011tf,Hassan:2011ea} for the most general case. 

A general feature of massive gravity is that the potential has to be formulated with respect to some reference metric, $f_{\mu\nu}$ as a function of $g^{\mu\al}f_{\al\nu}$. This reference metric is an absolute element of the theory, spoiling diffeomorphism invariance, one of the most attractive features of General Relativity.
As a consequence, one can show that a flat reference metric, $f_{\mu\nu}=\eta_{\mu\nu}$, does not allow for cosmological solutions (apart for the Milne Universe which also represents flat space)~\cite{deRham:2014zqa}. But there is no obstruction to render also the reference metric dynamical by introducing an Einstein-Hilbert term in the action also for this metric. This leads in a natural way to a ghost-free bimetric theory of gravity~\cite{Hassan:2011zd}.  Investigations of theoretical aspects of bimetric massive gravity can be found in~\cite{Akrami:2014lja, Hassan:2014vja, deRham:2014fha, Cusin:2014zoa, Noller:2014sta, Akrami:2013ffa, Comelli:2013tja, deRham:2014gla}.

To avoid the re-appearance of the ghost from the coupling to matter, one has to request that matter fields couple to one of the two metrics but not to both. The equivalence principle then implies that all matter fields couple to the same metric. A violation of this could be ruled out experimentally. For simplicity, in this work we therefore assume that matter couples only to one metric which we call the physical metric, $g_{\mu\nu}$. This assumption, however, does not affect our main result, which does not involve the matter couplings.  Relevant contribution to this bi-gravity sector are also given in~\cite{deRham:2014naa,Hassan:2014gta,deRham:2014fha,Heisenberg:2014rka,Heisenberg:2015iqa,Huang:2015yga,Melville:2015dba}. For a review of recent results in bimetric theories see~\cite{Schmidt-May:2015vnx}.

Cosmological solutions of bimetric theories can actually fit the expansion history of the accelerating Universe~\cite{Volkov:2011an, Comelli:2011zm, Konnig:2014dna, Tamanini:2013xia,Fasiello:2013woa}. Perturbation theory and observational tests of several  models of bigravity are presented in \cite{Solomon:2014dua,vonStrauss:2011mq,Berg:2012kn,2013JHEP...03..099A,Konnig:2013gxa}. The cosmology of bigravity in various cosmological settings is studied in \cite{Akrami:2015qga,Konnig:2015lfa} while in Refs.~\cite{Gumrukcuoglu:2015nua, Comelli:2015pua, Enander:2014xga}  the cosmology of models of bigravity where matter is coupled to a combination of the two metrics is investigated.

Previously, some of us have derived a general mass term for massive gravity perturbations on an arbitrary background and for an arbitrary reference metric~\cite{Guarato:2013gba}, see also~\cite{Bernard:2015mkk}. Here we want to generalize this work to bimetric gravity. In the bimetric context both the metrics admit fluctuations around their background configurations, $g_{\mu\nu}=\bar g_{\mu\nu}+ h_{\mu\nu}$ and  $f_{\mu\nu}=\bar f_{\mu\nu}+ \lc_{\mu\nu}$. The derivation of this mass term is the main goal of this paper. We then construct a convenient parametrization of the mass term for cosmological backgrounds and present the generic massive action for perturbations on a Friedmann-Lema\^\i tre spacetime. We express the mass term using the energy density and pressure as well as two new functions of  the parameters of the theory. Finally, to illustrate our result, we apply our finding to a specific model and discuss tensor perturbations in this example.

The remainder of this paper is structured as follows: in the next section we briefly present the general setting of massive bigravity. In section~\ref{Mmunu} we derive the mass term. This section contains our main result. In section~\ref{s4} we specialize to the cosmological setting and in section~\ref{backkk} we present an application. In section~\ref{s5} we conclude. Some lengthy calculations are deferred to several appendices.

{\bf Notation:} We set $c=\hbar=k_{\rm Boltzmann}=1$. $M_g=1/\sqrt{8\pi G}\equiv M_p\simeq 2.4\times 10^{18}$GeV is the reduced Planck mass. 
We work with the metric signature $(-, +, \dots, +)$ and we restrict to $D=4$ spacetime dimensions. With  $\cdot $ and  with $'$  we indicate derivatives with respect to physical time and to conformal time, respectively. We consider only one of the two metrics coupled to matter,  and we restrict to minimal couplings. We normalize the scale factor $a$ to be one at present time, i.e., $a_0=1$. %

\section{The bigravity action}\label{s2}

The action for Hassan and Rosen bigravity theory is given by~\cite{Hassan:2011zd}

\be\label{e:action}
S = \frac{M_g^2}{2}\int \mathrm{d}^{4}x \sqrt{-\det g}\,\left[R(g) - U(f,g) \right] + \frac{M_f^2}{2}\int \mathrm{d}^{4}x \sqrt{-\det f}\,R(f) + S_{\rm matter}(g,\Phi) \,,
\ee
where $R(g)$ and $R(f)$ are the Ricci scalar for the physical metric $g_{\mu\nu}$ and the reference metric $f_{\mu\nu}$, respectively, $M_g$ and $M_f$ are the respective Planck masses and $S_{\rm matter}(g,\Phi)$ denotes the matter action. We assume the matter fields $\Phi$ to be coupled to $g$ only. The potential $U(f,g)$ is given by
\be \label{e:pot3}
U(f,g) = 2 m^2\left[\beta_0 + \beta_1U_1(\mathbb{X})+ \beta_2U_2(\mathbb{X}) 
  + \beta_3U_3(\mathbb{X}) + \beta_4U_4(\mathbb{X}) \right] \,,  
\ee
where $\mathbb{X}\equiv \sqrt{ g^{-1}f}$, $\beta_i$ ($0\leq i \leq 4$) are arbitrary coefficients and 
\bea
U_1(\mathbb{X}) &=& [\mathbb{X}] \, , \label{e:U1} \\ 
U_2(\mathbb{X}) &=&  \frac{1}{2}\left([\mathbb{X}]^2-[\mathbb{X}^2]\right) \, ,\\
U_3(\mathbb{X}) &=&  \frac{1}{6}\left([\mathbb{X}]^3- 3[\mathbb{X}][\mathbb{X}^2] +2[\mathbb{X}^3]\right)\,,\\
U_4(\mathbb{X}) &=&  \frac{1}{24}\Big([\mathbb{X}]^4-6[\mathbb{X}]^2[\mathbb{X}^2] +3[\mathbb{X}^2]^2 + 8[\mathbb{X}][\mathbb{X}^3] -6[\mathbb{X}^4]\Big)=\det(\mathbb{X})\,.  \label{e:U4}
 \eea
Here we use the notation $[\mathbb{X}] =\tr \mathbb{X} = {\mathbb{X}^\mu}_\mu$,  $[\mathbb{X}^2] =\tr \mathbb{X}^2 = {\mathbb{X}^\mu}_\nu {\mathbb{X}^\nu}_\mu$ and so forth.

Varying this action with respect to the metrics $g$ and $f$ one obtains the equations of motion
\bea
G_{\mu\nu}(g) &=& 8\pi G T_{\mu\nu} + \sqrt{-g^{-1}}\frac{\dd\(\sqrt{-g}U(f,g)\)}{\dd g^{\mu\nu}}\,,\\
G_{\mu\nu}(f) &=&  \sqrt{-f^{-1}}\frac{\dd\(\sqrt{-g}U(f,g)\)}{\dd f^{\mu\nu}}\,.
\eea
Here $T_{\mu\nu}$ is the energy momentum tensor obtained by varying the matter action with respect to the metric $g^{\mu\nu}$, whereas $G_{\mu\nu}(g)$ and $G_{\mu\nu}(f)$ are the Einstein tensors for the metrics $g_{\mu\nu}$ and $f_{\mu\nu}$, respectively.

The Bianchi identities for the two metrics together with energy momentum conservation imply
\be \label{e:bianchi}
  \nabla_f^\nu\left(\frac{\dd(\sqrt{-g}U(f,g))}{\dd f^{\mu\nu}}\right)=
  \nabla_g^\nu\left(\frac{\dd(\sqrt{-g}U(f,g))}{\dd g^{\mu\nu}}\right) =0 \,,
\ee
where $ \nabla_f$ and  $ \nabla_g$ denotes the covariant derivatives w.r.t the metrics $f$ and $g$ respectively.
Equations (\ref{e:bianchi}) are called the Bianchi constraints for $f$ and $g$, respectively. One can show that both Bianchi constraints are equivalent.

Furthermore, using the fact that $\sqrt{ g^{-1}f}$ can be written in triangular form, it is easy to verify that
\begin{subequations} \label{eq:t_i}
\begin{align}
& t_1 \equiv U_1(\sqrt{ g^{-1}f}) = \sum_i \la_i^{1/2} \, , \\
& t_2 \equiv U_2(\sqrt{ g^{-1}f}) = \sum_{i<k} \la_i^{1/2} \la_k^{1/2} \, , \\
& t_3 \equiv U_3(\sqrt{ g^{-1}f}) = \sum_{i<k<l} \la_i^{1/2} \la_k^{1/2} \la_l^{1/2} \, , \\
& t_4 \equiv U_4(\sqrt{ g^{-1}f}) = \sqrt{\la_1\la_2\la_3\la_4} \, ,
\end{align}
\end{subequations}
where $\la_i$ are the eigenvalues of $ g^{-1}f$, and $1\le i,k,l\le 4$. In terms of the $t_i$, the potential defined in eq.~(\ref{e:pot3}) can be written as
\be\label{e:pot4}
U(f,g) = 2 m^2\left[\beta_0 + \beta_1t_1+ \beta_2t_2 + \beta_3t_3 + \beta_4t_4\right]\,.
\ee

We also introduce the quantities corresponding to the $t_i$ for the metric $g^{-1}f$ without  square root. They are easier to handle and will be used extensively in the rest of this work:
\begin{subequations} \label{eq:s_i}
\begin{align}
& s_1 \equiv U_1({ g^{-1}f}) = \sum_i \la_i \, , \\
& s_2 \equiv U_2({ g^{-1}f}) = \sum_{i<j} \la_i \la_j \, , \\
& s_3 \equiv U_3({ g^{-1}f}) = \sum_{i<j<k} \la_i \la_j \la_k \, , \\
& s_4 \equiv U_4({ g^{-1}f}) = \la_1\la_2\la_3\la_4 \, .
\end{align}
\end{subequations}
The relations between the $t_i$ and the $s_i$ defined in eq. (\ref{eq:t_i}) and eq. (\ref{eq:s_i}) respectively are, see also~\cite{D'Amico:2012pi}:
\begin{subequations} \label{eq:relations}
\begin{align}
t_1^2 &= s_1 + 2 t_2 \, , \\
t_2^2 &= s_2 - 2 \sqrt{s_4} + 2 t_1 t_3 \, , \\
t_3^2 &= s_3 + 2 t_2 \sqrt{s_4} \, , \\
t_4^2 &= s_4  \, .
\end{align}
\end{subequations}

These relations~(\ref{eq:relations}) are easily derived from eqs.~(\ref{eq:t_i}) and~(\ref{eq:s_i}), by using the fact that every matrix can be written in triangular form. In any way, 
since both sides of~(\ref{eq:relations}) involve only the coordinate independent quantities $t_i$ and $s_i$ they are of course generally true. This has also been shown explicitly in 
Ref.~\cite{Bernard:2015mkk}. 

\section{Metric perturbations}\label{Mmunu}

We now expand the action for bigravity  to second order in the metric perturbations, around generic background solutions $\bar g_{\mu\nu}$ and $\bar f_{\mu\nu}$. The perturbed metrics are defined as
\bea
g_{\mu\nu} &=& \bar g_{\mu\nu}+h_{\mu\nu}\,, \\ 
f_{\mu\nu} &=& \bar f_{\mu\nu}+\lc_{\mu\nu}\,.
\eea
From now on,  the indices of the tensor $h_{\mu\nu}$ will be raised and lowered with the physical background metric $\bar g_{\mu\nu}$, whereas the indices of the tensor $\lc_{\mu\nu}$ will be raised and lowered with the background metric $\bar f_{\mu\nu}$.

\subsection{Kinetic terms for $h_{\mu\nu}$ and for $\lc_{\mu\nu}$}

The kinetic term for $h_{\mu\nu}$ is given by the Lichnerowicz operator, $\EE^{\mu\nu\al\beta}(\bar g)$, in curved spacetime
\bea \label{e:RgE1}
\sqrt{\!-\det \!g}\,R(g)\! &=&\! \sqrt{\!-\det\bar g}\Big[R(\bar g)\!  - \!h_{\mu\nu}G^{\mu\nu}(\bar g)+ \,h_{\mu\nu}\EE^{\mu\nu\al\beta}(\bar g)h_{\al\beta}   +\nabla_\mu V_g^\mu \Big]+\OO(h^3)\,,
\eea
with
\begin{align}
\EE^{\mu\nu\al\beta}(\bar g) =&\frac{1}{4}\Big[\left(\bar g^{\mu\al}\bar g^{\nu\beta}-\bar g^{\mu\nu}\bar g^{\al\beta}\right)\Box+\left(\bar g^{\mu\nu}\bar g^{\al\rho}\bar g^{\beta\si}
+ \bar g^{\al\beta}\bar g^{\mu\rho}\bar g^{\nu\si} - \bar g^{\mu\beta}\bar g^{\nu\rho}\bar g^{\al\si}- \bar g^{\al\nu}\bar g^{\beta\rho}\bar g^{\mu\si}\right)\nabla_\rho\nabla_\si\Big] +\nn\\
&-\frac{R(\bar{g})}{8}\left(\bar{g}^{\mu\al}\bar{g}^{\nu\beta}+\bar{g}^{\nu\al}\bar{g}^{\mu\beta}-\bar{g}^{\mu\nu}\bar{g}^{\alpha\beta}\right)-\frac{1}{4}\left(\bar{g}^{\mu\nu}\,R^{\alpha\beta}(\bar{g})+\bar{g}^{\al\beta}\,R^{\mu\nu}(\bar{g})\right)+\nn\\
&+\frac{1}{4}\left(\bar{g}^{\mu\al}R^{\beta\nu}(\bar{g})+\bar{g}^{\mu\beta}R^{\al\nu}(\bar{g})+\bar{g}^{\nu\al}R^{\beta\mu}(\bar{g})+\bar{g}^{\nu\beta}R^{\alpha \mu}(\bar{g})\right) \,.
\label{e:EEg}
\end{align}
Here the covariant derivatives are taken with respect to the background metric $\bar g$ and $\Box =\bar g^{\rho\si}\nabla_\rho\nabla_\si$ is the d'Alembertian operator.
The kinetic term in square brackets in (\ref{e:EEg}) is  the curved spacetime version of the Lichnerowicz operator on flat space, see e.g.~\cite{deRham:2010ik}, and the terms proportional to $R(\bar{g})$ and $R^{\si\rho}$ contribute to the potential for $h_{\mu\nu}$ which vanishes on a flat background. This contribution has the form of a mass term (since it is quadratic in $h_{\mu\nu}$), which depends on the background solution.  While probably not new, we have not found this general form of the Lichnerowicz operator in the literature\footnote{The form given in eq.~(2.12) of Ref.~\cite{Schmidt-May:2015vnx} is equivalent but it does not look the same since it  uses $h^{\mu\nu}$ instead of $h_{\mu\nu}$ as independent variable.}. An expression for Einstein spaces, for which $R^{\mu\nu}(\bar g)=\bar{g}^{\mu\nu}R(\bar g)/4$ is given in Ref.~\cite{Hinterbichler:2011tt}, but also this does not quite agree with eq.~(\ref{e:EEg}), since it is written with a different index position like Ref.~\cite{Schmidt-May:2015vnx}.  Note also that these references consider a mass as well defined only for Einstein spaces, since in more general backgrounds one no longer has a local Poincar\'e symmetry and hence the mass as a Casimir of the Poincar\'e group is not well defined. We take here a more naive point of view and just call 'mass term' the term quadratic in the perturbations $h_{\mu\nu}$ which does not contain any derivatives,  by analogy to the scalar field case.
% which does not satisfy the Fierz-Pauli tuning, however, it is usually not harmful. \textcolor{red}{GC: I do not understand what harmful means}

$G^{\mu\nu}(\bar g)$ in eq.~(\ref{e:RgE1}) is the Einstein tensor which solves the background equations of motion for $g_{\mu\nu}$ while the total derivative $\nabla_\mu V_g^\mu$ is irrelevant for the equations of motion. (Such a term can actually be removed on the level of the action by adding a suitable Gibbons-Hawking boundary term.)

The kinetic term for $\lc_{\mu\nu}$ is  analogous to the one for $h_{\mu\nu}$,
\bea \label{e:RgE2}
\sqrt{\!-\det \!f}R(f)\! &=&\! \sqrt{\!-\det\bar f}\Big[R(\bar f)\!  - \!\lc_{\mu\nu}G^{\mu\nu}(\bar f) + \lc_{\mu\nu}\EE^{\mu\nu\al\beta}(\bar f)\lc_{\al\beta}   +\nabla_\mu V_f^\mu \Big]+\OO(\lc^3)\,.
\eea
In eq.~(\ref{e:RgE2}), covariant derivatives are taken with respect to the background metric $\bar f$ and $\Box =\bar f^{\rho\si}\nabla_\rho\nabla_\si$ is the d'Alembertian operator. $\EE^{\mu\nu\al\beta}(\bar f)$ is the curved spacetime version of the Lichnerowicz operator, given by the analogous of eq. (\ref{e:EEg}) for the $\bar{f}_{\mu\nu}$ background, 
$G^{\mu\nu}(\bar f)$ is the Einstein tensor which solves the background equations of motion for $f_{\mu\nu}$ while the total derivative $\nabla_\mu V_f^\mu$ is irrelevant for the equations of motion.

\subsection{The perturbed mass term}

Making use of the definitions~(\ref{eq:t_i}) and~(\ref{eq:s_i}) and of the relations~(\ref{eq:relations}), it is possible to write the perturbations of $t_i$ in terms of perturbations of $s_i$, which in turn can be obtained from the variation of $g^{-1}f\equiv g^{\mu\rho}f_{\rho\nu}$.  We keep only terms up second order in the metric perturbations. Therefore, for example, the fundamental quantity $g^{-1}f$ is expanded as
\be
g^{\mu\rho}f_{\rho\nu}=(\delta^{\mu}_{\alpha}-h^{\mu}_{\alpha}+h^{\mu}_{\gamma}h^{\gamma}_{\alpha})\bar g^{\alpha\rho}\bar f_{\rho\beta}(\delta^{\beta}_{\nu}+\lc^{\beta}_{\nu}) + \OO(h^3)\,.
\ee

%We want  to determine the second-order perturbation of the potential. 
Up to second order in the perturbations $h_{\mu\nu}$ and $\lc_{\mu\nu}$, the potential can be written as\footnote{We will always denote the indices of $h$ with the letters $\mu\nu$ and the indices of $\lc$ with the letters $\al\beta$ in the mixed term, $\MM^{\mu\nu\al\beta}_{\gc\fc}(\bar f,\bar g)h_{\mu\nu}\lc_{\al\beta}$.}
\begin{align}
\sqrt{-\det g}\,\,U(f,g) =& \sqrt{-\det\bar g}\,\Big[U(\bar f,\bar g)  +\MM^{\mu\nu}_{\gc}(\bar f, \bar g)h_{\mu\nu} +\MM^{\mu\nu}_{\fc}(\bar f, \bar g)\lc_{\mu\nu} +  \nn \\ 
 &+ \MM^{\mu\nu\al\beta}_{\gc\gc}(\bar f,\bar g)h_{\mu\nu}h_{\al\beta} + \MM^{\mu\nu\al\beta}_{\gc\fc}(\bar f,\bar g)h_{\mu\nu}\lc_{\al\beta} + \MM^{\mu\nu\al\beta}_{\fc\fc}(\bar f,\bar g)\lc_{\mu\nu}\lc_{\al\beta}\Big] \,,
\end{align}
where
\bea \label{e:Mmunu0}
\MM^{\mu\nu}_{\gc}(\bar f,\bar g) &\equiv & \frac{1}{\sqrt{-\det g}}\frac{\partial(\sqrt{-\det g}\,\,U(f,g))}{\partial g_{\mu\nu}}\bigg|_{g=\bar{g}, f=\bar{f}}\,, \\
\MM^{\mu\nu}_{\fc}(\bar f,\bar g) &\equiv &\frac{1}{\sqrt{-\det g}}\frac{\partial(\sqrt{-\det g}\,\, U(f,g))}{\partial f_{\mu\nu}}\bigg|_{g=\bar{g}, f=\bar{f}}\,,\\ 
\MM^{\mu\nu\al\beta}_{\gc\gc}(\bar f,\bar g) &\equiv & \frac{1}{2}\frac{1}{\sqrt{-\det g}}\frac{\partial^{2}(\sqrt{-\det g}\,\, U(f,g))}{\partial g_{\mu\nu}\partial g_{\al\beta}}\bigg|_{g=\bar{g}, f=\bar{f}}\,, \\
\MM^{\mu\nu\al\beta}_{\gc\fc}(\bar f,\bar g) &\equiv & \frac{1}{\sqrt{-\det g}}\frac{\partial^{2}(\sqrt{-\det g}\,\, U(f,g))}{\partial g_{\mu\nu}\partial f_{\al\beta}}\bigg|_{g=\bar{g}, f=\bar{f}}\,, \\
\MM^{\mu\nu\al\beta}_{\fc\fc}(\bar f,\bar g) &\equiv & \frac{1}{2}\frac{1}{\sqrt{-\det g}}\frac{\partial^{2}(\sqrt{-\det g} \,\,U(f,g))}{\partial f_{\mu\nu}\partial f_{\al\beta}}\bigg|_{g=\bar{g}, f=\bar{f}}\,.\label{e:Mmunuf}
\eea

Since we are considering perturbations around solutions of the background equations of motion, the terms linear in $h_{\mu\nu}$ and in $\lc_{\mu\nu}$ in the Lagrangian cancel on shell and in our discussion they can be omitted.

The explicit calculation of the matrix elements (\ref{e:Mmunu0})-(\ref{e:Mmunuf}) is cumbersome but rather straightforward. We present here only the final result in a compact form, collecting the details of the calculation and the open expressions of the final result in Appendix~\ref{a:comp}. We find that the mass term expanded to quadratic order in the perturbation variables can be written as 
\be\label{e:action_pert}
 \sqrt{-\det g}\,\,U(f,g) = \sqrt{-\det\bar g}\,\biggl[U(\bar f,\bar g) +  \MM^{\mu\nu\al\beta}_{\gc\gc}(\bar f,\bar g)h_{\mu\nu}h_{\al\beta} + \MM^{\mu\nu\al\beta}_{\gc\fc}(\bar f,\bar g)h_{\mu\nu}\lc_{\al\beta} + \MM^{\mu\nu\al\beta}_{\fc\fc}(\bar f,\bar g)\lc_{\mu\nu}\lc_{\al\beta} \biggr]\,,
 \ee
 with
 \bea 
\MM^{\mu\nu\al\beta}_{\gc\gc} &=&  m^2\left[\beta_0\MM_{0,\gc\gc}^{\mu\nu\al\beta} +\beta_1\MM_{1,\gc\gc}^{\mu\nu\al\beta} +\beta_2\MM_{2,\gc\gc}^{\mu\nu\al\beta} +\beta_3\MM_{3,\gc\gc}^{\mu\nu\al\beta} \right] \,, \label{e:MMhh}\\
\MM^{\mu\nu\al\beta}_{\gc\fc} &=&  m^2\left[\beta_1\MM_{1,\gc\fc}^{\mu\nu\al\beta} +\beta_2\MM_{2,\gc\fc}^{\mu\nu\al\beta} +\beta_3\MM_{3,\gc\fc}^{\mu\nu\al\beta} +\beta_4\MM_{4,\gc\fc}^{\mu\nu\al\beta} \right] \,, \label{e:MMhl}\\
\MM^{\mu\nu\al\beta}_{\fc\fc} &=&  m^2\left[\beta_1\MM_{1,\fc\fc}^{\mu\nu\al\beta} +\beta_2\MM_{2,\fc\fc}^{\mu\nu\al\beta} +\beta_3\MM_{3,\fc\fc}^{\mu\nu\al\beta} +\beta_4\MM_{4,\fc\fc}^{\mu\nu\al\beta} \right] \,, \label{e:MMll1}\\
\MM_{0,\gc\gc}^{\mu\nu\al\beta} &=&  \frac{1}{\sqrt{-\det g}}\frac{\dd^2 \sqrt{-\det g}}{\dd g_{\mu\nu}\dd g_{\al\beta}} = \frac{1}{4}\bar g^{\mu\nu}\bar g^{\al\beta} -\frac{1}{4}\biggl(\bar g^{\mu\al}\bar g^{\nu\beta}+\bar g^{\mu\beta}\bar g^{\nu\al}\biggr)\,, \label{e:MMhh0}\\
\MM_{i,\gc\gc}^{\mu\nu\al\beta} &=& \bar t_i\MM_{0,\gc\gc}^{\mu\nu\al\beta} +\frac{1}{2}\left(\bar g^{\mu\nu} t_{i,\gc}^{\al\beta}+\bar g^{\al\beta} t_{i,\gc}^{\mu\nu}\right) + 2 t_{i,\gc\gc}^{\mu\nu \al\beta}\,,\label{e:MMhhi} \\
\MM_{i,\gc\fc}^{\mu\nu\al\beta} &=& \bar g^{\mu\nu} t_{i,\fc}^{\al\beta} + 2 t_{i,\gc\fc}^{\mu\nu \al\beta}\,, \\     \label{e:MMhli}
\MM_{i,\fc\fc}^{\mu\nu\al\beta} &=& 2 t_{i,\fc\fc}^{\mu\nu \al\beta}\label{e:MMll}\,, 
\eea
where $i=1,2,3,4$ and to shorten the notation we have defined the following derivatives which are calculated explicitly in  Appendix~\ref{a:comp}:
\bea\label{vart}
 && t_{i,\gc}^{\mu\nu} = \left.\frac{\dd t_i}{\dd g_{\mu\nu}}\right|_{g=\bar g,f=\bar f}\;, \qquad
   t_{i,\fc}^{\mu\nu} = \left.\frac{\dd t_i}{\dd f_{\mu\nu}}\right|_{g=\bar g,f=\bar f}\;, \qquad 
  t_{i,\gc\gc}^{\mu\nu\al\beta}~ = ~\left.\frac{1}{2}\frac{\dd^2 t_i}{\dd g_{\mu\nu}\dd g_{\al\beta}}\right|_{g=\bar g,f=\bar f} \,,\\
 && t_{i,\gc\fc}^{\mu\nu\al\beta}~ = ~\left.\frac{\dd^2 t_i}{\dd g_{\mu\nu}\dd f_{\al\beta}}\right|_{g=\bar g,f=\bar f}\;, \qquad
 t_{i,\fc\fc}^{\mu\nu\al\beta}~ = ~\left.\frac{1}{2}\frac{\dd^2 t_i}{\dd f_{\mu\nu}\dd f_{\al\beta}}\right|_{g=\bar g,f=\bar f}\, . \label{vartfin}
\eea
All these objects are computed explicitly in terms of the background metrics in Appendix~\ref{a:comp}. It is easy to check that this potential correctly reduces to the massive gravity one calculated in~\cite{Guarato:2013gba}, once the limits $\lc_{\mu\nu}\rightarrow0$ and $f_{\mu\nu}\rightarrow\bar f_{\mu\nu}$ are taken.  Furthermore, one can check that 
\be\label{e:sym}
\sqrt{-\det \bar g}\,\MM_{i,\fc\fc}^{\mu\nu\al\beta}(\bar f,\bar g)  = \sqrt{-\det \bar f}\,\MM_{4-i,\gc\gc}^{\mu\nu\al\beta}(\bar g,\bar f) \quad \mbox{ and } \quad
\sqrt{-\det \bar g}\,\MM_{i,\gc\gc}^{\mu\nu\al\beta}(\bar f,\bar g)  = \sqrt{-\det \bar f}\,\MM_{4-i,\fc\fc}^{\mu\nu\al\beta}(\bar g,\bar f) \,.
\ee
This is a consequence of the fact that for $M_g=M_f$, the gravitational action (\ref{e:action}) is invariant under the simultaneous exchange $f\leftrightarrow g$ and $\beta_i\ra \beta_{4-i}$.

Using the results above, we can write the general expression for the perturbed action, quadratic in the metric perturbations  $h_{\mu\nu}$ and $\ell_{\mu\nu}$
\begin{align}\label{eqcov}
S^{(2)} =&\frac{M_g^2}{2}\int d^4x \,\sqrt{-\det \bar g}\, h_{\mu\nu} \mathcal{E}^{\mu\nu\alpha\beta}(\bar{g})h_{\alpha\beta}+\frac{M_f^2}{2}\int d^4x\,\,\sqrt{-\det \bar f}\,\, \lc_{\mu\nu} \mathcal{E}^{\mu\nu\alpha\beta}(\bar{f})\lc_{\alpha\beta}\nn \\
-&\frac{M_g^2}{2}\int d^4x\,\sqrt{-\det \bar g}\,\Big[\MM^{\mu\nu\alpha\beta}_{\gc\gc}(\bar f,\bar g)h_{\mu\nu}h_{\alpha\beta} + \MM^{\mu\nu\alpha\beta}_{\gc\fc}(\bar f,\bar g)h_{\mu\nu}\lc_{\alpha\beta} +\MM^{\mu\nu\alpha\beta}_{\fc\fc}(\bar f,\bar g)\lc_{\mu\nu}\lc_{\alpha\beta}\Big] \,.
\end{align}

Eq.~(\ref{eqcov}) together with the expressions for the mass terms $\MM^{\mu\nu\alpha\beta}_{\bullet\bullet}$ in Appendix~A and the Einstein operator given in eq.~(\ref{e:EEg}) are the main result of our paper. It allows to write down the perturbation equations of bimetric massive gravity on an arbitrary background. In the following we  apply it to cosmology.

\section{Mass term on cosmological backgrounds}\label{s4}

In this and the following sections, we specify to background solutions where both the metrics exhibit spatial isotropy and homogeneity. For simplicity, we assume that both the metrics have a flat spatial section, $K=0$. Then, modulo time re-parameterizations,  the most general form for the metrics (in conformal time $\tau$) is
\be\label{FRWg}
\bar{g}_{\mu\nu}dx^{\mu}dx^{\nu}=a^2(\tau)\left(-d\tau^2+\delta_{ij}dx^idx^j\right)\,,
\ee
\be\label{FRWf}
\bar{f}_{\mu\nu}dx^{\mu}dx^{\nu}=b^2(\tau)\left(-c^2(\tau) d\tau^2+\delta_{ij}dx^idx^j\right)\,.
\ee
It is convenient to define the conformal Hubble parameter ($\HH$) and the physical one ($H$) for both metrics:
\be
H=\frac{\mathcal{H}}{a}=\frac{a'}{a^2}\,,\hspace{0.5 cm} H_f=\frac{\mathcal{H}_f}{b}=\frac{b'}{b^2\,c}\,,
\ee
where with $'$ we denote the derivative with respect to the conformal time $\tau$. We introduce also the ratio between the two conformal scale factors
\be
r=\frac{b}{a}\,.
\ee
The Friedmann equations can then be written as
\bea
&3 H^2 =8\pi G\,(\rho+\rho_g)\,,\hspace{1 em} &\text{with}\hspace{0.5 em}\rho_g=\frac{m^2}{8\pi G}\left(\beta_{0}+3\beta_{1}r+3\beta_{2}r^{2}+\beta_{3}r^{3}\right)\label{eqF1}\,,\\
&3 H_f^2=8\pi G\,\rho_f\,, &\text{with}\hspace{0.5 em}\rho_f=\frac{m^2}{8\pi G M_*^2}\left(\beta_{4}+3\beta_{3}r^{-1}+3\beta_{2}r^{-2}+\beta_{1}r^{-3}\right)  \label{eqF2}\,,
\eea
\bea
&3H^{2}+2{\dot H}  =  -8\pi G\,\left(p+p_{g}\right)\,, \hspace{0.5 em} &\text{with} \hspace{0.5 em}p_{g}  =- \frac{m^2}{8\pi G} \left(\beta_{0}+\beta_{1}(c+2)r+\beta_{2}\left(2c+1\right)r^{2}+\beta_{3}cr^{3}\right) \label{epg}\\
&3c{H}_{f}^{2} + 2{\dot H}_{f} = -8\pi G\,p_{f} \,,  &\text{with} \hspace{0.5 em}
p_{f}  =  -\frac{m^2}{8\pi G M_*^2}\left(c\beta_{4}+\beta_{3}(2c+1)r^{-1}+\beta_{2}(c+2)r^{-2}+\beta_{1}r^{-3}\right)\
.\label{ac2}
\eea
Here $(8\pi G)^{-1/2}=M_P=M_g$ is the physical Planck mass and $M_*$ is the ratio between the two Planck masses, $M_*\equiv M_f/M_g$, while $\rho$ and $p$ are the usual matter energy density and pressure. The quantities $\rho_g,~p_g$ and $\rho_f,~p_f$ play the role of gravitational 'energy densities' and 'pressure'. They come from the mass term.

With the ansatz (\ref{FRWg}, \ref{FRWf}) for the background metrics, homogeneity and isotropy request that the mass tensor in eq. (\ref{e:action_pert}) admit the following general parametrization. For the $\gc\gc$ and $\fc\fc$ terms
\begin{align}
\MM^{0000}_{\bullet \bullet}(\bar f,\bar g)&=m^2a^{-4} \alpha_{\bullet}(\tau)\,,\\
\MM^{ij00}_{\bullet \bullet}(\bar f,\bar g)&= \MM^{00ij}_{\bullet \bullet}(\bar f,\bar g)= m^2a^{-4}  \gamma_{\bullet}(\tau) \delta^{ij}\,,\\
\MM^{i0j0}_{\bullet \bullet}(\bar f,\bar g)&=\MM^{0i0j}_{\bullet \bullet}(\bar f,\bar g)=m^2a^{-4} \epsilon_{\bullet}(\tau) \delta^{ij}\,,\\
\MM^{ijkl}_{\bullet \bullet}(\bar f,\bar g)&=m^2a^{-4}\left\{\eta_{\bullet}(\tau)\delta^{ij}\delta^{kl}+\frac{\sigma_{\bullet}(\tau)}{2}\left(\delta^{ik}\delta^{jl}+\delta^{il}\delta^{jk}\right)\right\}\,.
\end{align}
where $\bullet$ stands for either $\gc$ or $\fc$. And for the mixed terms $\gc\fc$ the parametrization takes the form
\begin{align}
\MM^{0000}_{\gc\fc}(\bar f,\bar g)&=m^2a^{-4} \alpha_{\gc\fc}(\tau)\,,\\
\MM^{ij00}_{\gc\fc}(\bar f,\bar g)&=m^2a^{-4}  \gamma_{\gc\fc}(\tau) \delta^{ij}\,,\\
\MM^{00ij}_{\gc\fc}(\bar f,\bar g)&=m^2a^{-4} \gamma_{\fc\gc}(\tau) \delta^{ij} \,,\\
\MM^{i0j0}_{\gc\fc}(\bar f,\bar g)&=\MM^{0i0j}_{\gc\fc}(\bar f,\bar g)=m^2a^{-4} \epsilon_{\gc\fc}(\tau) \delta^{ij}\,,\\
%\MM^{j0i0}_{hl}(\bar f,\bar g)&=-m^2 \delta^{ij}\epsilon_{lh}(\tau)\,,\\
\MM^{ijkl}_{\gc\fc}(\bar f,\bar g)&=m^2a^{-4}\left\{\eta_{\gc\fc}(\tau)\delta^{ij}\delta^{kl}+\frac{\sigma_{\gc\fc}(\tau)}{2}\left(\delta^{ik}\delta^{jl}+\delta^{il}\delta^{jk}\right)\right\}\,.
\end{align}

The functions $\alpha_{\bullet}$, $\gamma_{\bullet}$, $\epsilon_{\bullet}$, $\sigma_{\bullet}$ and $\eta_{\bullet}$ (with $\bullet=\gc, \fc, \gc\fc$ or $\fc\gc$) depend on conformal time through the ratio between the two scale factors, $r$, and the lapse function $c$.  Their explicit expressions are  given in Appendix~\ref{FRW}. Note that contrary to $\gc\gc$ and $\fc\fc$, $\MM^{ij00}_{\gc\fc}\neq \MM^{00ij}_{\gc\fc}$ and we have introduced  $\gamma_{\gc\fc}\neq\gamma_{\fc\gc}$.

Given this parametrization it is straightforward to write the mass term for any type of perturbations on a cosmological background, 

\begin{equation}\label{general_mass_ term_FRW}
\mathcal{S}_{m}^{(2)}=-\frac{M_g^2}{2}\int d^4x\,\left[\mathcal{L}_{\gc\gc}^{(2)}+\mathcal{L}_{\fc\fc}^{(2)}+\mathcal{L}_{\gc\fc}^{(2)}\right],
\end{equation}

\begin{align}
\mathcal{L}_{\gc\gc}^{(2)} & = m^{2}\left[\alpha_{\gc}h_{00}^2+\gamma_{\gc}h_{00}h_{ij}\delta^{ij}+2\epsilon_{\gc}h_{0i}h_{0j}\delta^{ij}+\eta_{\gc}h_{ij}h_{kl}\delta^{ij}\delta^{kl}+\frac{\sigma_{\gc}}{2}h_{ij}h_{kl}\left(\delta^{ik}\delta^{jl}+\delta^{il}\delta^{jk}\right)\right] \\
\mathcal{L}_{\fc\fc}^{(2)}& = m^{2}\left[\alpha_{\fc}\lc_{00}^2+\gamma_{\fc}\lc_{00}\lc_{ij}\delta^{ij}+2\epsilon_{\fc}\lc_{0i}\lc_{0j}\delta^{ij}+\eta_{\fc}\lc_{ij}\lc_{kl}\delta^{ij}\delta^{kl}+\frac{\sigma_{\fc}}{2}\lc_{ij}\lc_{kl}\left(\delta^{ik}\delta^{jl}+\delta^{il}\delta^{jk}\right)\right]\\
\mathcal{L}_{\gc\fc}^{(2)}& = m^{2}\left[\alpha_{\gc\fc}h_{00}\lc_{00}+\gamma_{\fc\gc}h_{00}\lc_{ij}\delta^{ij}+\gamma_{\gc\fc}\lc_{00}h_{ij}\delta^{ij}+2\epsilon_{\gc\fc}h_{0i}\delta^{ij}\lc_{0j}+\eta_{\gc\fc}h_{ij}\lc_{kl}\delta^{ij}\delta^{kl}\right.\\
& +\left.\frac{\sigma_{gf}}{2}h_{ij}\lc_{kl}\left(\delta^{ik}\delta^{jl}+\delta^{il}\delta^{jk}\right)\right]\nn
\end{align}

Even though this parametrization contains 16 functions of the background, these functions can be expressed solely in terms of two combinations of the $\beta_i$ parameters, which we denote as $\sigma_1$ and $\sigma_2$ together with the 'densities' and 'pressures' $\rho_g$, $\rho_f$ and $p_g,~p_f$. This is because one can use the Friedmann equations to simplify the mass term by combining it with contributions from the Einstein-Hilbert and the matter action. We present  this in detail in  Appendix~\ref{FRW}.

\section{Cosmological application: Algebraic branch}\label{backkk}

As an illustration of our formalism, we now apply our general result for the mass term to a specific cosmological setting, the algebraic branch of solutions of bigravity with matter minimally coupled to the $g$ metric.

The Bianchi constraint~(\ref{e:bianchi}) for the cosmological ansatz takes the form
\be\label{Bianchic}
m^2\left(\beta_1+2\beta_2\,r+\beta_3\,r^2\right)\,(\HH-\HH_f)=0\,.
\ee
We distinguish two branches of solutions, depending on how the Bianchi constraint (\ref{Bianchic}) is solved. Either there is an algebraic constraint for $r$
\be\label{Bianchic1}
\mbox{ Branch I}\qquad\qquad \left(\beta_1+2\beta_2\,r+\beta_3\,r^2\right)=0\,,
\ee
or 
\be
\mbox{ Branch II}\hspace{3.84cm} \mathcal{H}_f=\mathcal{H}\,.
\ee
At the background level the first branch, also called \emph{algebraic} branch, is equivalent to GR with an effective cosmological constant, while the solution II gives rise to a richer cosmology.

Since branch~II has been studied in detail in Refs.~\cite{Lagos:2014lca,Cusin:2014psa,Johnson:2015tfa,Cusin:2015pya}, here we
 consider branch~I.  We shall show, that the claim which can be found in the literature~\cite{vonStrauss:2011mq,Lagos:2014lca}, that in this branch not only the background but also the perturbations are identical to $\La$CDM is not strictly true, but that tensor perturbations, i.e. gravitational waves in this branch actually develop a tachyonic instability at late time. The Bianchi constraint in branch~I is realized by setting $r=\bar{r}=$const. such that
\be\label{Bianchi}
\beta_3\,\bar{r}^2+2\beta_2\,\bar{r}+\beta_1=0\,,
\ee
hence 
\be\label{63}
H_f=\frac{1}{\bar{r}c}\,H\,.
\ee
Solving the Bianchi constraint, eq. (\ref{Bianchi}),  $\bar{r}$ can be expressed as a functions of the $\beta_i$
\be\label{r}
\bar{r}=\frac{-\beta_2\pm \sqrt{\beta_2^2-\beta_1\beta_3}}{\beta_3}\,, \qquad \mbox{for $\beta_3\neq 0$ or,  if } \beta_3=0, \quad \bar r=-\frac{\beta_1}{2\beta_2}.
\ee

In this case, the gravitational 'energy densities' in eqs. (\ref{eqF1}) and (\ref{eqF2}) become constant,
\be\label{Laeff}
\La_{\rm eff}= 8\pi G\rho_g(\bar r) \,, \qquad \La_c = 8\pi G\rho_f(\bar r) \,.
\ee
It is easy to check that  (independently of the value of $c$), eq.~(\ref{Bianchi}) implies $p_g(\bar r) = -\rho_g(\bar r)$ and the physical sector at the background level is equivalent to GR with an effective cosmological constant. 

From the Friedmann equation (\ref{eqF2}), we can extract the lapse function $c$, with the  result\footnote{In the decoupling limit $M_*^2\rightarrow \infty$,  eq. (\ref{eqF2}) implies that $H_f^2\rightarrow0$ and, since $r$ is a constant, from eq. (\ref{63}) it follows $H^2/c^2\rightarrow 0$ which is a consequence of $c\ra\infty$. The Friedmann eq.~(\ref{eqF1}) is not affected by the decoupling limit.}
\be
c^2=\frac{8\pi G\rho+\Lambda_{\rm eff}}{\bar r^2\Lambda_c}  =\frac{H^2}{\bar r^2H_f^2}\propto H^2\,.
\ee

In order to have a viable background evolution, we have to impose conditions on the parameters  such that $\Lambda_{\rm eff}\geq 0$ and $c^2\geq0$, i.e., respectively 
\be \label{cond1}
\beta_{0}+3\beta_{1}r+3\beta_{2}r^{2}+\beta_{3}r^{3}\geq 0\,,\hspace{1.5 em}\text{and}\hspace{1.5 em}
\beta_{4}+3\beta_{3}r^{-1}+3\beta_{2}r^{-2}+\beta_{1}r^{-3}\geq 0 \,, 
\ee
and we have to request that $\bar{r}$ given by eq. (\ref{r}) is non-vanishing
\be\label{cond3}
\frac{-\beta_2\pm \sqrt{\beta_2^2-\beta_1\beta_3}}{\beta_3}\neq 0\,,
\hspace{1.5 em}\text{or if } \quad \beta_3=0
\hspace{1.5 em} \frac{\beta_1}{2\beta_2}\neq 0\,.
\ee

We want now to find a minimal model (i.e. with the minimal number of non-vanishing parameters) in this branch satisfying the conditions (\ref{cond1}) and (\ref{cond3}) above. We restrict to the case $\beta_0=0$, since this parameter simply represents a \emph{traditional} cosmological constant, and we know that the model $\beta_0=\La/m^2$, $\beta_i=0$ for $i\neq 0$ works well, it is simply the standard $\La$CDM cosmology which we do not want to investigate here. Single-parameter models are not viable since they give $\bar{r}=0$ or $\bar{r}=\infty$. (The case where only $\beta_4$ or $\beta_0$ is non-vanishing leads to no coupling between the two metrics.) It is easy to check that none of the $\beta_3\beta_i$ models satisfy the viability requirements.  Also the $\beta_1\beta_2$ model alone is not feasible since in this case eq.~(\ref{cond1}) requires at the same time $\beta_2<0$ and $\beta_2>0$. All three-parameter models, $\beta_i\beta_j\beta_k$ are viable  if the parameters are chosen to satisfy the following conditions:
\begin{align} 
  &\beta_1\beta_2\beta_4& \bar r= \frac{-\beta_1}{2\beta_2}\,, \quad & 0< -\beta_2<\left(\frac{\beta_1^2\beta_4}{4}\right)^{1/3},\\
 &\beta_1\beta_3\beta_4&\bar r=-\sqrt{\frac{-\beta_1}{\beta_3}}\,, \quad &0\leq \sqrt{-\beta_1\beta_3}\leq\frac{1}{2}\frac{\beta_1\beta_4}{\beta_3}\,,
\qquad\text{or}&\bar r=\sqrt{\frac{-\beta_1}{\beta_3}}\,,\quad & \frac{1}{2}\frac{\beta_1\beta_4}{\beta_3}\leq \sqrt{-\beta_1\beta_3}\leq0\,,\\
&\beta_2\beta_3\beta_4&\bar r=-2 \frac{\beta_2}{\beta_3}\,,\quad&0\leq \frac{3}{4}\frac{\beta_3^2}{\beta_4}\leq \beta_2\,.\label{nostre}
\end{align}

\begin{figure}[ht!]
    \centering
    \subfigure[\label{Hdef}]
      {\includegraphics[scale=0.38]{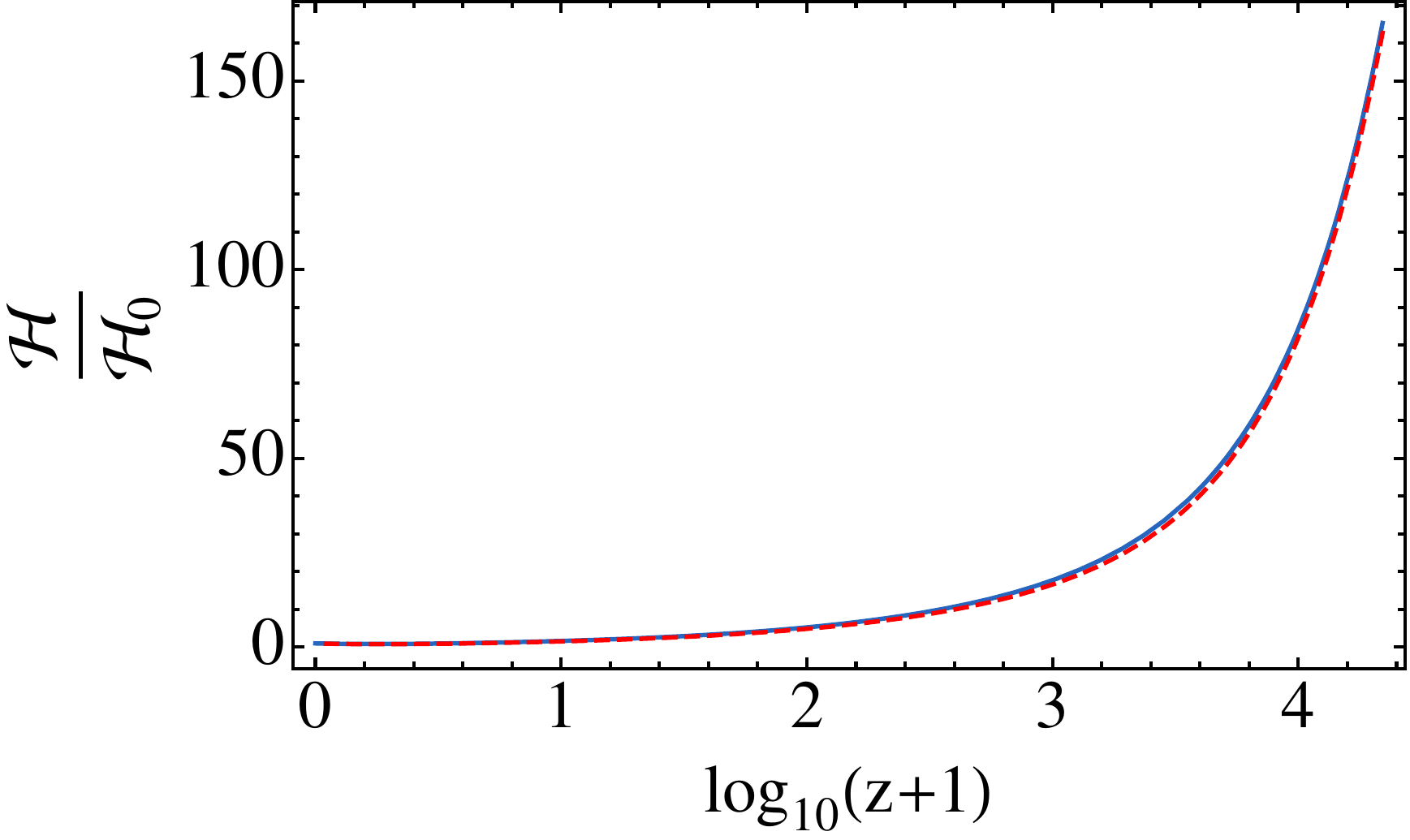}}\qquad
 \subfigure[\label{c}]
      {\includegraphics[scale=0.38]{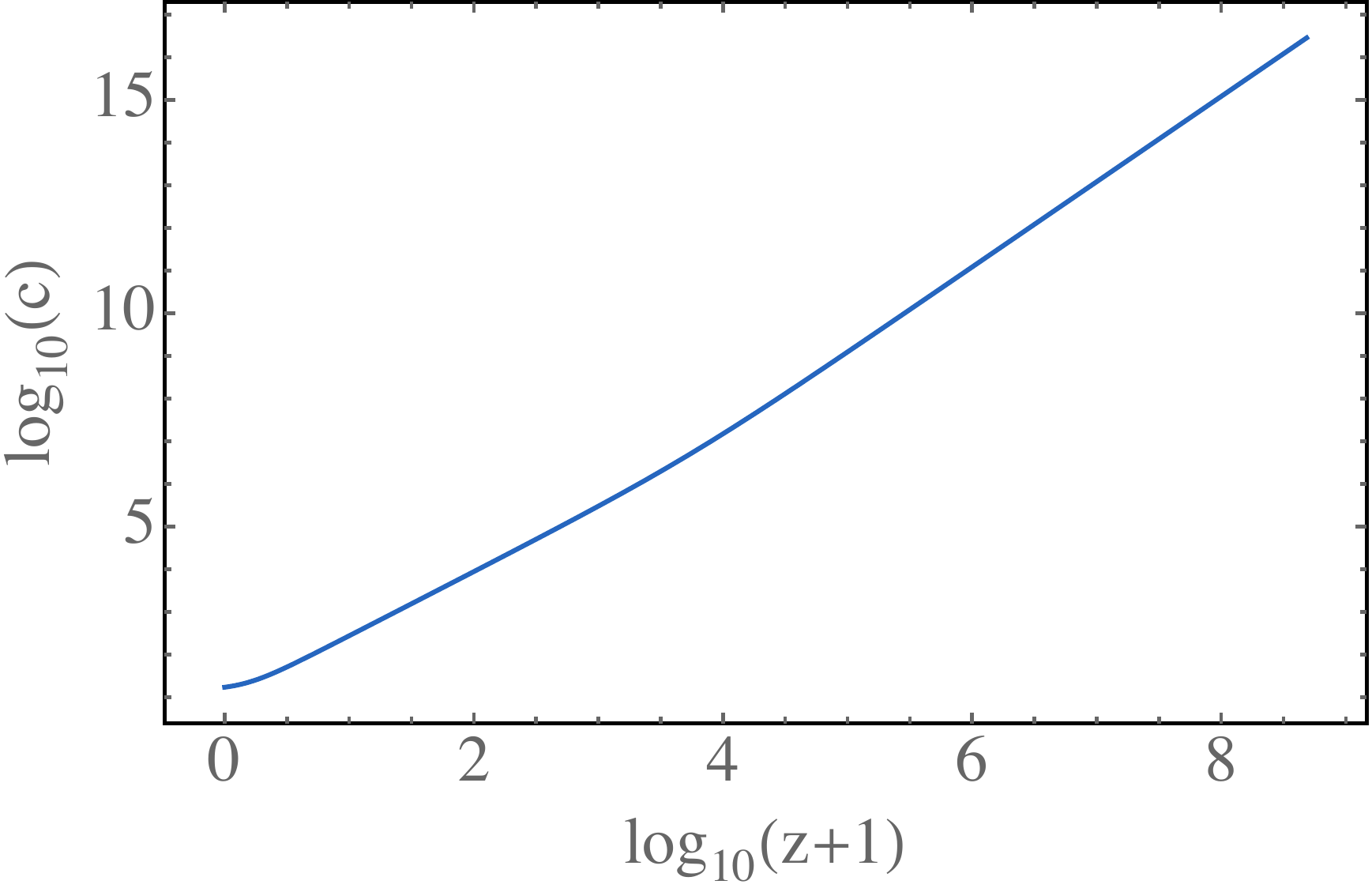}}
\caption{\label{back} Evolution of the lapse function $c$ and of the conformal Hubble parameter in the $\beta_2\beta_3\beta_4$ bigravity model with $m^2\beta_2=0.997\,\mathcal{H}_0^2$\,, $m^2\beta_3=1.408\,\mathcal{H}_0^2$ and $m^2\beta_4=1.496\,\mathcal{H}_0^2$. In Fig \ref{Hdef} the evolution of the conformal Hubble parameter in bigravity is plotted together with the one of  $\Lambda CDM $ with $\Omega_{\Lambda}=0.73$ (in red). The two quantities perfectly agree. }
 \end{figure}

In what follows we set $M_*=1$. This is not a restriction since a finite $M_*$ can always be absorbed in the normalisation of the metric $f_{\mu\nu}$ and the parameters $\beta_i$. The three 'minimal' models above have equations for the background and for the perturbations which are of the same structure, hence we expect them to be qualitatively similar both at the level of the background and of the perturbations. For definitiveness, but no other more physical motivation, in the rest of this work, we  focus on the $\beta_2\beta_3\beta_4$ model. We choose $\beta_i$'s satisfying the inequalities~(\ref{nostre}) and such that $\Omega_{\Lambda}=\frac{\La_{\rm eff}}{3H_0^2}$ has the observed value of about 0.73. We set $\beta_3^2=2\,\beta_2^{3}$ and 
$\beta_4\geq 3/2\,\beta_2$.  With the choice $\beta_4\simeq 3/2\,\beta_2$, we obtain $\Lambda_c\simeq 3/2 \beta_2(1-\beta_2)\,m^2$.  Setting $m^2\beta_2=0.997\,\mathcal{H}_0^2$, this  results in  $\bar{r}\simeq-1.4$.\footnote{With this choice of parameters, at late times $c^2\simeq \Lambda_{\rm eff}/(\bar r^2\Lambda_{c})\gg 1$, hence, as we will see in the equations for tensor perturbations, the coupling between the two propagating tensor modes will be not negligible at late times. }  
With this choice of parameters, the background evolution of the physical quantities is  the one of $\Lambda CDM$ with $\Omega_{\Lambda}=0.73$. In Fig.~\ref{back} we show the evolution of the conformal Hubble parameter and of the lapse $c$ as functions of the redshift. 
   
The equations for the perturbations in this background branch of bigravity are derived in \cite{Comelli:2012db}. We recall that this branch is affected by a strong coupling problem: neglecting matter, from the canonical analysis 7 degrees of freedom are expected (one massive and one massless graviton) but only 4 are accounted for. The helicity zero and helicity 1 modes of the massive graviton are not present and they are probably infinitely strongly coupled, i.e. have vanishing kinetic terms. The same problem of infinite strong coupling appears in the analogous branch in massive gravity. 
%This fact is not very surprising since the condition (\ref{Bianchic1}) requests $a=$constant, hence  a flat background. As a consequence, the extra degrees of freedom are frozen at the linear level. 

Recently, in \cite{DeFelice:2015hla}, a 'minimal' model of massive gravity  has been proposed which  propagates at the fully non linear level the same number of degrees of freedom as linearized dRGT massive gravity in the algebraic branch. In this model,  at every order in the perturbative expansion, only the helicity-2 modes are propagating. In principle it may be possible to extend the formalism developed in \cite{DeFelice:2015hla} to the bigravity case, building a \emph{minimal model of bigravity}: this model would then propagate at every order in the perturbative expansion only 4 degrees of freedom, two belonging to the massless graviton and two the massive one. In particular, 
this model would have the same perturbation equations as the present algebraic bigravity branch, without being affected by the strong coupling problem. % In particular, the tensor sector would propagate $2+ 2$ degrees of freedom. 

In the following we focus on the study of the tensor sector in the algebraic branch, showing how the machinery developed in the first part of this paper can be applied to this case. We recall that the results of this study have to be taken \emph{cum grano salis}, as we do not take into account problems which may be present as a consequence of the infinitely  strong coupling of the helicity 0 and 1 modes. Our study would be realistic if these modes do not affect the evolution of the helicity 2 modes, e.g. because they remain small or because they even vanish as in a \emph{would be minimal model of bigravity}. 

\subsection{New formalism applied to the study of tensor perturbations}

 In the following, we study  tensor perturbations in the algebraic branch, both numerically and analytically. Making use of the formalism developed in this work, we  show that the late time instability that we find in the tensor sector is due to the fact that an eigenvalue of the mass matrix becomes negative during the cosmological evolution. 

 Tensor perturbations of a given $\bk$-mode are composed of two independent helicity modes,
 \be
 h^{TT}_{ij} =  h^+e^{(+2)}_{ij}  + h^{-}e^{(-2)}_{ij}
 \ee
 where $+$ and $-$ denote the two helicity-2 modes of the gravitational wave. For a plane wave with wave vector $\bk$ in the  orthonormal system $(\widehat{\bk},\bfe^{(1)},\bfe^{(2)})$ we have
 \be \bfe^{\pm} =\frac{1}{\sqrt{2}}\left(\bfe^{(1)}\pm i\bfe^{(2)}\right) \quad \mbox{ and }
 \quad e^{(+2)}_{ij} =\bfe^{+}_i\bfe^+_j \,, \quad  e^{(-2)}_{ij} =\bfe^{-}_i\bfe^-_j \,. 
 \ee
 
 For parity invariant perturbations 
 $$\langle h^+(\bk)(h^+(\bk'))^*\rangle =\langle h^-(\bk)(h^-(\bk'))^*\rangle = \delta(\bk-\bk')2\pi^2P_h(k)\,,$$
  and $\langle h^+h^-\rangle =0$.  Here $\langle\cdots\rangle$ denotes a stochastic expectation value. We consider a stochastic ensemble of parity invariant Gaussian gravitational waves. We discuss just one mode, say $ h^+=hG_h $ for the $g$ tensor perturbations and $\lc^+=\lc G_{\lc}$ for the $f$ tensor perturbations. Here $G_a$\,, with $a=h\,,\lc$ are independent Gaussian random variables with vanishing mean and with variance 
  $\langle G_a(\bk)G_b(\bk')\rangle = \de_{ab}\delta(\bk-\bk')2\pi^2$, so that $h, \,\lc$ is the square root of the power spectrum. 
  All what follows is also valid for the modes $h^-,\,\lc^-$ which are not correlated with  $h^+,\, \lc^+$ in the parity symmetric situation which we consider.
 
 In the tensor sector, for the first order modified Einstein equations with a perfect fluid source, i.e. no anisotropic stress, in the algebraic background branch,  from eq. (\ref{eqcov}) we obtain
 \be\label{total action}
S^{(\pm 2)}=\frac{1}{2}M_g^2\int d^4x\,a^2\left\{\left(h'\right)^2+\frac{\bar{r}^2}{c}\left(\lc'\right)^2-k^2 h^2-k^2c\,\bar{r}^2 \lc^2+a^2m^2 \sigma_2 \bar{r} \left(h-\lc\right)^2\right\}\,,
\ee
where 
\be\label{def:D}
m^2 \bar r  \sigma_2=m^2\left[\beta_2 \bar r^2+\beta_1 \bar r+\left(\beta_3 \bar r^3+\beta_2 \bar r^2\right)c(\tau)\right]\equiv m^2\left[C+D\,c(\tau)\right]\,,
\ee 
is the mass term for the helicity 2 modes. Notice that the mass term in the above action is much simpler than the \textit{original} one obtained from the action (\ref{general_mass_ term_FRW}) for tensor perturbations. This is a result of replacing the Friedmann equations, which simplify contributions of the \textit{original} mass term by combining them with contributions from the kinetic and the matter parts of the action, as explained in Appendix~\ref{FRW}.  This simplification is general for the helicity-2 mode in cosmological solutions of bigravity and does not depend on our choice of branch. While the kinetic term is as expected diagonal in $h$ and $\lc$, the mass term acts on $h-\lc$.

Varying this action with respect to $h$ and $\lc$, we find the following equations of motion for the two tensor modes
\be\label{peg}
h^{''}+2\mathcal{H}\,h^{'}+k^2 h+m^2a^2 \sigma_2 \bar r \left(h-\lc\right)=0\,,
\ee
\be\label{pef}
\lc^{''}+\left(2\mathcal{H}-\frac{c'}{c}\right)\,\lc^{'}+c^2 k^2\,\lc-m^2\,\sigma_2 \,\frac{c\, a^2}{\bar r}\, \left(h-\lc\right)=0\,. 
\ee
The kinetic term in eq. (\ref{total action}) is diagonal and, for the background under study, positive definite. The tensor sector is therefore free of (Higuchi) ghost instabilities~\cite{Fasiello:2013woa,Cusin:2014psa}.

 We introduce canonically normalized variables associated with each of the two tensor modes 
\be\label{can}
Q_h\equiv M_g\,a\, h\,,\hspace{2 em}Q_\lc\equiv M_g\,a\, \frac{\bar r}{\sqrt{c}} \lc\,.
\ee
The action (\ref{total action}) for the canonically normalized variables can be written in matrix form as
\be\label{total action 2}
S^{(\pm 2)} =\frac{1}{2}\int d^4 x \left\{
\left(
\begin{array}{cc}
Q_h'\,,&Q_\lc'\\
\end{array}
\right)
\left(
\begin{array}{cc}
1&0\\
0&1\\
\end{array}
\right)
\left(
\begin{array}{c}
Q_h'\\
Q_\lc'\\
\end{array}
\right)-
\left(
\begin{array}{cc}
Q_h\,,&Q_\lc\\
\end{array}
\right)\mathcal{M}
\left(
\begin{array}{c}
Q_h\\
Q_\lc\\
\end{array}
\right)
\right\}\,,
\ee
where the  $2\times 2$ matrix $\mathcal{M}$ is given by
\be
\mathcal{M}=\left(
\begin{array}{cc}
\mathcal{E}(a)+k^2+a^2\,m^2\,\sigma_2 \bar r &-a^2m^2 \sigma_2\sqrt{c}\\
&\\
-a^2m^2 \sigma_2\sqrt{c}&\mathcal{F}(a,c)+k^2c^2+a^2m^2\sigma_2\frac{c}{\bar r}\\
\end{array}
\right)\,,
\ee
and 
 \be
 \mathcal{E}(a)=-\frac{a''}{a}\,,\hspace{2.5 em}\mathcal{F}(a,c)=-\frac{a''}{a} +\frac{1}{2}\frac{c''}{c}+\frac{c'}{c}\frac{a'}{a}-\frac{3}{4}\left(\frac{c'}{c}\right)^2\,.
 \ee
The action for the canonically normalized modes, eq. (\ref{total action 2}), is the action for two coupled harmonic oscillators with potential given by the matrix $\mathcal{M}$. From now on, we will refer to this matrix as the \emph{mass matrix}, since it has dimension of mass square. This matrix collects the contributions from the original mass term in the bigravity action (i.e. the terms proportional to $m^2$), the gradient terms (proportional to $k^2$) and contributions generated by the change of variable (\ref{can}) from the original action (\ref{total action}) which canonically normalizes the variables and removes all damping terms (first order derivatives). 

In matrix notation, in terms of $Q=(Q_h, Q_\lc)^T$, the equations of motion now reduce to
\be\label{e:can-eom}
 Q'' = -\MM Q \,.
\ee
By construction, the kinetic matrix is simply the identity. The fact that we can choose it positive means that there is no ghost present. The eigenvalues of the mass matrix are 
\be
\lambda_{1,2}=p \pm q\,,
\ee
with 
\be
p=\frac{1}{2 \bar r^2}\left[\left(k^2(c^2+1) +\mathcal{E}+\mathcal{F}\right)\bar r^2+a^2 m^2 \sigma_2 \bar r \left(c+\bar r^2\right)\right]\,,
\ee
\be
q=\frac{1}{2 \bar r^2}\left(\left(k^2( c^2 -1) -\mathcal{E}+\mathcal{F}\right)^2 \bar r^4+2 a^2 m^2\, \sigma_2\, \bar r^3 \left(k^2 c^2 -k^2 -\mathcal{E}+\mathcal{F}\right)\left(c-\bar r^2\right)+a^4 m^4 (\sigma_2 \bar r)^2\left(c+\bar r^2\right)^2\right)^{1/2}\,.
\ee
For  our choice of parameters with $m^2\beta_i\simeq \mathcal{H}_0^2$ and  $c\gg \bar r^2\simeq 2$,  $c^2=H^2/(\bar r H_f)^2 =3\HH^2/(a^2\bar r^2\La_c)$, in the various epochs the eigenvalues  can be approximated in compact form as 
\bea
\lambda_1 &\equiv p-q \simeq & k^2+\mathcal{E}(a)=k^2-\frac{a''}{a} =k^2-\frac{\al(\al-1)}{\tau^2}\,,
\\
\lambda_2 &\equiv p+q \simeq& k^2c^2+\mathcal{F}(a,c)+\frac{a^2}{\bar r^2} m^2 D c^2 = k^2c^2 +\frac{1}{\tau^2}\left(\left(-\frac{9}{4}+\frac{\beta_3\bar r+\beta_2}{3(1-\beta_2)}\right)\al^2+2\al-\frac{3}{4}\right)\,,
\eea
where $D$ is defined in eq. (\ref{def:D}) and  we have considered a power law background expansion, $a\propto \tau^\al$ in the last equalities for $\lambda_1$ and $\lambda_2$. We call $Q_1$ and $Q_2$ the mass eigenstates associated to $\lambda_1$ and $\lambda_2$, respectively. 
The eigenvalue $\lambda_1$ is the expression for the squared frequency of the standard canonically normalized tensor mode in GR. Hence, one of the two mass eigenstates is exactly the massless graviton of GR and, as is well known, it is (marginally) stable. On super Hubble scales $Q_1\propto a$, such that $Q_1/a=$constant while for sub-Hubble scales $Q_1$ oscillates at constant amplitude.  The more interesting eigenvalue is $\lambda_2$. For our choice of the parameters $\beta_i$ and  $m$ , we obtain
\be
\lambda_2  \simeq  k^2c^2 -\frac{359\al^2-2\al+3/4}{\tau^2} \,.
\ee

For the stability of the second mass eigenstate we request $\lambda_2\geq\la_1$.  In the various cosmological epochs, from the expressions for $\mathcal{H}$  one can easily derive
\be\label{condition}
k^2-\frac{\al(\al-1)}{\tau^2}\lsim k^2c^2-\frac{\xi}{\tau^2}\,,
\ee
where $\xi=357+3/4\,, ~1432+3/4\,, ~361+3/4$ in radiation ($\al=1$), matter ($\al=2$) and de\,Sitter ($\al=-1$) eras respectively. 

At early times, condition (\ref{condition}) is satisfied for all modes of cosmological interest, $k^2\gsim H_0^2\simeq 1/\tau_0^2$, since $c\gg1$. At low redshift, instead, a tachyonic instability appears which sets-in earlier for lower frequency modes. 

 Both tensor modes of the $g$- and $f$- metrics develop an instability since they are given by a mixture of the two mass eigenstates. To quantify the effects of the instability on the physical $g$ sector, we need to go beyond this qualitative analysis and to explicitly solve the equations for perturbations in the various cosmological epochs. 
 
In Appendix~\ref{analytic tens} we determine analytical solutions during a de\,Sitter like inflationary stage, during radiation domination, matter domination and during a late  de\,Sitter phase in the limit where the couplings between $h$ and $\lc$ can be neglected.  We see that in this case, no instabilities occur. We also  find  that  for an inflationary Hubble parameter $H_I\gg H_0$
the power spectrum of the $\lc$-mode is severely suppressed with respect to the one of $h$ for all modes of cosmological interest. Therefore, we expect that at  high redshift, the evolution of the physical tensor mode $h$ will be not affected by the coupling with $\lc$.  We explicitly verify this, by solving analytically the decoupled differential equations describing the evolution of gravitational waves in the radiation- and matter-dominated epochs and by estimating the amplitude of the coupling. 
During the radiation and matter era at sufficiently high redshift $h$ evolves like in $\Lambda$CDM while $\lc$ has a constant and a decaying mode. Only at late times the coupling term in the equation for the physical tensor mode becomes relevant and deviation from the GR evolution can be expected. 

The scale of inflation may lie above the strong coupling scale\footnote{The scale of self-interactions of the helicity-0 mode of the massive graviton, at which perturbativity breaks down.}. For massive gravity on a flat background this scale is $\La_3 \equiv (m^2 M_{Pl})^{1/3}$ but for  cosmological backgrounds it is not known neither for massive gravity nor for bigravity. For this reason the results for the inflationary power spectra may not be representative. In the following numerical treatment we shall not consider this point and we do not make any assumptions on the initial conditions but just vary them and study their effect.  

\subsection{Numerical results}

In this section we report the results of a numerical integration of the tensor mode  evolution equations. In particular we want to test the appearance of a tachyonic instability at late times for low-$k$ modes. For this, we numerically evolve the system of differential equations (\ref{peg}), (\ref{pef}), starting from the redshift of matter-radiation equality. We focus on modes that become sub-horizon during the late matter-dominated era.

Since during inflation and radiation, the physical tensor mode $h$ evolves like in GR, see Appendix~\ref{analytic tens}, we choose GR-like initial conditions for $h$ at the redshift of equality
\be
h(z_{eq})=1\,,\hspace{2 em}h'(z_{eq})=0\,.
\ee
The mode $\lc$ at the end of inflation is suppressed with respect to $h$ by a factor $(k/H_0)(H_0/H_I)^{1/2}$ and during radiation $\lc$ has a constant and a decaying mode, which oscillates with a very high frequency. In our numerical study we choose generic initial condition for $\lc$, parametrized by two constants $(A,B)$ as
\be
\lc(z_{eq})=A\,,\hspace{2 em}\lc'(z_{eq})=\HH_0\,B\,,
\ee
and we explore how the result of the numerical integration is affected changing the parameters $A$ and $B$.

\begin{figure}[ht!]
      \vspace{-0.3cm}
    \centering
 \subfigure[\label{h1k10}]
      {\includegraphics[scale=0.35]{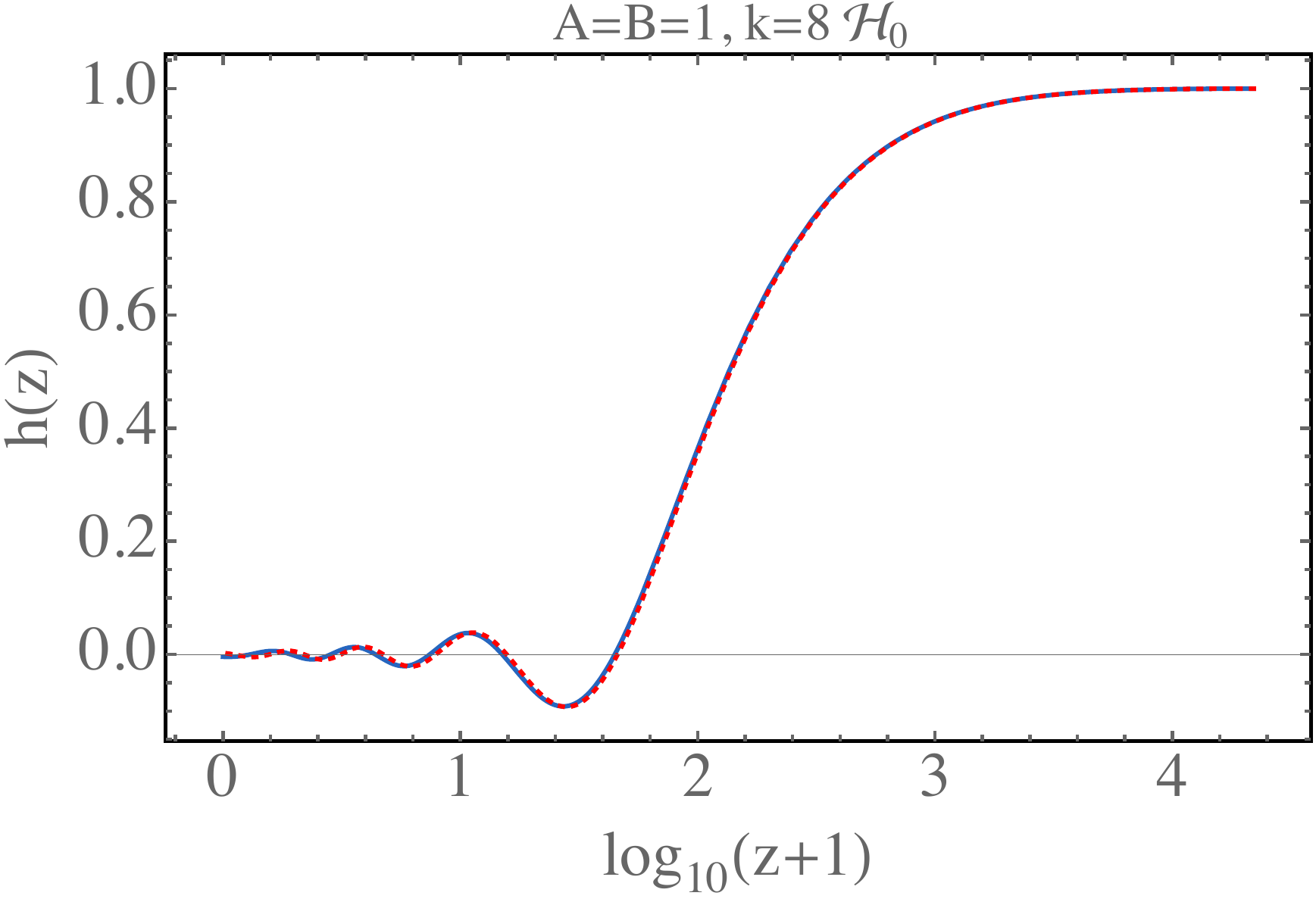}}\qquad\qquad
 \subfigure[\label{h2k10}]
      {\includegraphics[scale=0.35]{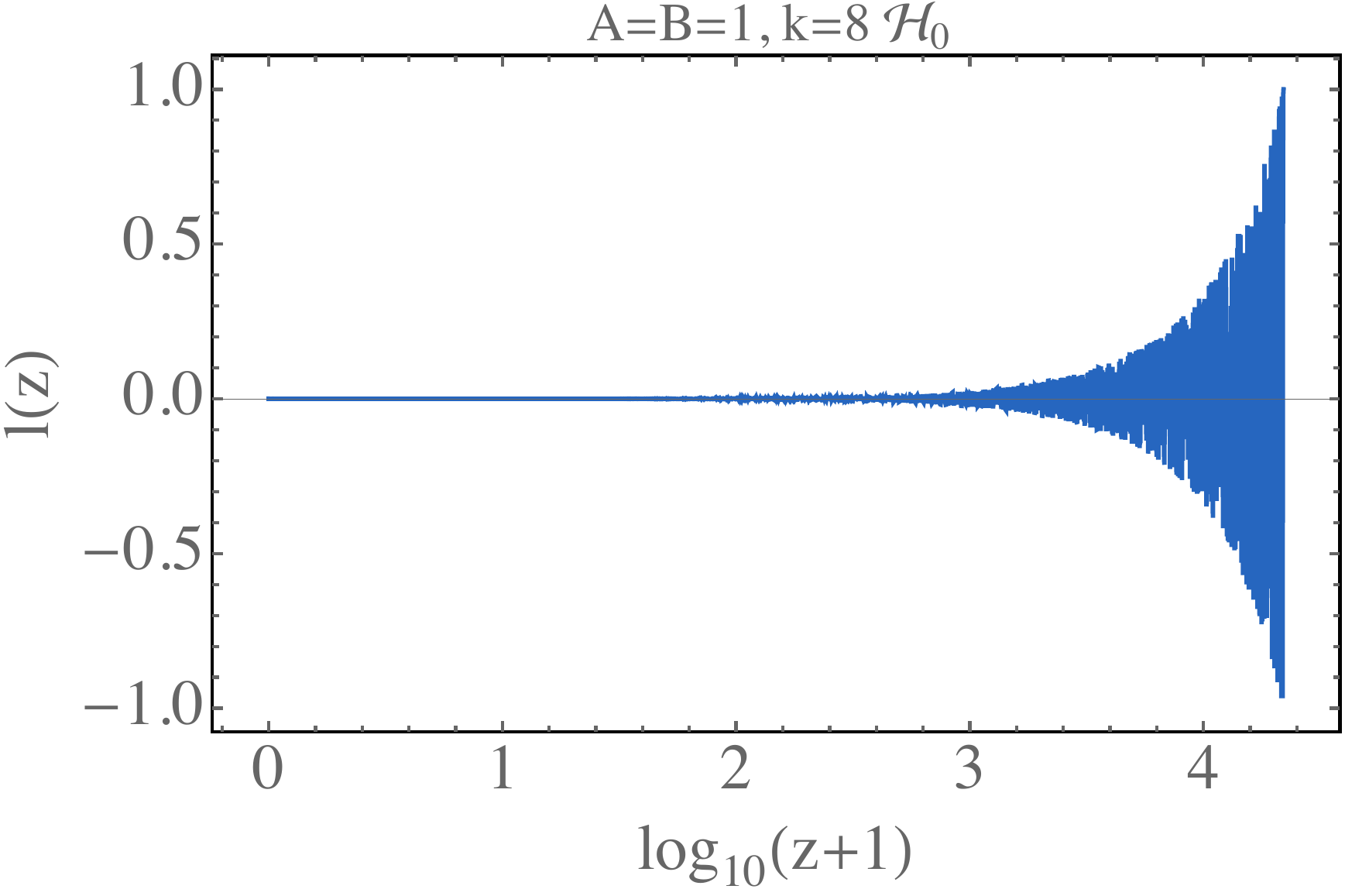}}\qquad
 \subfigure[\label{h1k5}]
      {\includegraphics[scale=0.35]{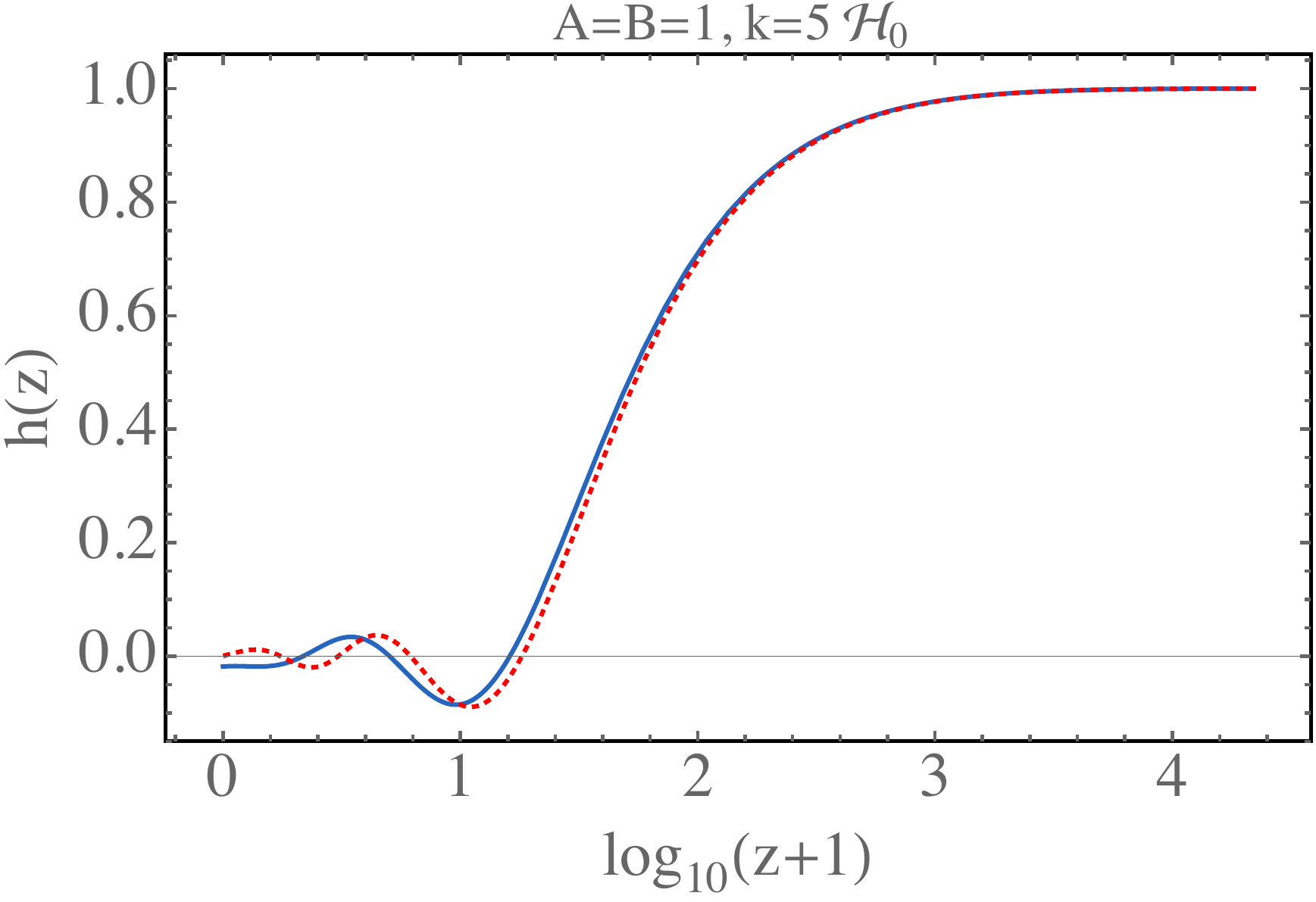}}\qquad\qquad
\subfigure[\label{h2k5}]
     {\includegraphics[scale=0.35]{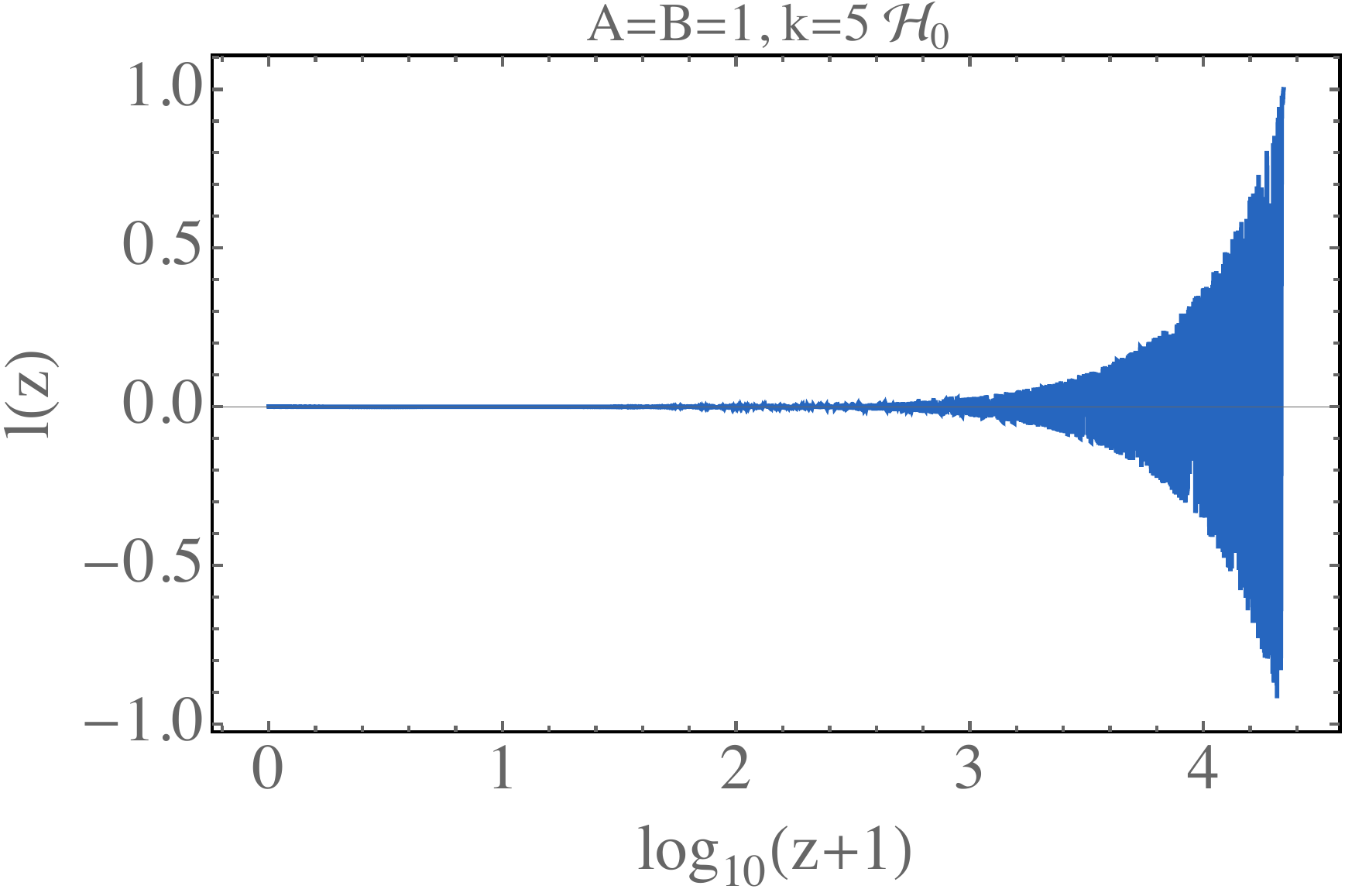}}\qquad
 \subfigure[\label{h1k3}]
      {\includegraphics[scale=0.35]{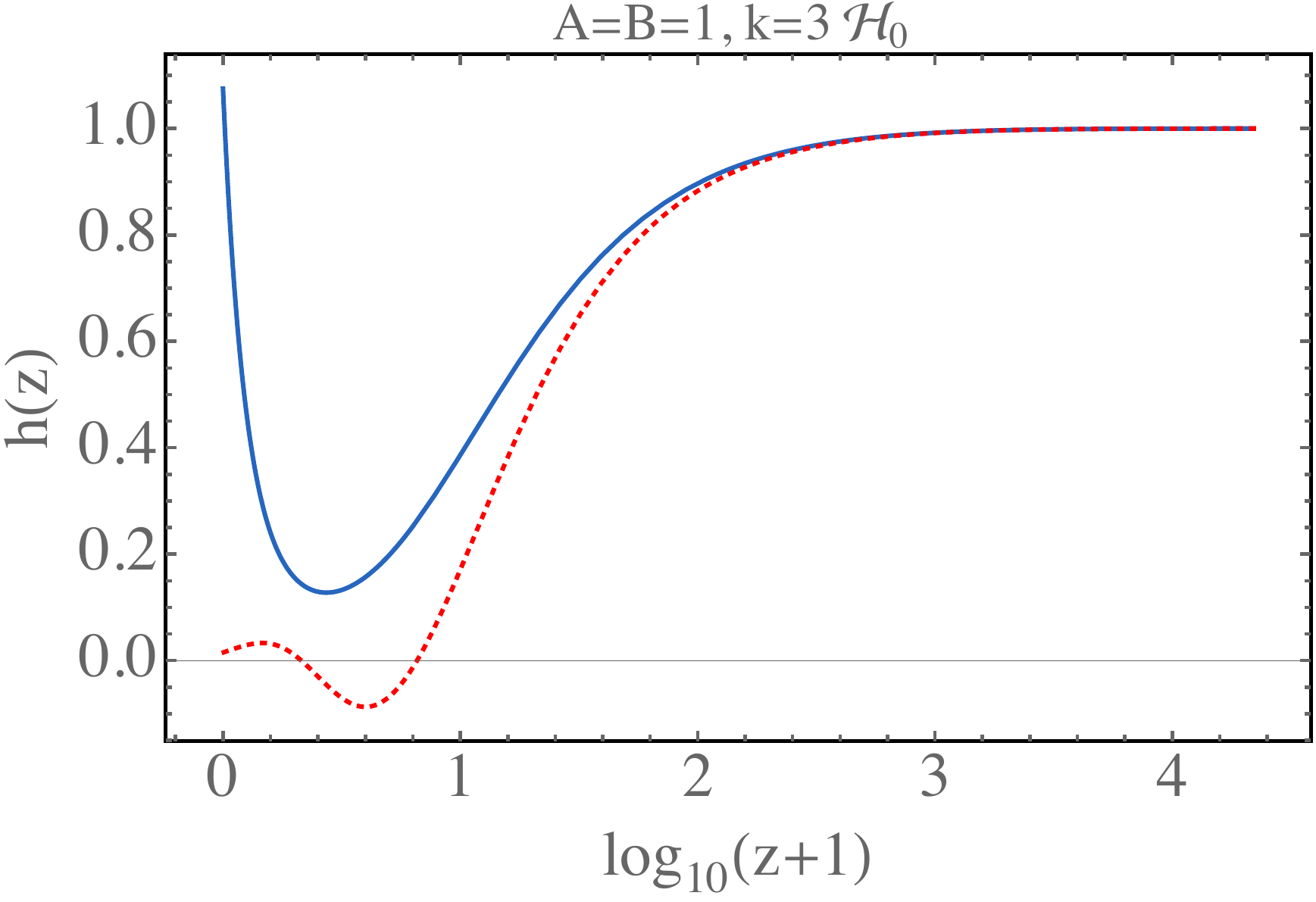}}\qquad\qquad
 \subfigure[\label{h2k3}]
      {\includegraphics[scale=0.35]{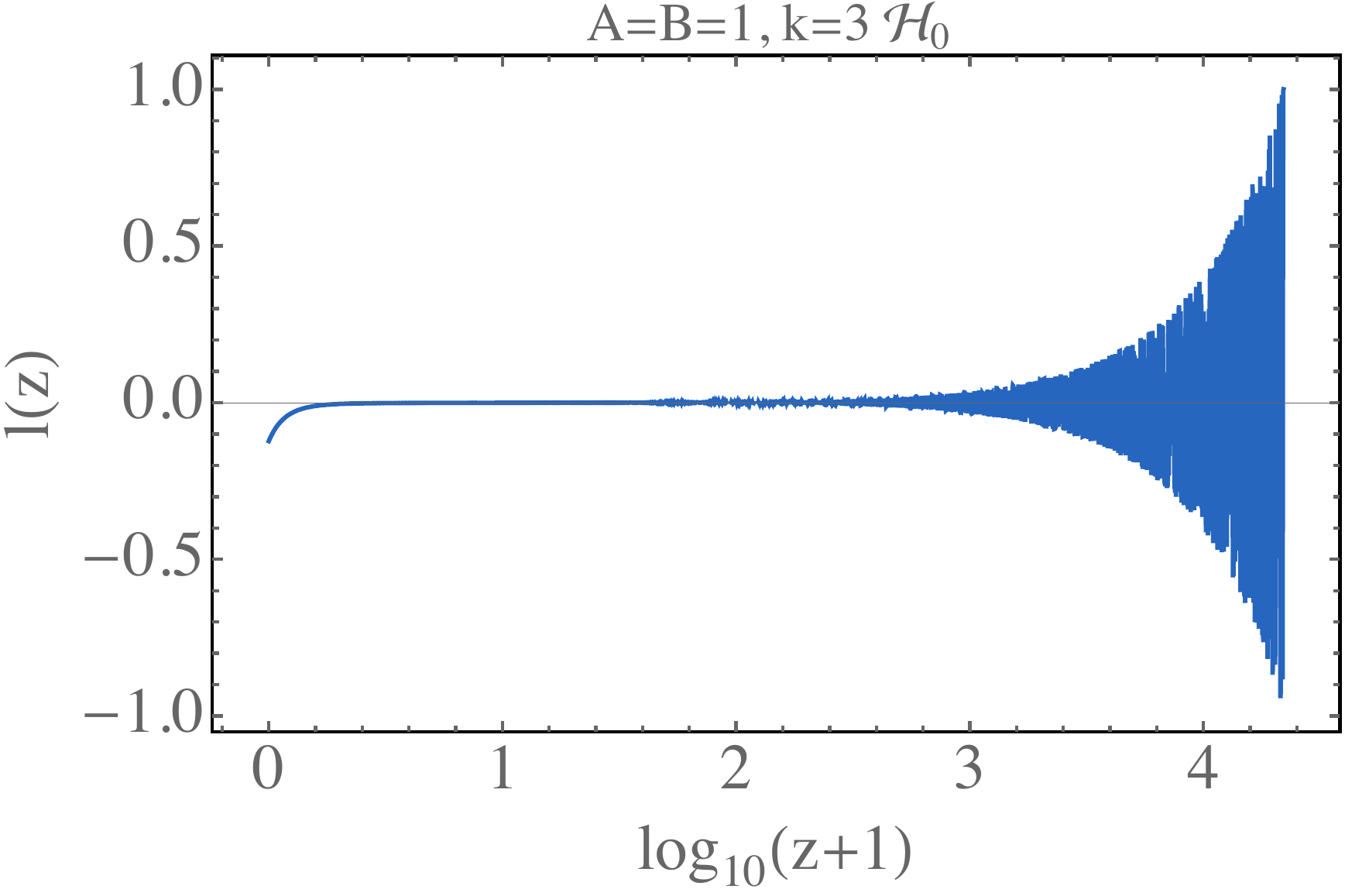}}\qquad
 %\subfigure[\label{h1k2}]
  %    {\includegraphics[scale=0.40]{h1k2}}\qquad
% \subfigure[\label{h2k2}]
 %     {\includegraphics[scale=0.40]{h2k2}}\qquad
       \subfigure[\label{h1k1}]
      {\includegraphics[scale=0.35]{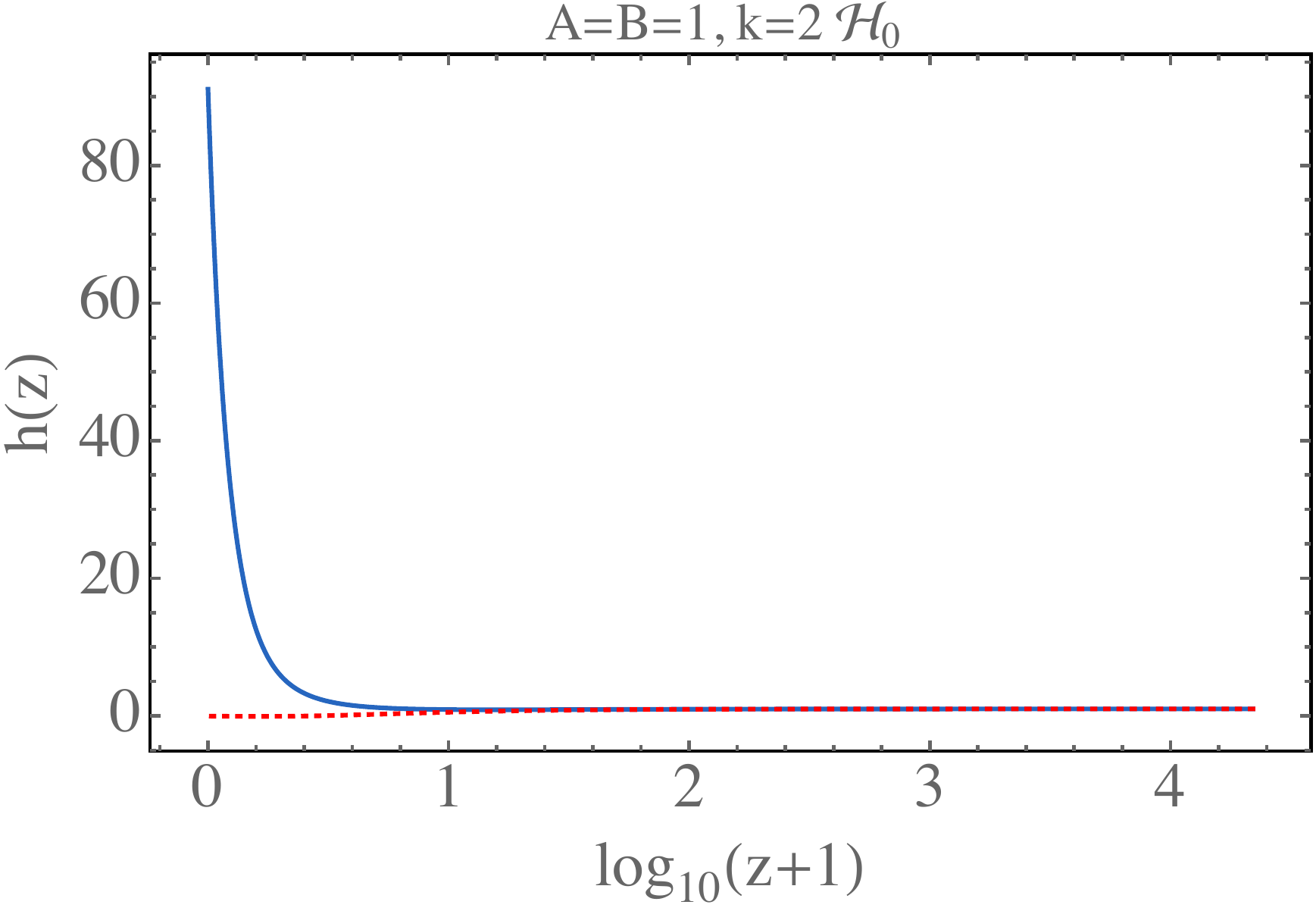}}\qquad\qquad
 \subfigure[\label{h2k1}]
      {\includegraphics[scale=0.35]{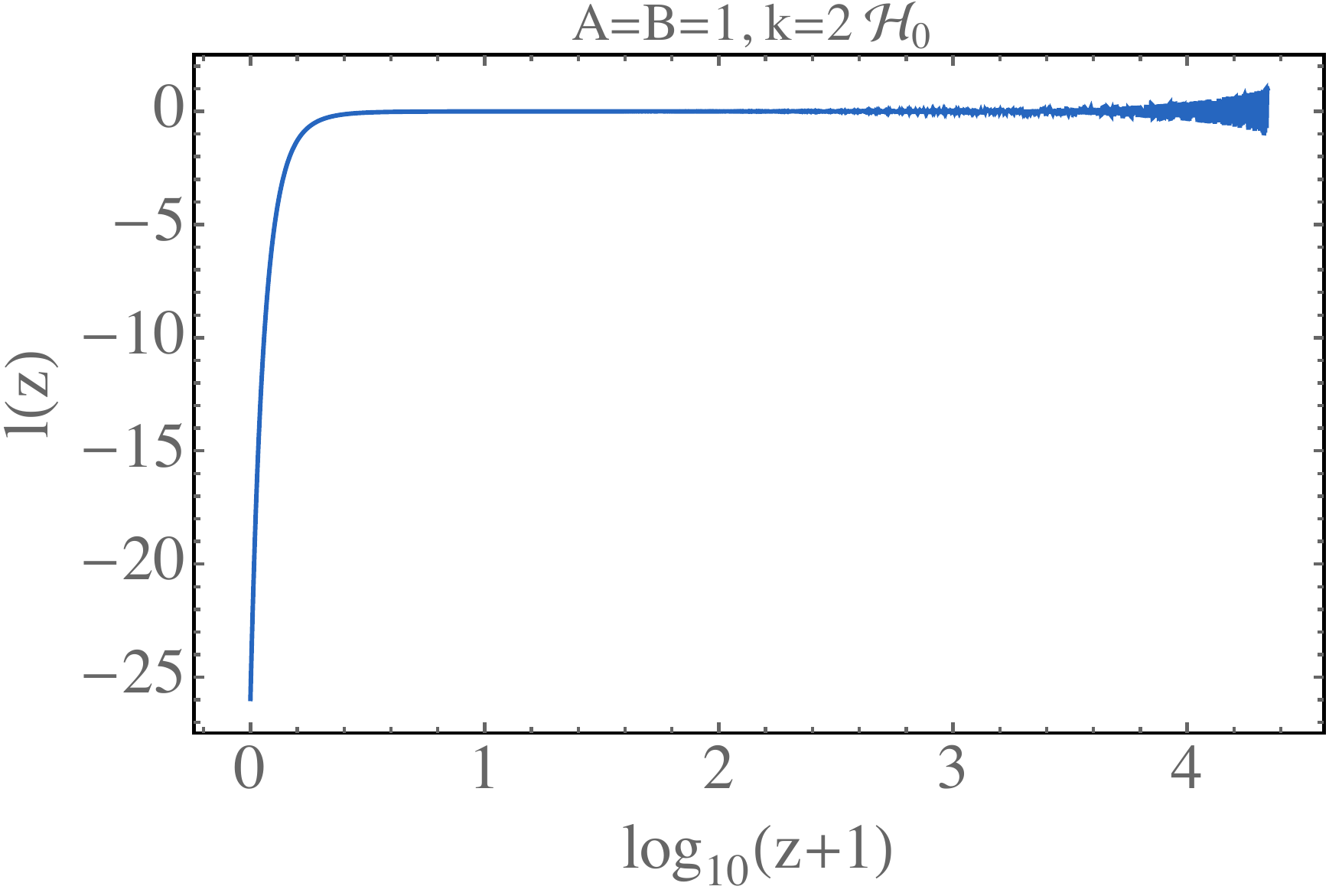}}\qquad
\caption{\label{fig2} Tensor perturbations for GR-like type initial conditions in the physical sector, $h(z_{eq})=1\,,\,\, h'(z_{eq})=0$ and $\lc(z_{eq})=A\,,\,\, \lc'(z_{eq})=H_0\,B$. The evolution of tensor perturbations in the physical sector is plotted together with the one of $\Lambda CDM$ with the same initial conditions (red, dotted line).}
 \end{figure}

\begin{figure}[ht!]
    \centering
% \subfigure[\label{h1k5}]
      %{\includegraphics[scale=0.40]{h1k5}}
% \subfigure[\label{h2k5}]
    %  {\includegraphics[scale=0.40]{h2k5}}\qquad
 \subfigure[\label{futureh1k5}]
      {\includegraphics[scale=0.40]{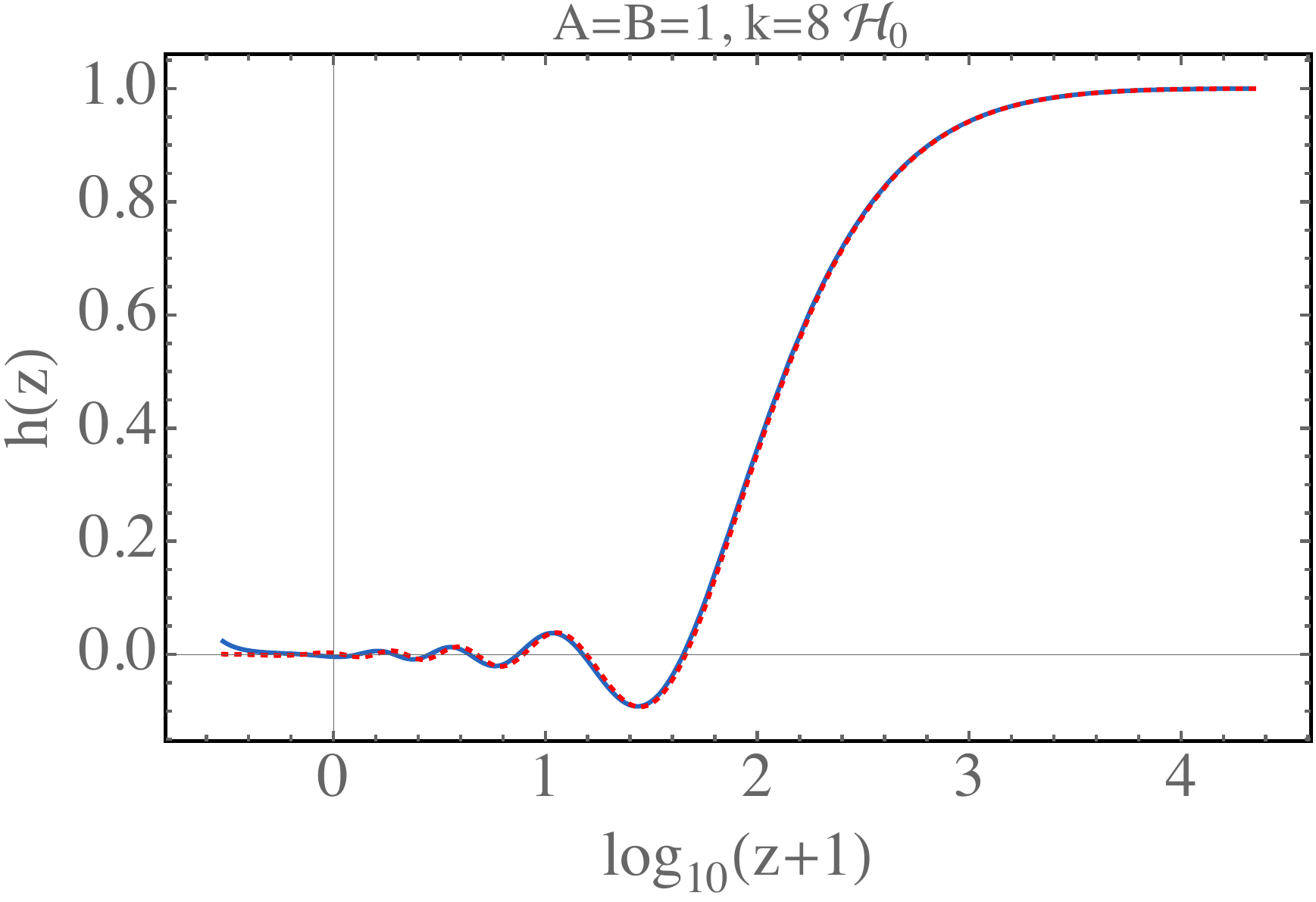}}\qquad
 \subfigure[\label{futureh2k5}]
      {\includegraphics[scale=0.40]{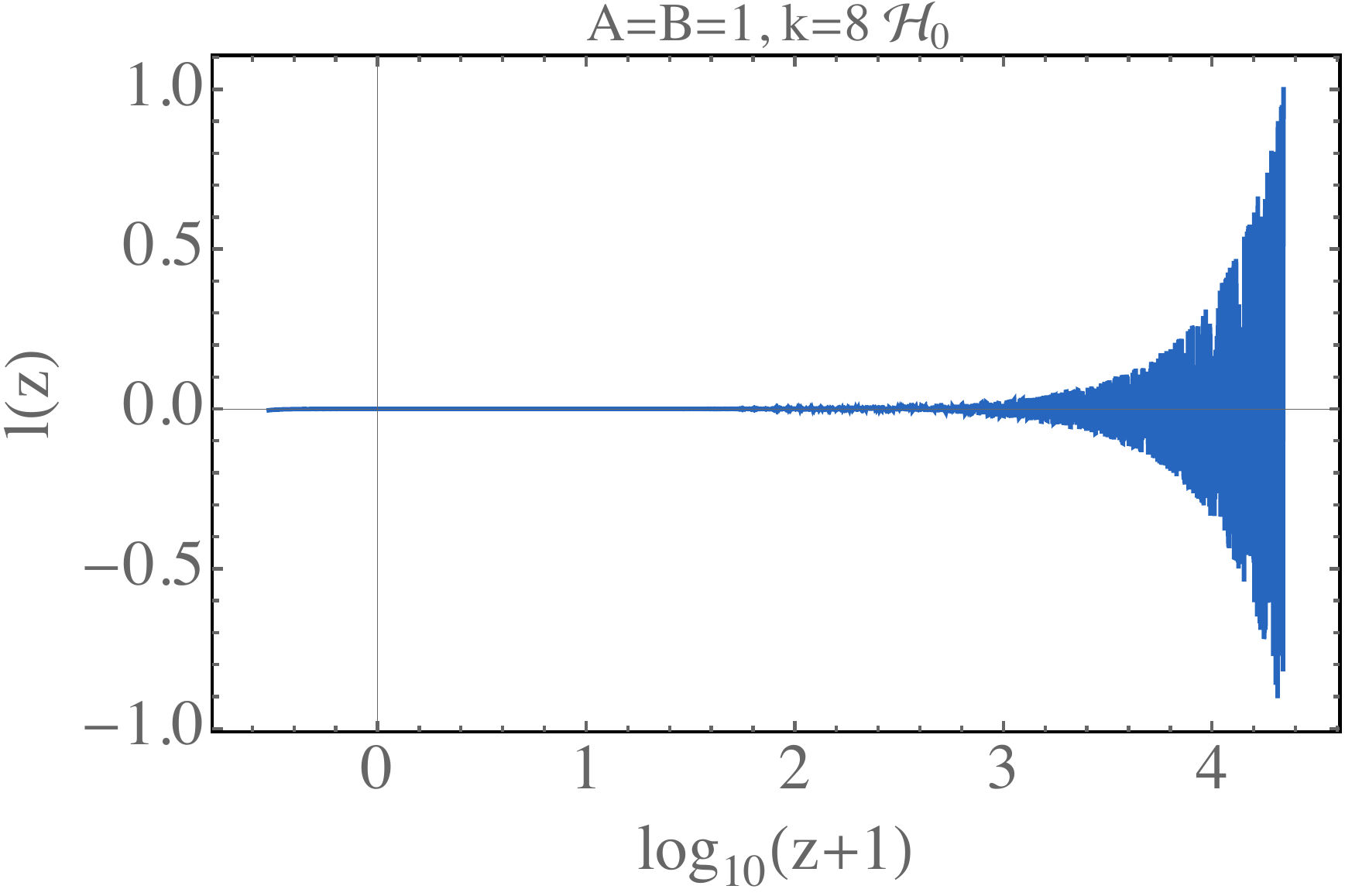}}\qquad
       \subfigure[\label{futureh1k3}]
      {\includegraphics[scale=0.40]{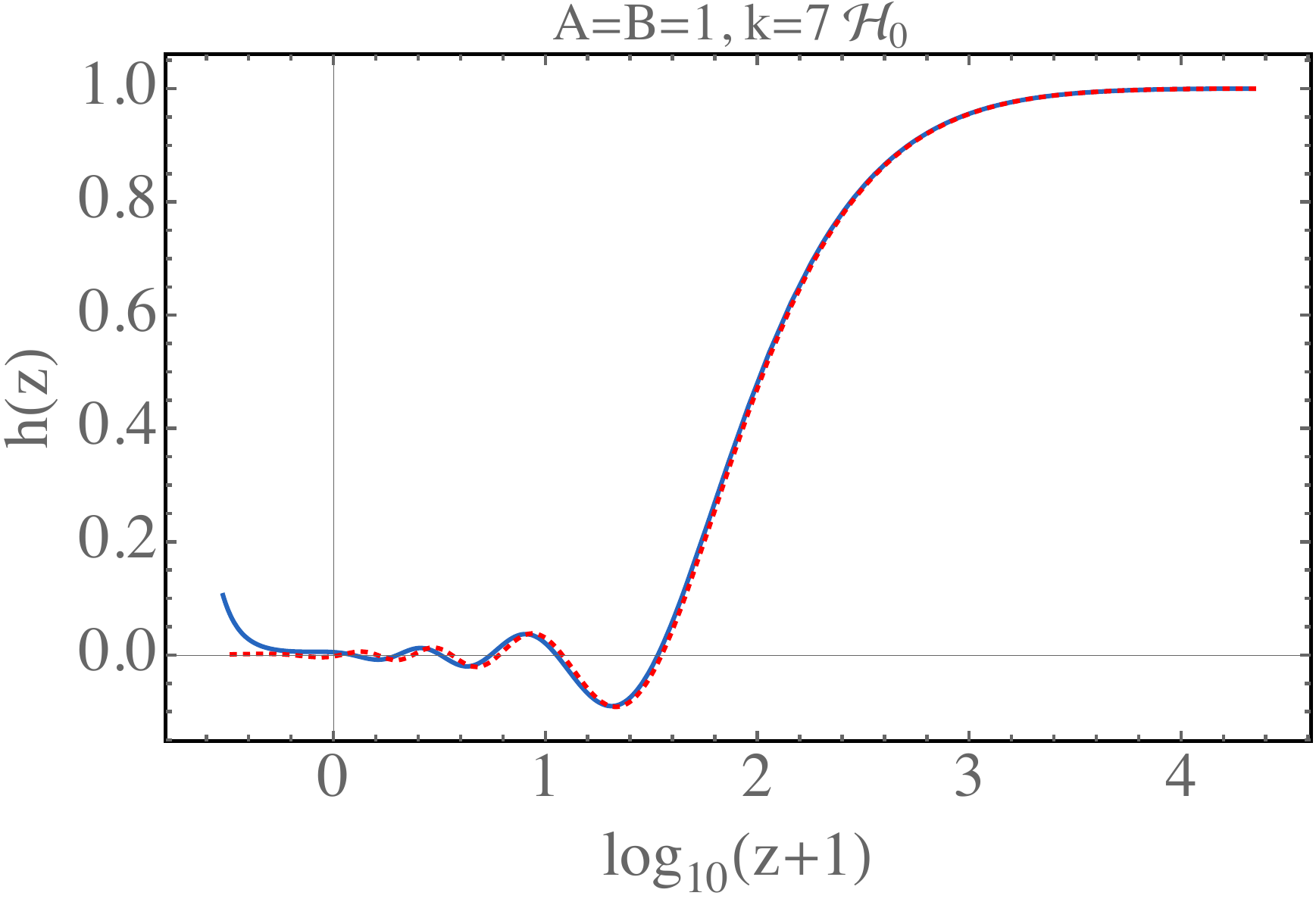}}\qquad
 \subfigure[\label{futureh2k3}]
      {\includegraphics[scale=0.40]{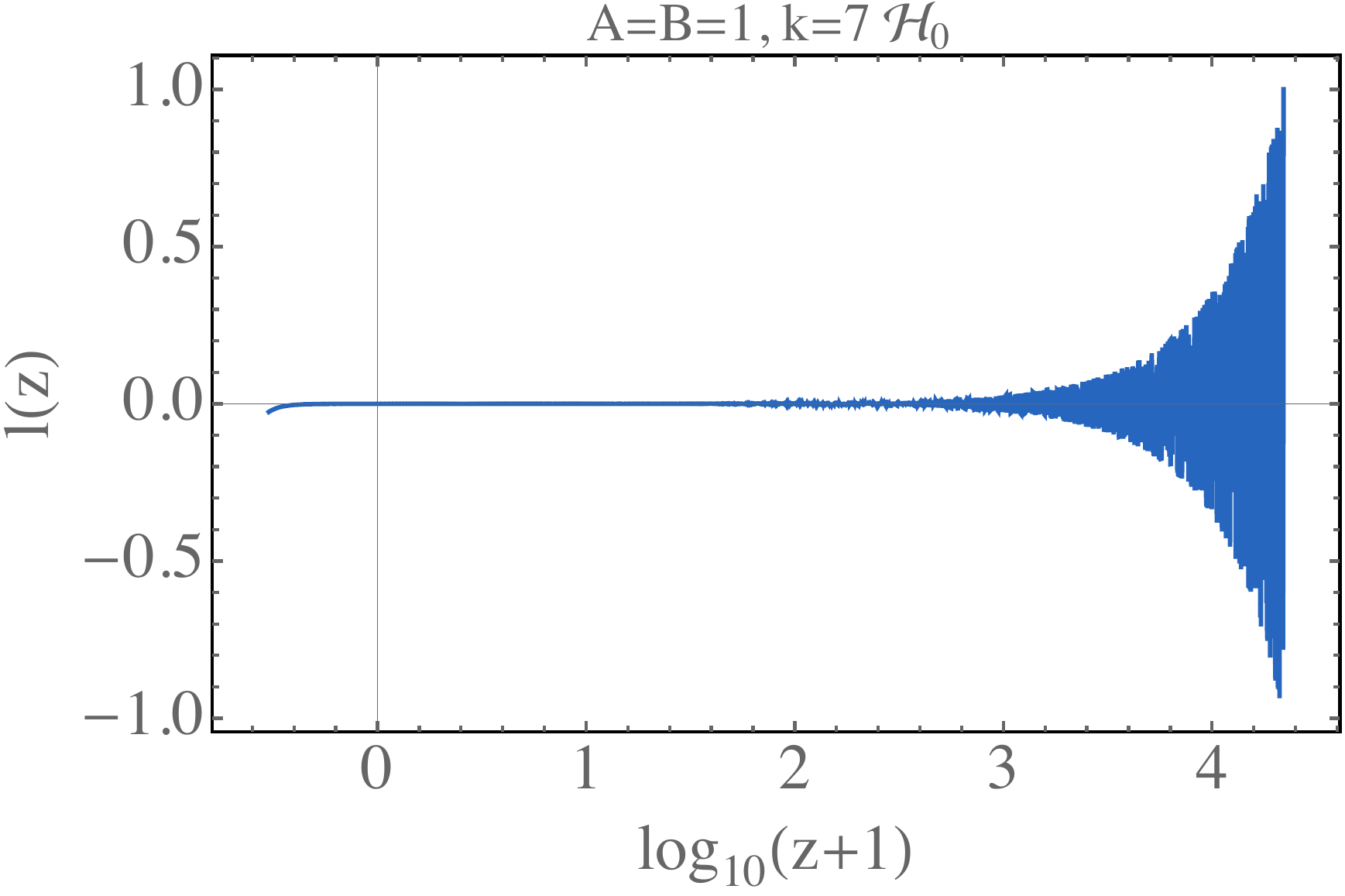}}\qquad
       \subfigure[\label{futureh1k2}]
      {\includegraphics[scale=0.40]{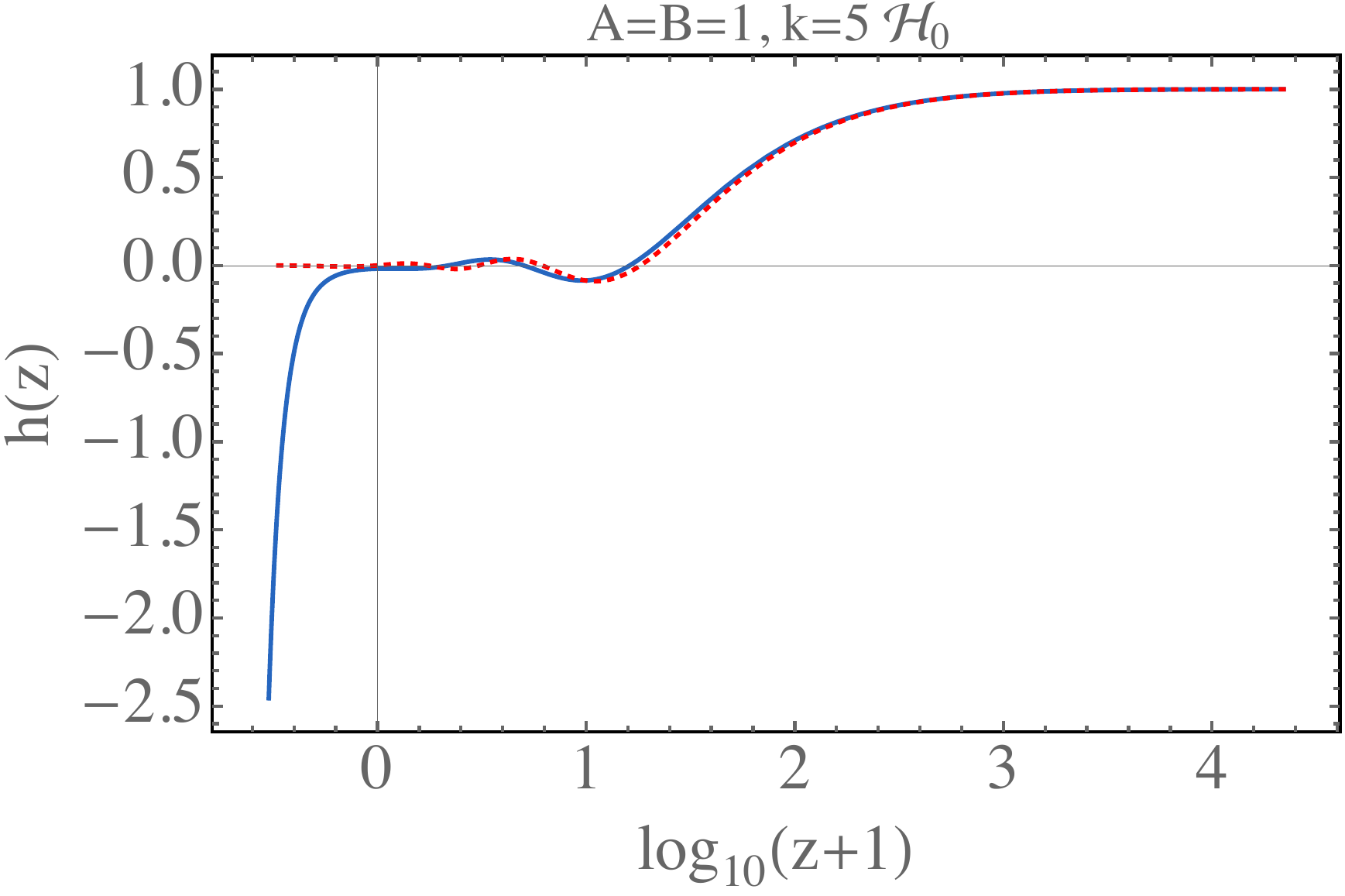}}\qquad
 \subfigure[\label{futureh2k2}]
      {\includegraphics[scale=0.40]{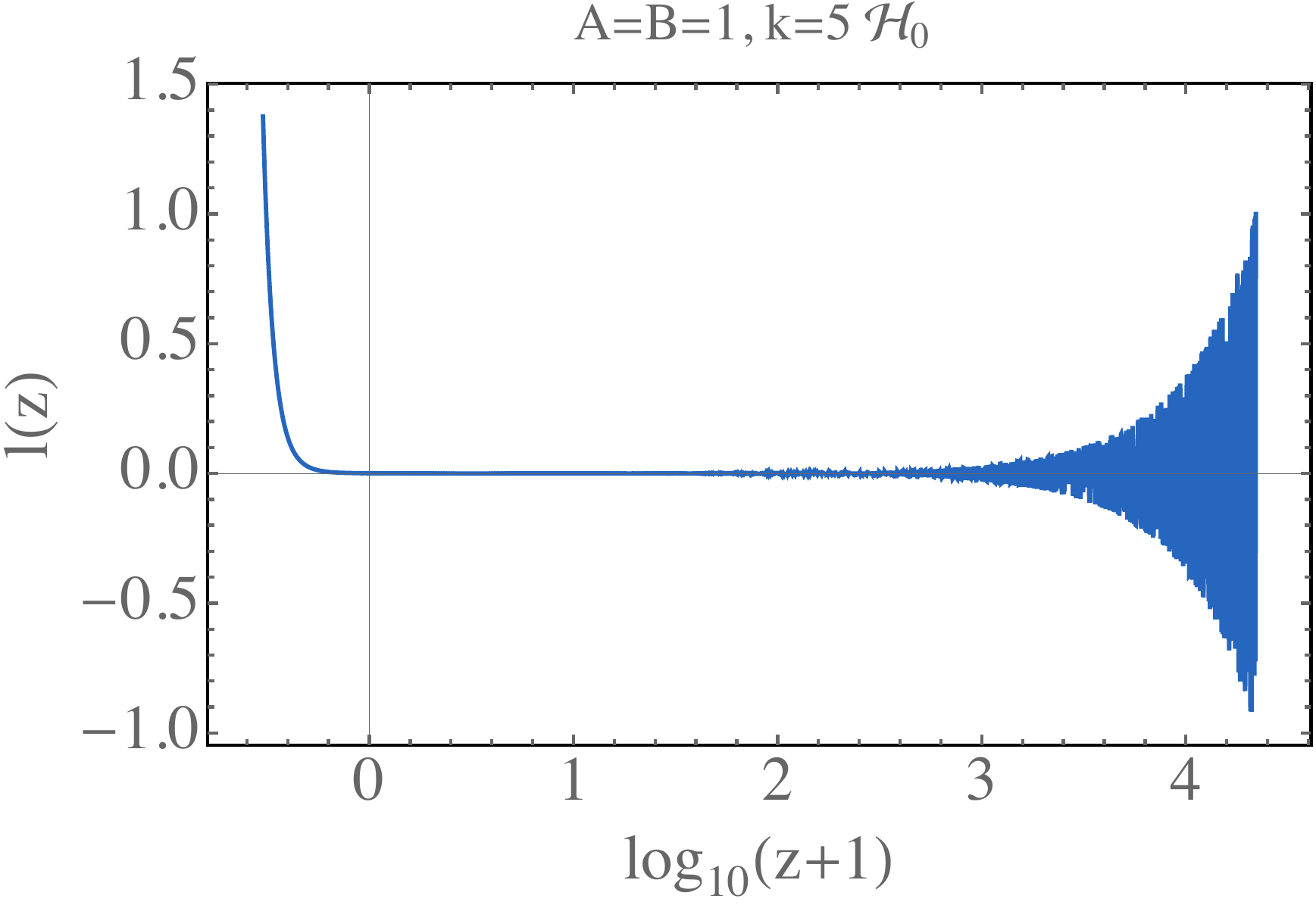}}\qquad
\caption{\label{fig3} Tensor perturbations for GR-like type initial conditions in the physical sector, $h(z_{eq})=1\,,\,\, h'(z_{eq})=0$ and $\lc(z_{eq})=A\,,\,\, \lc'(z_{eq})=H_0\,B$. The evolution of tensor perturbations in the physical sector is plotted together with the one of $\Lambda CDM$ with the same initial conditions (red, dotted line). The system is evolved into the future  to show the appearance of an instability for $k\simeq \mathcal{H}$.}
 \end{figure}

Fig.~\ref{fig2} shows the result of the numerical integration for different modes for the case $A=B=1$. The evolution of the physical tensor mode $h$ is plotted together with the corresponding result in $\Lambda CDM$ (red, dotted line). 
A deviation from the $\Lambda CDM$ evolution appears at late times and becomes more important for small values of $k/\mathcal{H}_0$. This is in line with what we find in Appendix~\ref{analytic tens}: at late times the coupling terms in the equations for the tensor modes (\ref{ds1}) and (\ref{ds2}) grow like $\mathcal{H}^2$ and become relevant when they reach the order of magnitude of the $k$-term. Therefore, the effect of the coupling  becomes relevant earlier for larger modes. In other words, at late times, modes with $k\simeq \mathcal{H}_0$ will experience a larger deviation from the $\Lambda CDM$ evolution. This is visible in Fig.~\ref{fig3}, where for different values of $k/\mathcal{H}_0$ the evolution of tensor perturbations is plotted into the future, until negative redshift. We see that the tensor sector exhibits an instability and the instant at which this instability shows up in the physical sector is pushed towards future time for smaller wave lengths, $k \gg \mathcal{H}_0$. 

\begin{figure}[ht!]
    \vspace{-0.3cm}
    \centering
             \subfigure[\label{h1k5A7B7}]
     {\includegraphics[scale=0.35]{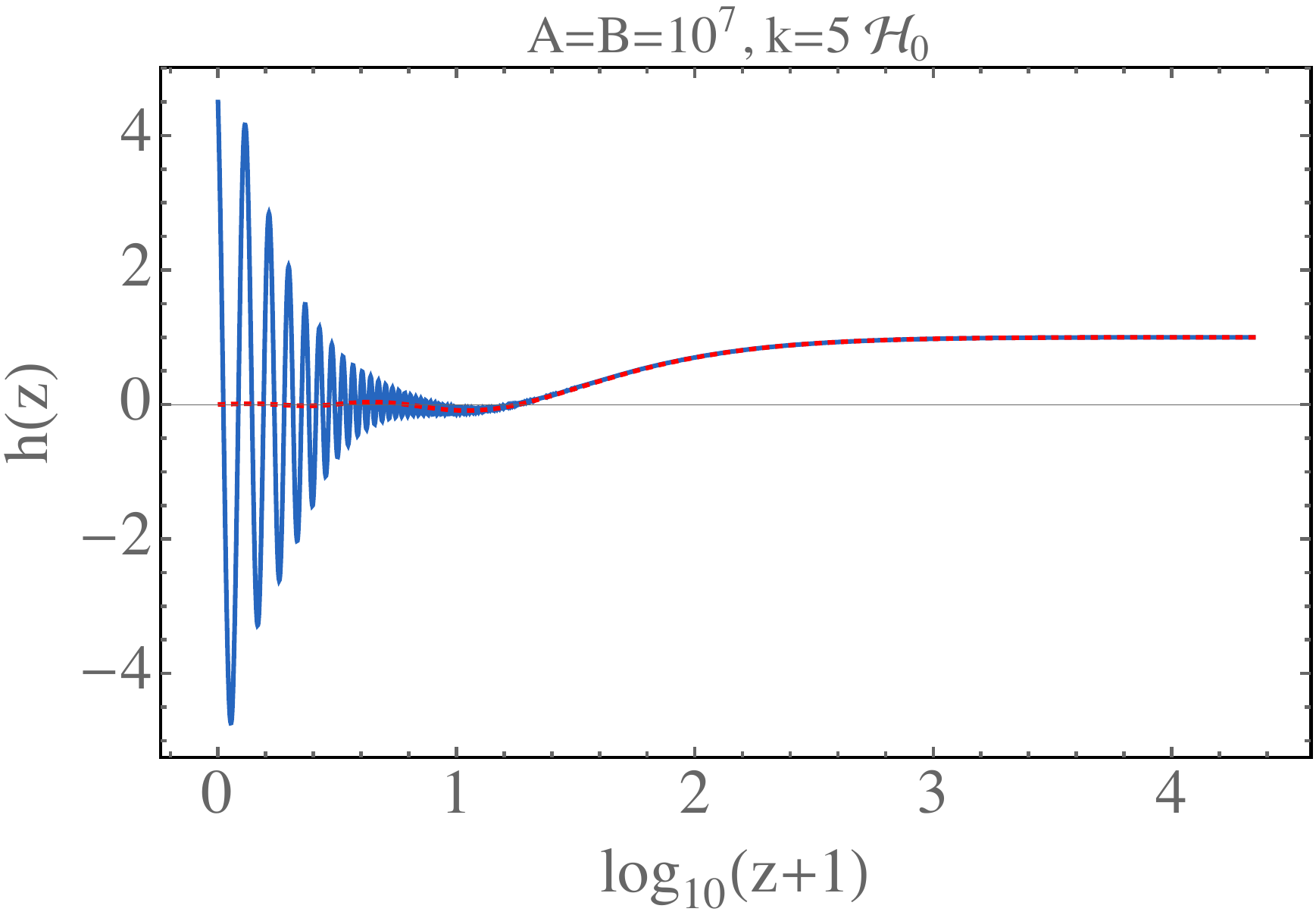}}\qquad\qquad
 \subfigure[\label{h2k5A7B7}]
     {\includegraphics[scale=0.35]{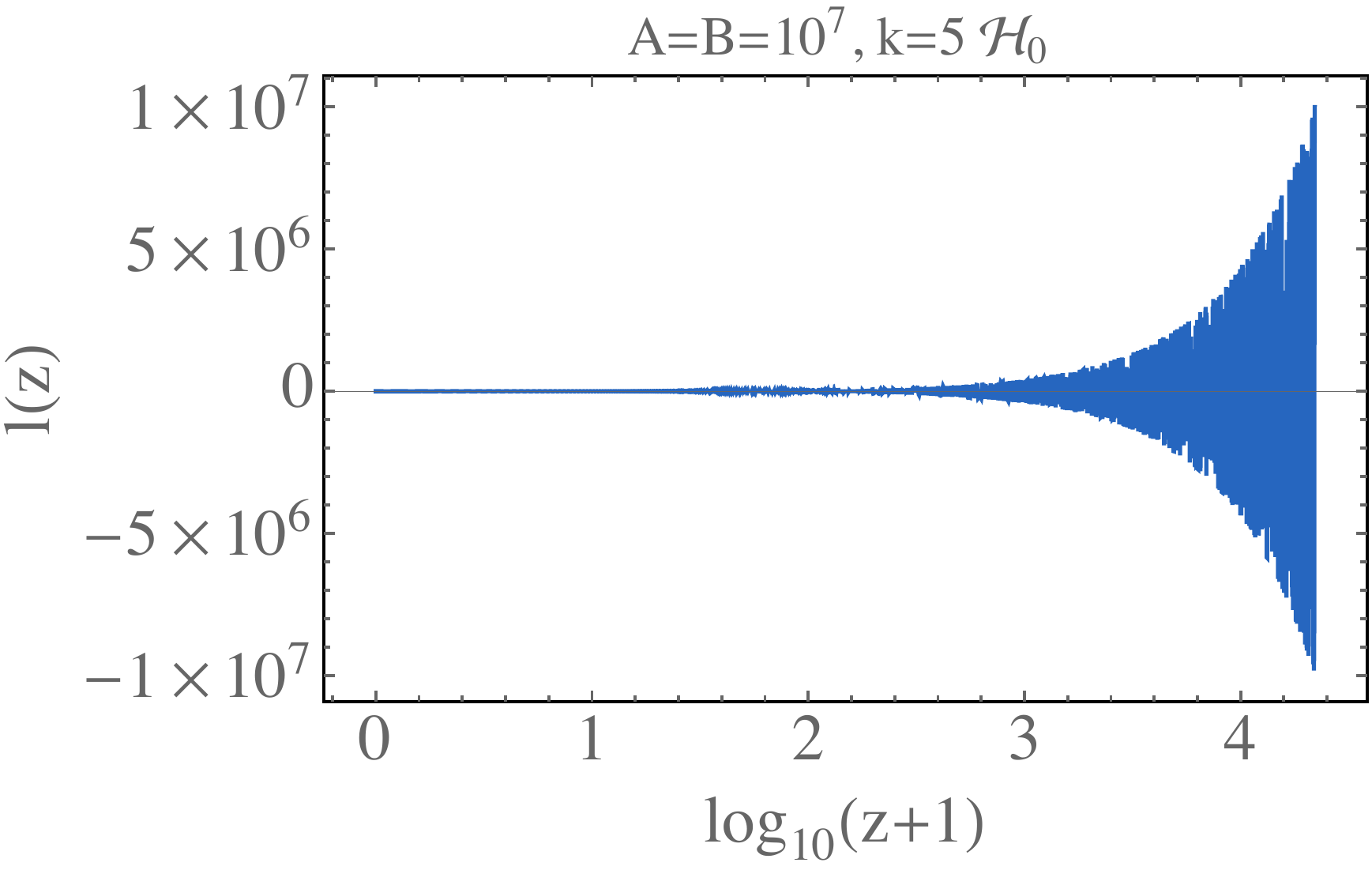}}
         \subfigure[\label{h1k5A5B5}]
     {\includegraphics[scale=0.35]{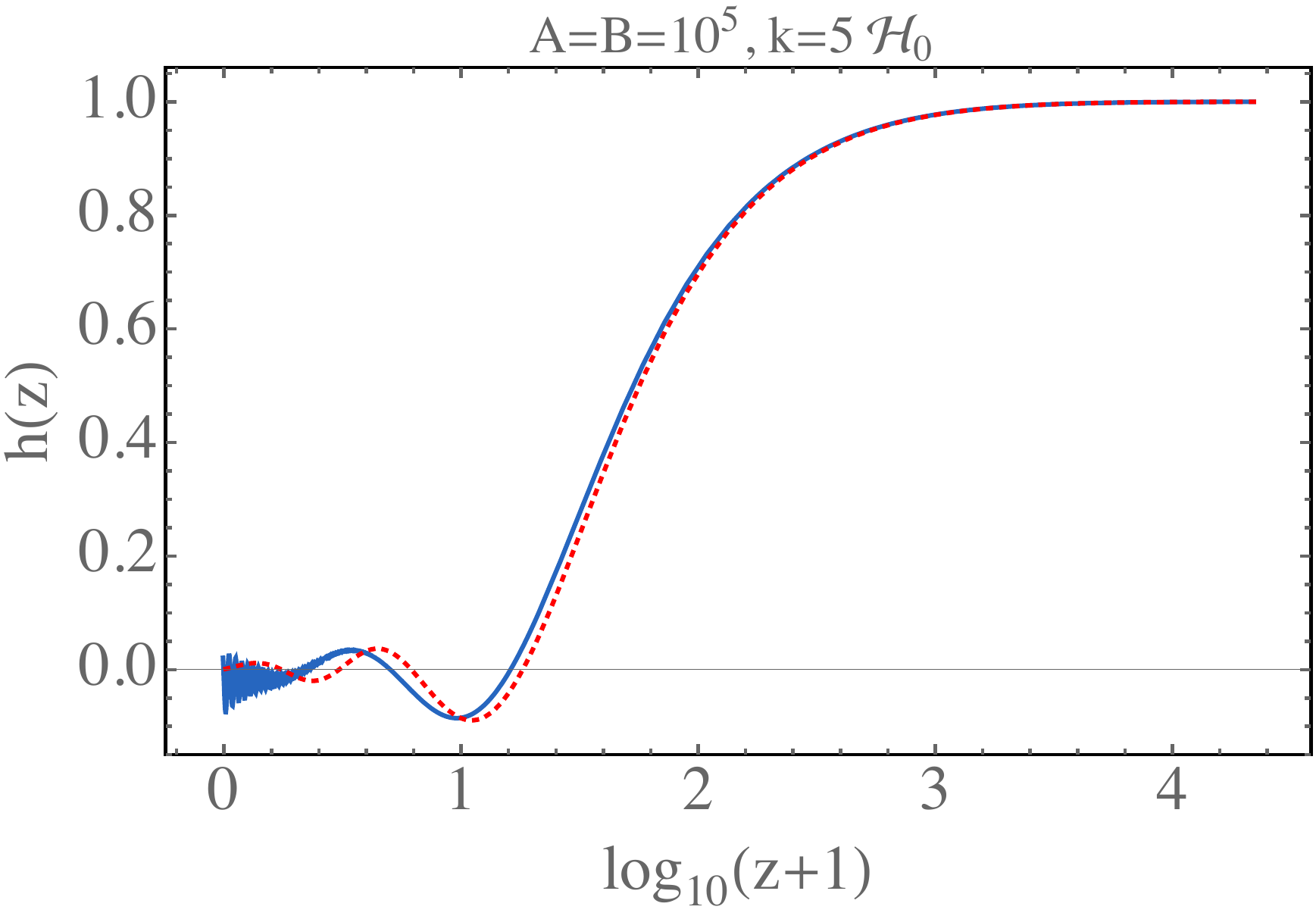}}\qquad\qquad
 \subfigure[\label{h2k5A5B5}]
     {\includegraphics[scale=0.35]{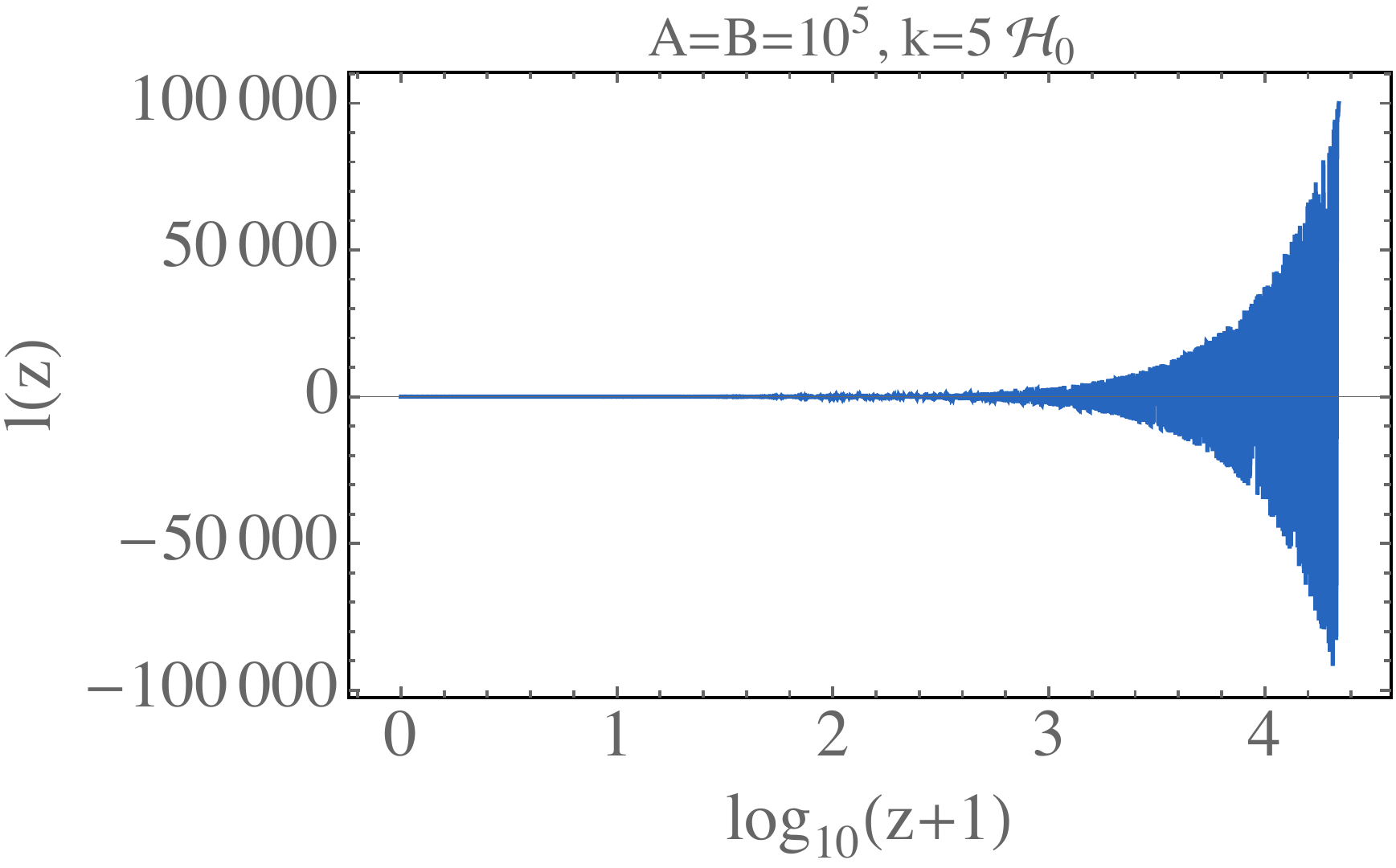}}
 \subfigure[\label{h1k5A3B3}]
      {\includegraphics[scale=0.35]{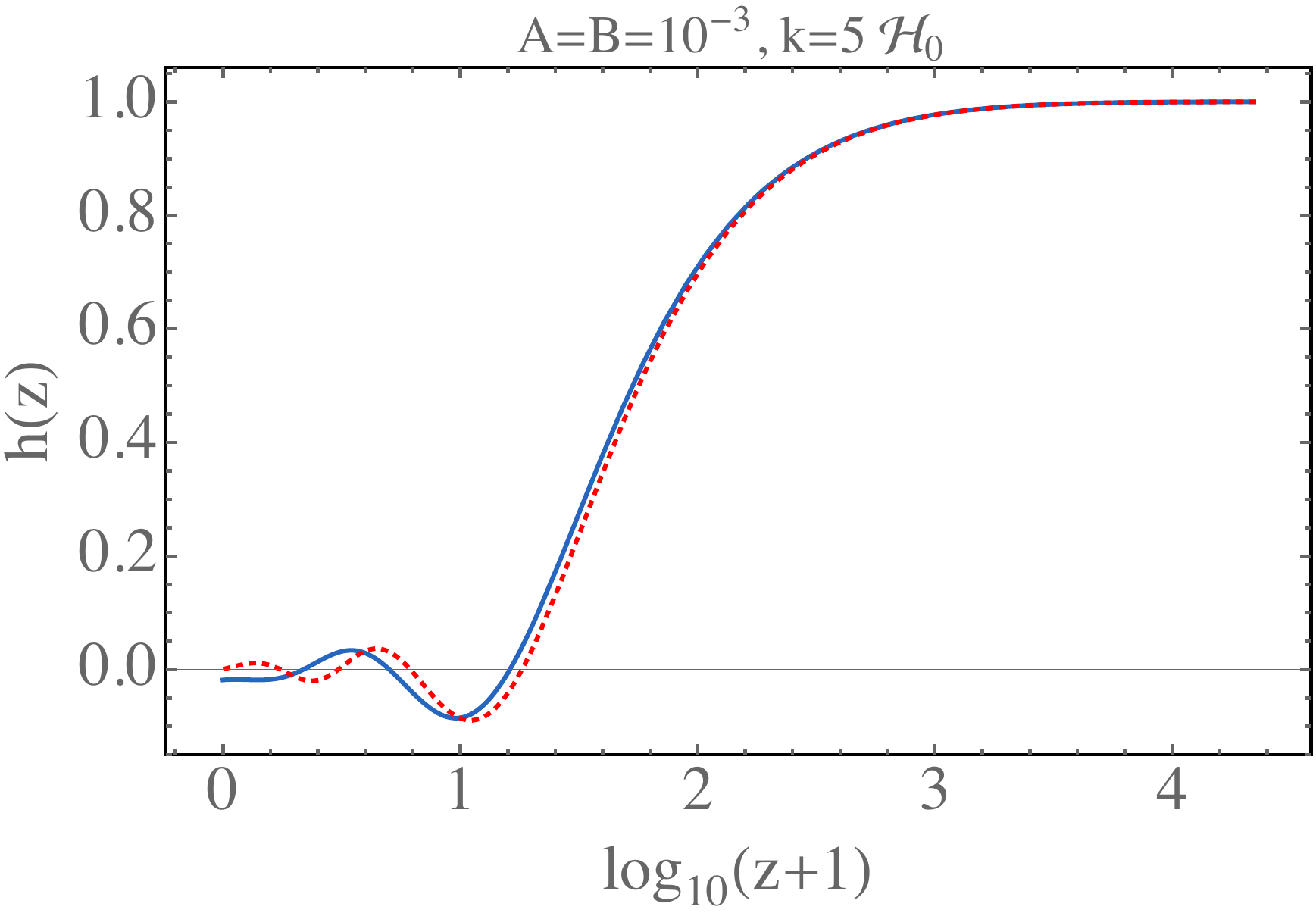}}\qquad\qquad
 \subfigure[\label{h2k5A3B3}]
      {\includegraphics[scale=0.35]{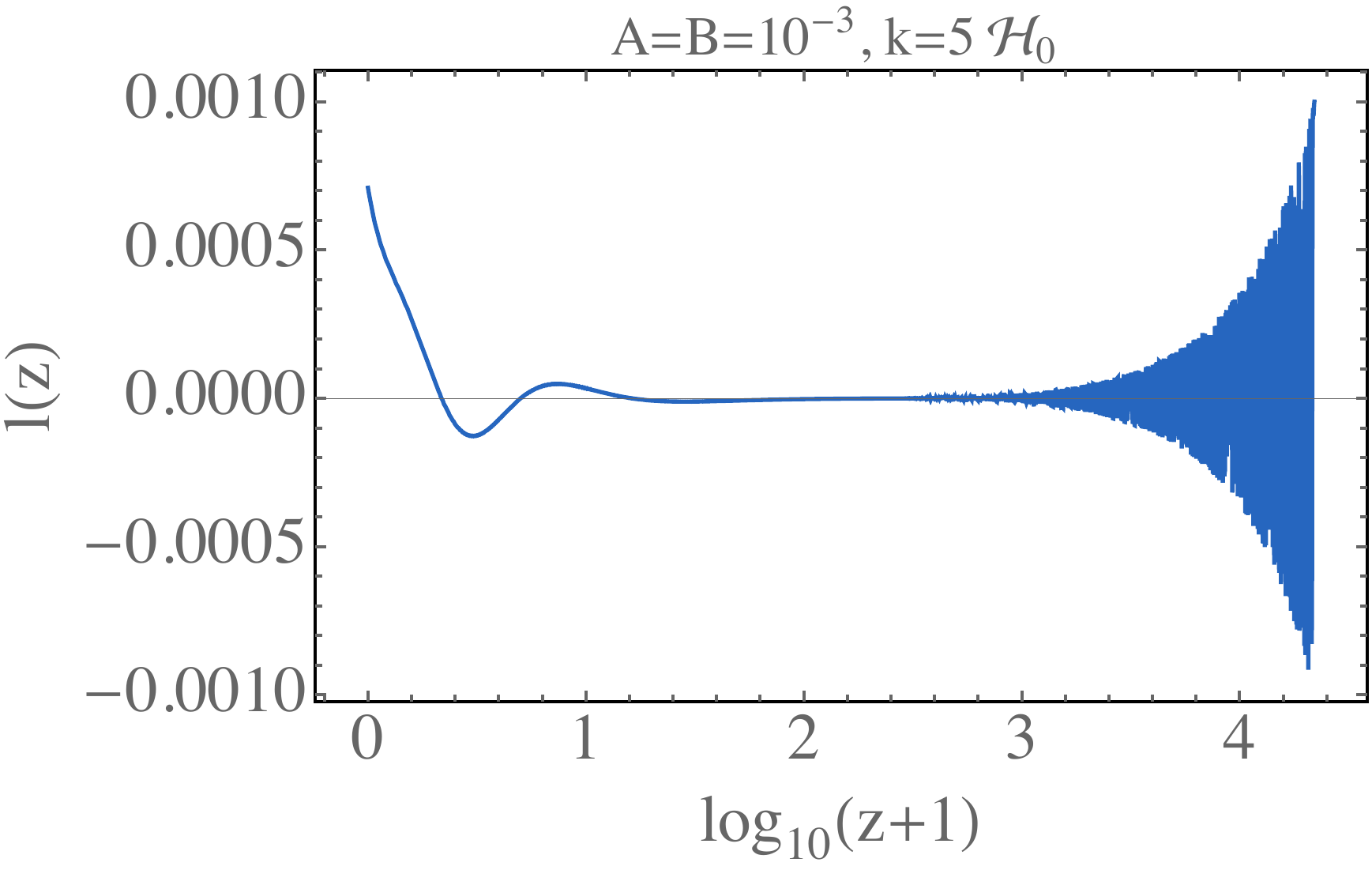}}
% \subfigure[\label{h1k5A4B4}]
  %    {\includegraphics[scale=0.40]{h1k5A4B4}}\qquad
 %\subfigure[\label{h2k5A4B4}]
   %   {\includegraphics[scale=0.40]{h2k5A4B4}}\qquad
       \subfigure[\label{h1k5A0B0}]
      {\includegraphics[scale=0.35]{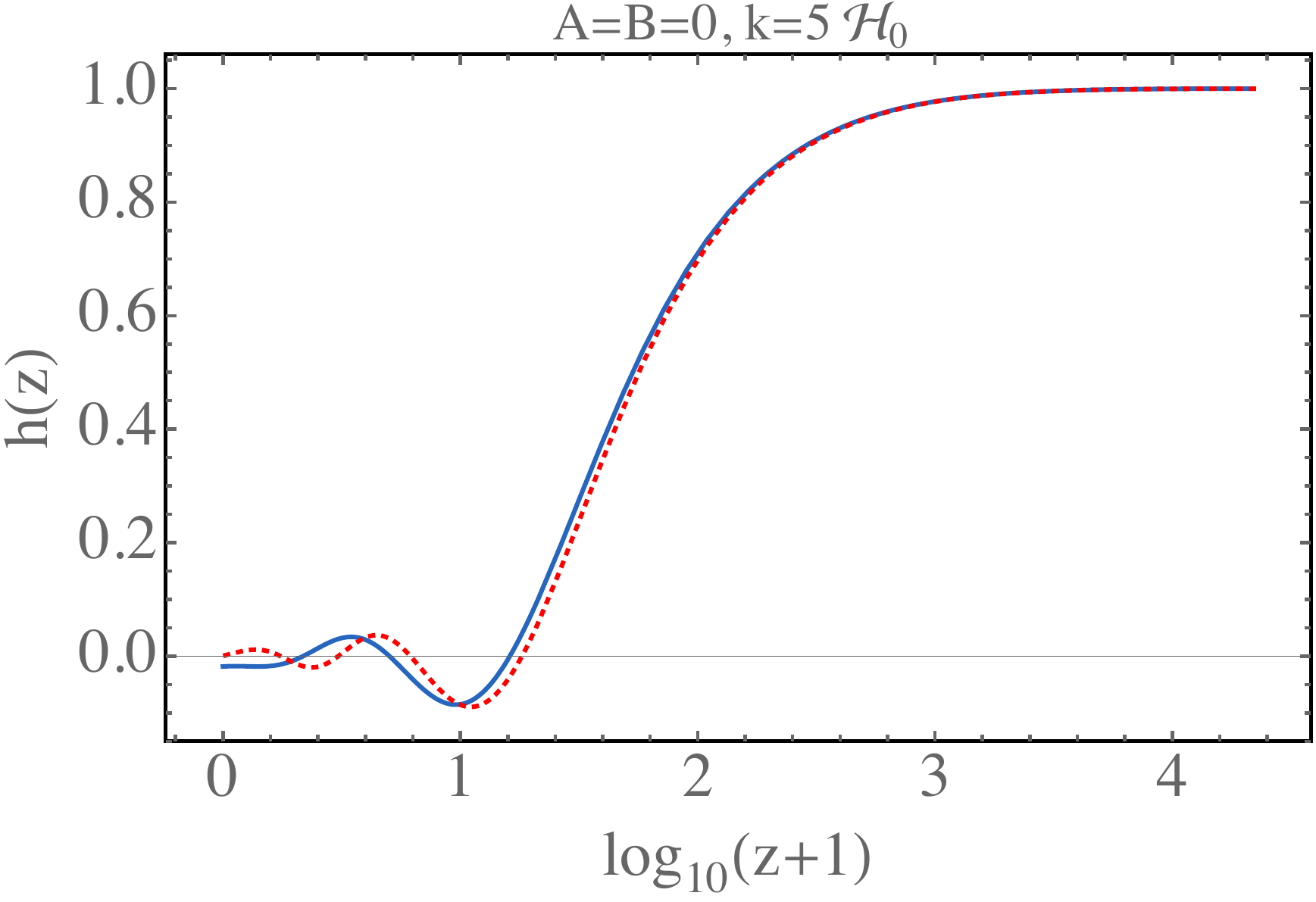}}\qquad\qquad
 \subfigure[\label{h2k5A0B0}]
      {\includegraphics[scale=0.35]{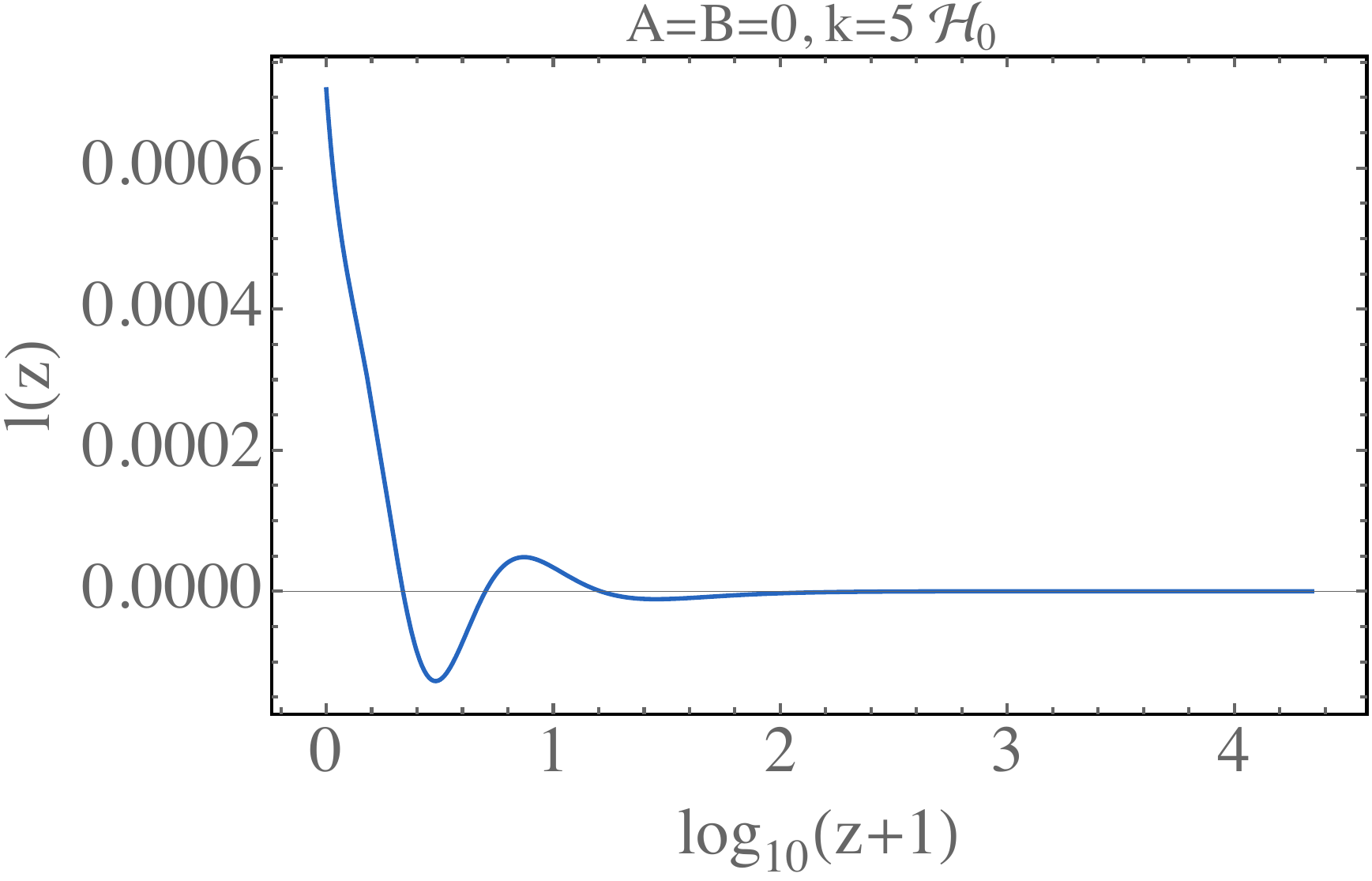}}
 \vspace{-0.2cm}
 \caption{\label{fig4} Tensor perturbations for initial conditions $h(z_{eq})=1\,,\,\, h'(z_{eq})=0$ and $\lc(z_{eq})=A\,,\,\, \lc'(z_{eq})=H_0\,B$. The evolution of tensor perturbations in the physical sector is plotted together with the $\Lambda CDM$ one with the same initial conditions (red, dotted line).  The result of the numerical integration is shown  for various choices of $A\simeq B$.  Note that even for $A=B=0$, the coupling generates an $\lc$-mode at late time, which has a (small) effect on the physical gravitational wave amplitude $h(z)$.}
 \end{figure}

In Fig.~\ref{fig4} we show the evolution of tensor perturbations keeping the mode $k$ fixed and we vary the initial conditions for the $\lc$ mode,  $A$ and $B$.  For $h$ we again choose for GR-like initial conditions $h(z_{eq})=1$ and $h'(z_{eq})=0$. 
The presence of the coupling of the $\lc$ mode to $h$ affects the evolution of the physical mode only at late time and only if the ratio $\lc/h$ at the beginning of matter domination is not too small(i.e. $A$ and $B \gsim 1$), see Figs. \ref{h1k5A7B7} and \ref{h1k5A5B5}. The evolution of $h$ at the beginning of matter domination is not affected by the coupling, even if initially $\lc/h\gg1$. 
Figs. \ref{h2k5A3B3} and \ref{h2k5A0B0} show that when $h$ enters the horizon and starts oscillating, this oscillation is transferred to the $\lc$ sector trough the coupling between the two modes \footnote{In the cases examined in Figs. \ref{h2k5A3B3} and \ref{h2k5A0B0}, the oscillation in the $\lc$ sector after horizon-crossing  is visible since $\lc$ is initially suppressed.}.


\section{Discussion and conclusion}\label{s5}

In this work we have developed a new formalism that allowed us to write the general expression for the bi-gravity action, perturbed to second order in the metric perturbations around generic backgrounds.  We explicitly give this mass term of the perturbations in terms of the arbitrary background metrics $\bar g$ and $\bar f$. Even though we use a special form (upper triangular) to derive the expressions, we finally give them in terms of the metric components valid in arbitrary coordinates.

 We then apply our formalism to cosmology. We present the generic action of cosmological perturbation which can be expressed in terms of the energy densities and pressures, $\rho_g,~\rho_h$ and $p_g,~p_f$ and two additional functions $\si_1$ and $\si_2$ of the background variables $r$ and $c$. 
 
 We finally study the evolution of tensor perturbations in the algebraic branch of bigravity cosmology.  Analyzing the positivity condition of the mass matrix for the canonically normalized tensor modes, we find that at late times, the tensor sector is affected by a tachyonic instability, which sets in earlier for smaller wave numbers. This result has been checked both analytically, solving the equations for tensor perturbations in the various cosmological epochs, and numerically. We find that the evolution of tensor perturbations in the algebraic branch manifests a late time instability in both the $g-$ and the $f-$ sectors. The order of magnitude of the instability strongly depends on the ratio between the two tensor modes at the beginning of matter domination. In particular if $\lc$ is of the same order as $h$, this instability manifests itself at present time in the physical sector for $k$ of the order of the Hubble parameter today. If instead initially $\lc$ is significantly suppressed with respect to $h$ the initial time of the instability is pushed towards the future. This last situation $\lc(z_e)\ll h(z_e)$ is the one which appears in our analytic study of the dynamics of the tensor sector during de\,Sitter inflation.  However, since inflation lies most probably in the strong coupling regime where the effective bimetric massive gravity theory studied in this paper is no longer valid, we have discussed the instability for arbitrary initial conditions.  Furthermore, independent of the initial conditions, for every mode $k$, there is a redshift (in the past or in the future) at which this mode will become unstable. 

Finally, some comments on our choice of the $\beta_i$ parameters are in order. We have chosen the values of the $\beta_i$ such that the constraints for the viability of the background evolution are satisfied (see section \ref{backkk}) and such that they reproduce the observed late time evolution of the Universe.  The remaining choice of the values for the $\beta_i$ does not affect the qualitative result found for the evolution of tensor perturbations and the appearance of a late-time instability. Indeed, in our analysis the physical quantity which is sensitive to a change in the parameters is the redshift at which a given mode becomes unstable, given by eq. (\ref{condition}) for $m^2\beta_i\simeq \mathcal{H}_0^2$. The redshift at which an instability shows up depends on the values of the $\beta_i$ and only if we choose a special tuning, e.g., such that the right hand side of eq. (\ref{condition}) is vanishing, the tensor sector is stable on all scales and redshifts. 

Of course, the analysis presented here is a linear perturbation analysis and it cannot  determine what happens at higher order. Whether a positive $h^4$ term or couplings to scalar and vector perturbations re-install stability cannot be decided with our analysis.

Another point in our analysis where the values of the $\beta_i$ enter is the expression for $c$ at the end of inflation. Our estimate $c\simeq \sqrt{H_I/H_0}\gg1$ is valid for $m^2\beta_i\simeq \mathcal{H}_0^2$. The fact that $c$ is so large at the end of inflation determines the suppression of the power spectrum of the $f$ tensor mode with respect to the one of the physical mode and it is crucial to push the beginning of the instability towards future times. This suppression can be reduced by tuning the value of $m^2\beta_i$ to lower the value of $c$ at the end of inflation, e.g. $c\simeq 1$ for $m^2\beta_i\simeq H_I^2$. However, this choice of the $\beta_i$ looks rather contrived. Indeed,  if we take $m^2 \beta_i\simeq H_I^2 \gg H_0^2$,  in eq. (\ref{Laeff}) a very significant fine tuning  is needed to obtain the correct  dark energy density today. 

 The mass term computed in this paper can be used for all future application of massive bigravity for cosmology or to discuss perturbations on an arbitrary background solution. It can also be used to investigate whether tachyonic modes are present for a given solution, rendering the theory unstable.
 
 \FloatBarrier
%\newpage

\section{Acknowledgements}
We thank Antonio De Felice, Tomi Koivisto, Nima Khosravi, Macarena Lagos, Giovanni Marozzi, Ermis Mitsou and Ignacy Sawicki for useful suggestions and discussions. This work is supported by the Swiss National Science Foundation.
\vspace{3cm}

%\newpage

 \appendix
 
 \section{Details of the mass term}\label{a:comp}
 
In this appendix, we present the details of the computation of the mass term $\MM^{\mu\nu\al\beta}(\bar g, \bar f)$ and the final result given in section~\ref{Mmunu}, written explicitly in terms of the background metrics. 

In the main text the mass matrix elements are given in terms of first and second derivatives of the functions $t_i(g,f)$ in eqs.~(\ref{e:MMhh} -- \ref{e:MMll}). Here we compute these derivatives.
 
Making use of eq.~(\ref{eq:relations}), we can express the first- and second-order perturbations of $t_i$ in terms of those of $s_i$. The variables $t_i$ and $s_i$ are defined in eqs. (\ref{eq:t_i}) and (\ref{eq:s_i}), respectively. We first introduce the  derivatives of  $s_i$ in the same way as those for the functions $t_i$ given in eqs.~(\ref{vart}) and (\ref{vartfin}),
\begin{align}
s_{i,\gc}^{\mu\nu} = \left.\frac{\dd s_i}{\dd g_{\mu\nu}}\right|_{g=\bar g, f=\bar f}&\,,\hspace{2.5 em}
 s_{i,\fc}^{\mu\nu} = \left.\frac{\dd s_i}{\dd f_{\mu\nu}}\right|_{g=\bar g, f=\bar f}\,,\label{sprime}\\
 s_{i,\gc\gc}^{\mu\nu\al\beta} = \left.\frac{1}{2}\frac{\dd^2 s_i}{\dd g_{\mu\nu}\dd g_{\al\beta}}\right|_{g=\bar g, f=\bar f}\,,\hspace{1.5 em}
 &s_{i, \gc\fc}^{\mu\nu\al\beta} = \left.\frac{\dd^2 s_i}{\dd g_{\mu\nu}\dd f_{\al\beta}}\right|_{g=\bar g, f=\bar f}\,,\hspace{1.5 em}
 s_{i,\fc\fc}^{\mu\nu\al\beta} = \left.\frac{1}{2}\frac{\dd^2 s_i}{\dd f_{\mu\nu}\dd f_{\al\beta}}\right|_{g=\bar g, f=\bar f}\,,\label{ssecond}
\end{align}
for $i\in \{1,2,3,4\}$. 

The first derivatives of $t_i$, which have been defined in eq. (\ref{vart}), can then be written as
\begin{align} \label{e:deltatmunu}
 &t_{1,\bullet}^{\mu\nu} = A\left\{ \bar{t}_4 \left(\bar{t}_1
   \bar{t}_4-\bar{t}_2 \bar{t}_3\right) s^{\mu\nu} _{1,\bullet}  - \bar{t}_3 \bar{t}_4 s^{\mu\nu} _{2,\bullet} - \bar{t}_1
   \bar{t}_4 s^{\mu\nu} _{3,\bullet}   + \left(\bar{t}_3-\bar{t}_1 \bar{t}_2\right) s^{\mu\nu} _{4,\bullet}  \right\} \, , \nn \\
 &t_{2,\bullet}^{\mu\nu} = A\left\{   - \bar{t}_3^2 \bar{t}_4  s^{\mu\nu} _{1,\bullet}   - \bar{t}_3 \bar{t}_4 \bar{t}_1 s^{\mu\nu} _{2,\bullet}  
   - \bar{t}_4 \bar{t}_1^2 s^{\mu\nu} _{3,\bullet} + \left(\bar{t}_3-\bar{t}_1\bar{t}_2\right) \bar{t}_1  s^{\mu\nu} _{4,\bullet}   \right\} \, , \nn \\  
 &t_{3,\bullet}^{\mu\nu} = A\left\{ - \bar{t}_3
   \bar{t}_4^2  s^{\mu\nu} _{1,\bullet}  - \bar{t}_1 \bar{t}_4^2 s^{\mu\nu} _{2,\bullet}  + \left(\bar{t}_3-\bar{t}_1 \bar{t}_2\right)\bar{t}_4 s^{\mu\nu} _{3,\bullet} + \left(\bar{t}_2 \bar{t}_3+\bar{t}_1
   \left(\bar{t}_4-\bar{t}_2^2\right)\right) s^{\mu\nu} _{4,\bullet}  \right\} \, , \nn \\  
 &t_{4,\bullet}^{\mu\nu} = \frac{s^{\mu\nu} _{4,\bullet}}{2 \bar{t}_4} \, ,
\end{align}
where
\be
A=\left[2\,\bar{t}_4\left(\bar{t}_3^{2}+\bar{t}_1^{2}\bar{t}_4-\bar{t}_1 \bar{t}_2 \bar{t}_3 \right)\right]^{-1} \,.
\ee
Here $\bullet$ denotes $\gc$ or $\fc$.

The mass term is well defined only if $A$ is finite, i.e., $\bar{t}_3^{2}+\bar{t}_1^{2}\bar{t}_4-\bar{t}_1 \bar{t}_2 \bar{t}_3\neq 0$ ($\bar{t}_4$, being the determinant of $\sqrt{\bar{g}^{-1}\bar{f}}$ is of course non-zero). If the eigenvalues $\la_i$ of $\bar{g}^{-1}\bar{f}$ are all positive, it is easy to check that $A$ is always negative\footnote{To do so, we can use the expressions of the $\bar{t}_{i}$ functions in terms of the $\lambda_{i}^{1/2}$ given in Eqs. (\ref{eq:t_i}): it turns out that $A$ is just the sum of negative terms.}.
This means that we have to require that the time directions of $f$ and of $g$ are sufficiently closely aligned such that $\bar{g}^{-1}\bar{f}$  is positive definite. This is also requested for our formalism to make sense, as otherwise the $\la_i$ might not have real roots.

To obtain an explicit expression for the second derivatives of $t_i$ in terms of these first derivatives and the second derivatives of $s_i$ eq. (\ref{ssecond}), we have to derive the relation between $s_i$ and $t_i$, eq. (\ref{eq:relations}), a second time.
A rather cumbersome but straightforward calculation leads finally to the following expressions:
\begin{subequations} \label{eq:delta2t}
\begin{align}
%t_1
 t^{\mu\nu\al\beta}_{1,\bullet\bullet} = & ~ A\biggl\{\bar{t}_4 \left(\bar{t}_1 \bar{t}_4-\bar{t}_2 \bar{t}_3\right) S_{1,\bullet\bullet}^{\mu\nu\al\beta}-\bar{t}_3 \bar{t}_4 S_{2,\bullet\bullet}^{\mu\nu\al\beta} -\bar{t}_1 \bar{t}_4 S_{3,\bullet\bullet}^{\mu\nu\al\beta} + \left(\bar{t}_3-\bar{t}_1 \bar{t}_2\right) S_{4,\bullet\bullet}^{\mu\nu\al\beta} \biggr\}  %OK
   \, , \\
   %t_2
 t^{\mu\nu\al\beta}_{2,\bullet\bullet} = & ~ A\biggl\{ - \bar{t}_3^2 \bar{t}_4 S_{1,\bullet\bullet}^{\mu\nu\al\beta}-\bar{t}_1 \bar{t}_3 \bar{t}_4 S_{2,\bullet\bullet}^{\mu\nu\al\beta}- \bar{t}_1^2 \bar{t}_4 S_{3,\bullet\bullet}^{\mu\nu\al\beta} + \bar{t}_1 \left(\bar{t}_3-\bar{t}_1\bar{t}_2\right)S_{4,\bullet\bullet}^{\mu\nu\al\beta}\biggr\} \, , \\
% t_3
 t^{\mu\nu\al\beta}_{3,\bullet\bullet} = & ~ A\biggl\{-\bar{t}_3 \bar{t}_4^2 S_{1,\bullet\bullet}^{\mu\nu\al\beta}-\bar{t}_1 \bar{t}_4^2 S_{2,\bullet\bullet}^{\mu\nu\al\beta}
   + \left(\bar{t}_3-\bar{t}_1 \bar{t}_2\right) \bar{t}_4
   S_{3,\bullet\bullet}^{\mu\nu\al\beta}+\left[\bar{t}_2 \bar{t}_3+\bar{t}_1 \left(\bar{t}_4-\bar{t}_2^2\right)\right]S_{4,\bullet\bullet}^{\mu\nu\al\beta}\biggr\} \, , \\
% t_4       
 t^{\mu\nu\al\beta}_{4,\bullet\bullet} = & ~ \frac{S_{4,\bullet\bullet}^{\mu \nu \alpha \beta }}{2 \bar{t}_4}  \,,    
  \end{align}     
   \end{subequations}
where $\bullet\bullet$ denotes $\gc\gc$, $\fc\fc$ or $\gc\fc$, and where we have introduced
\be
S_{q,\bullet\bullet}^{\mu\nu\al\beta}\equiv s_{q,\bullet\bullet}^{\mu\nu\al\beta}-\xi\cdot\sum_{p\in \mathcal{I}}(-1)^{q+p}t_{p,\bullet}^{\mu\nu}t_{2q-p,\bullet}^{\al\beta} \,.
\ee
The above sum goes over $\mathcal{I}\subset \{1,2,3,4\}$ such that both, $p$ and $2q-p$ are in  $\{1,2,3,4\}$ and  
%$\mathcal{I}\equiv \{\max[q-1, 1], q, \min[q+1,4]\}$
the  parameter $\xi$ is $\xi=1$ when $\bullet\bullet=\gc\gc$ or $\bullet\bullet=\fc\fc$ (since the symmetrization with respect to the exchange of the indices $(\mu, \nu)  \leftrightarrow (\al, \beta)$ applies in this case), whereas $\xi=2$ when $\bullet\bullet=\gc\fc$ (since the same symmetrization does not apply in this case). 

With this we have expressed the derivatives of the variables $t_i$ in terms of those of the $s_i$. 
Recalling the definition of the $s_i$, eq. (\ref{eq:s_i}) and the definition of the potential, eqs.~(\ref{e:U1})-(\ref{e:U4}), the derivatives of $s_i$  can be obtained directly by expanding the matrix 
\be
g^{-1}f = \left(\bar{g}(1+h)\right)^{-1} \bar{f} (1+\lc) = (1-h +h^2)\bar g^{-1}\bar{f} (1+\lc)  + {\cal O}(h^3)\,, 
\ee
where $h$ denotes $({h^\mu}_\nu) = (\bar g^{\mu\al}h_{\al\nu})$ and $\lc$ denotes $({\lc^\mu}_\nu) = (\bar f^{\mu\al}\lc_{\al\nu})$.  

We have also used the variations of the metric determinant, which appear in expressions (\ref{e:MMhh0}-\ref{e:MMhhi}):
\begin{eqnarray}
\left.\frac{\partial^{2}\sqrt{-\det g}}{\partial g_{\mu\nu}\partial g_{\alpha\beta}}\right|_{g=\bar{g},f=\bar{f}} & \equiv & \left(\sqrt{-\det g}\right),_{gg}^{\mu\nu\alpha\beta}=\frac{1}{4}\sqrt{-\det g}\left[\bar g^{\mu\nu}\bar g^{\al\beta} -\biggl(\bar g^{\mu\al}\bar g^{\nu\beta}+\bar g^{\mu\beta}\bar g^{\nu\al}\biggr)\right] \\
\left.\frac{\partial\sqrt{-\det g}}{\partial g_{\mu\nu}}\right|_{g=\bar{g},f=\bar{f}} & \equiv & \left(\sqrt{-\det g}\right),_{g}^{\mu\nu}=\frac{1}{2}\sqrt{-\det g}\: g^{\mu\nu}
\end{eqnarray}

To shorten the notation in what follows, we define $\bar{\Sigma}^{\mu}_{\phantom{\mu}\nu}\equiv \bar{g}^{\mu\rho}\bar{f}_{\rho\nu}$, and also $\bar{\Sigma}^{\mu\nu}\equiv \bar{\Sigma}^{\mu}_{\phantom{\mu}\rho}\bar{g}^{\rho\nu}$, $\left(\bar{\Sigma}^{2}\right)^{\mu\nu}\equiv \bar{\Sigma}^{\mu}_{\phantom{\mu}\rho_1}\bar{\Sigma}^{\rho_1}_{\phantom{\mu}\rho_2}\bar{g}^{\rho_{2}\nu}$, and generalizing $\left( \bar{\Sigma}^{k}\right)^{\mu\nu}\equiv \bar{\Sigma}^{\mu}_{\phantom{\mu}\rho_1}\bar{\Sigma}^{\rho_1}_{\phantom{\mu}\rho_2}\dots\bar{\Sigma}^{\rho_{k-1}}_{\phantom{\mu}\rho_k}\bar{g}^{\rho_{k}\nu}$. As usual, square brackets denote the trace,  e.g. $[\bar{\Sigma}] \equiv \mathrm{Tr}\bar{\Sigma}=\bar{\Sigma}^{\mu}_{\phantom{\mu}\mu}$.

A direct evaluation of $s_i$ and their first and second derivatives leads to\footnote{We use the well known identity  
$\frac{\dd F}{\dd g_{\mu\nu}} = -g^{\mu\al}g^{\nu\beta}\frac{\dd F}{\dd g^{\al\beta}}\,,$ valid for an arbitrary function $F(g^{-1})$.}
%\FloatBarrier

\begin{align} \label{eq:lista}
\bar{s}_1 &= \[\bar{\Sigma} \]\,,\\
s^{\mu\nu}_{1,g} &= -\bar{\Sigma}^{\mu\nu}\,,\\
s^{\mu\nu}_{1,f} &= \bar g^{\mu\nu} \, , \\
s^{\mu\nu\al\beta}_{1,gg} &=  \frac{1}{8}\left\{ \left[\bar{g}^{\al\nu}\bar{\Sigma}^{\mu\beta} + (\mu \leftrightarrow \nu) + (\al \leftrightarrow \beta)  +  (\mu \leftrightarrow \nu) (\al \leftrightarrow \beta) \right]  + \left[ \cdots \right]\big( (\mu,\nu) \leftrightarrow (\al,\beta)\big)\right\}   \nn\\
&\equiv   {\rm Sym}\left\{ \bar{g}^{\al\nu}\bar{\Sigma}^{\mu\beta} \right\}\,\\
s^{\mu\nu\al\beta}_{1,gf} &= -\frac{1}{2}\left(\bar g^{\mu\al}\bar g^{\nu\beta}+\bar g^{\mu\beta}\bar g^{\nu\al}\right) \,,\\
s^{\mu\nu\al\beta}_{1,ff} &= 0\,, 
\end{align}
\begin{align}
\bar{s}_2 &=  \frac{1}{2}\left( \[ \bar{\Sigma}\]^{2} - \[\bar{\Sigma}^{2}\]  \right)\,,\\   
s^{\mu\nu}_{2,g} &=  \left( \bar{\Sigma}^{2} \right)^{\mu\nu} - \[ \bar{\Sigma} \] \bar{\Sigma}^{\mu\nu}\,,\\
s^{\mu\nu}_{2,f} &= \bar{g}^{\mu\nu}\[\bar{\Sigma}\] - \bar{\Sigma}^{\mu\nu}\,,\\
s^{\mu\nu\al\beta}_{2,gg} &= {\rm Sym}\left\{ \bar{g}^{\mu\al}\left( \[\bar{\Sigma}\]\bar{\Sigma}^{\beta\nu} -\left(\bar{\Sigma}^{2}\right)^{\beta\nu} \right)+ \frac{1}{2}\bar{\Sigma}^{\mu\nu}\bar{\Sigma}^{\al\beta}  - \frac{1}{2}\bar{\Sigma}^{\al\mu}\bar{\Sigma}^{\beta\nu} \right\}\,,\\
 s^{\mu\nu\al\beta}_{2,gf} &= \frac{1}{4} \[ \bar{g}^{\beta\mu}\bar{\Sigma}^{\al\nu} + \bar{g}^{\al\nu}\bar{\Sigma}^{\beta\mu} - \bar{g}^{\al\beta}\bar{\Sigma}^{\mu\nu} - \bar{g}^{\al\mu}\bar{g}^{\beta\nu} \[\bar{\Sigma}\] + (\mu \leftrightarrow \nu) + (\al \leftrightarrow \beta)  +  (\mu \leftrightarrow \nu) (\al \leftrightarrow \beta) \]  \nn\\
 &\equiv   {\rm sym}\left\{ \bar{g}^{\beta\mu}\bar{\Sigma}^{\al\nu} + \bar{g}^{\al\nu}\bar{\Sigma}^{\beta\mu} - \bar{g}^{\al\beta}\bar{\Sigma}^{\mu\nu} - \bar{g}^{\al\mu}\bar{g}^{\beta\nu} \[\bar{\Sigma}\]  \right\}\, , \\
 s^{\mu\nu\al\beta}_{2,ff} &= \frac{1}{2}\bar g^{\al\beta}\bar g^{\mu\nu} - \frac{1}{4}\left( \bar g^{\al\nu}\bar g^{\beta\mu} + \bar g^{\al\mu}\bar g^{\beta\nu}  \right)   \, , \\  &  \nonumber 
 \end{align}
\begin{align}
  \bar{s}_3 &= \frac{1}{6} \left( \[ \bar{\Sigma}\]^{3} -3\[\bar{\Sigma} \] \[\bar{\Sigma}^{2}\] +2\[\bar{\Sigma}^{3} \] \right) \, , \\
 s^{\mu\nu}_{3,g} &= \[ \bar{\Sigma} \] \left( \bar{\Sigma}^{2} \right)^{\mu\nu} - \left(\bar{\Sigma}^{3} \right)^{\mu\nu} + \frac{1}{2}\bar{\Sigma}^{\mu\nu} \left( \[\bar{\Sigma}^{2}\] - \[\bar{\Sigma} \]^{2} \right)   \, , \\
 s^{\mu\nu}_{3,f} &= \left(\bar{\Sigma}^{2}\right)^{\mu\nu} - \bar{\Sigma}^{\mu\nu} \[\bar{\Sigma}\] +\frac{1}{2}\bar{g}^{\mu\nu}\left(\[\bar{\Sigma} \]^{2} - \[\bar{\Sigma}^{2} \] \right)\, , \\
s^{\mu\nu\al\beta}_{3,gg} &= {\rm Sym} \biggl\{ \bar{\Sigma}^{\mu\al}\left( \bar{\Sigma}^{2}\right)^{\beta\nu} - \left(\bar{\Sigma}^{2} \right)^{\mu\nu}\bar{\Sigma}^{\al\beta} + \frac{1}{2}\[\bar{\Sigma}\]\left(\bar{\Sigma}^{\mu\nu}\bar{\Sigma}^{\al\beta}-\bar{\Sigma}^{\mu\al}\bar{\Sigma}^{\nu\beta} \right) \nn\\
&\phantom{{\rm sym}000}+ \bar{g}^{\nu\al}\left( \left( \bar{\Sigma}^{3} \right)^{\mu\beta} - \left(\bar{\Sigma}^{2}\right)^{\mu\beta}\[ \bar{\Sigma}\] \right) +\frac{1}{2}\bar{g}^{\mu\al}\bar{\Sigma}^{\nu\beta}\left( \[\bar{\Sigma} \]^{2} - \[\bar{\Sigma}^{2} \] \right)   \biggr\}  \, , \\
s^{\mu\nu\al\beta}_{3,gf} &= {\rm sym}\biggl\{ 2 \bar g^{\nu\beta} \left(\bar{\Sigma}^{\mu\al} \[\bar{\Sigma} \] - \left( \bar{\Sigma}^{2}\right)^{\mu\al}\right) +\bar{g}^{\al\beta}\left( \left( \bar{\Sigma}^{2} \right)^{\mu\nu}-\bar{\Sigma}^{\mu\nu}[\bar{\Sigma} ]  \right) + \bar{\Sigma}^{\mu\nu}\bar{\Sigma}^{\al\beta} - \bar{\Sigma}^{\mu\beta}\bar{\Sigma}^{\al\nu}\nn \\
&+ \frac{1}{2}\bar{g}^{\mu\al}\bar{g}^{\nu\beta}\left( \[\bar{\Sigma}^{2} \]- \[\bar{\Sigma}\]^{2} \right)  \biggr\}    \, , \\%
s^{\mu\nu\al\beta}_{3,ff} &= {\rm Sym}\biggl\{ \frac{1}{2}\bar{g}^{\mu\nu}\bar{g}^{\al\beta}\[ \bar{\Sigma}\]-\frac{1}{2}\bar{g}^{\mu\beta}\bar{g}^{\nu\al}\[ \bar{\Sigma}\] + \bar{g}^{\mu\al}\bar{\Sigma}^{\nu\beta}-\bar{g}^{\al\beta}\bar{\Sigma}^{\mu\nu} \biggr\}  \, ,\nn 
\end{align}
\begin{align}
\bar{s}_4 &= \det\left(\bar g^{-1}\bar{f} \right) \,, \\
s^{\mu\nu}_{4,g} &=  -\bar{s}_4 \bar g^{\mu\nu}  \, , \\
s^{\mu\nu}_{4,f} &=  \bar{s}_4 \bar f^{\mu\nu}  \, , \\
s^{\mu\nu\al\beta}_{4,gg} &= \frac{\bar{s}_4}{2}\left(\bar g^{\mu\nu}\bar g^{\al\beta} + \frac{1}{2}\bar g^{\mu\al}\bar g^{\nu\beta}+ \frac{1}{2}\bar g^{\nu\al}\bar g^{\mu\beta}\right) \, , \\
s^{\mu\nu\al\beta}_{4,gf} &= - \bar{s}_4\bar g^{\mu\nu}\bar f^{\al\beta} \, , \\
s^{\mu\nu\al\beta}_{4,ff} &= \frac{\bar{s}_4}{2}\left( \bar f^{\mu\nu}\bar f^{\al\beta} - \frac{1}{2}\bar f^{\mu\al}\bar f^{\nu\beta} - \frac{1}{2}\bar f^{\mu\beta}\bar f^{\nu\al} \right)\,.
 \end{align}
The operator ${\rm Sym}\{\cdots \}$ indicates symmetrization in $ (\mu \leftrightarrow \nu)$, $(\al \leftrightarrow \beta)$, $(\mu \leftrightarrow \nu) (\al \leftrightarrow \beta)$ and  $(\mu, \nu)  \leftrightarrow (\al, \beta)$, whereas the operator ${\rm sym}\{\cdots \}$ indicates symmetrization in $ (\mu \leftrightarrow \nu)$, $(\al \leftrightarrow \beta)$ and $(\mu \leftrightarrow \nu) (\al \leftrightarrow \beta)$ but not  $(\mu, \nu)  \leftrightarrow (\al, \beta)$.

These are the expressions for the derivatives of the functions $s_i$ which enter the expressions for the derivatives of the $t_i$, eqs. (\ref{eq:delta2t}), which in turn enter in the expression for 
$\MM^{\mu\nu\al\beta}$, eqs. (\ref{e:MMhh})-(\ref{e:MMll}). Not surprisingly, the expressions for the second derivatives of $s_2$ and $s_3$ are lengthy. The $s_{i,g}^{\mu\nu}$ and the $s_{i,gg}^{\mu\nu\al\beta}$ have already been computed in Ref.~\cite{Guarato:2013gba} and the $s_{i,f}^{\mu\nu}$ and the $s_{i,ff}^{\mu\nu\al\beta}$ are less complicated. The terms   $s_{i,gf}^{\mu\nu\al\beta}$ are quite cumbersome and they are new.

\section{Parametrization of the cosmological mass term}\label{FRW}
 
We give here the explicit expressions for the functions which parametrize the mass tensor on cosmological  backgrounds, as presented in Section \ref{s4}:

\begin{eqnarray} 
\alpha_{\gc} & = & -\frac{1}{2}\left(\beta_{0}+\beta_{3}r^{3}+3\beta_{2}r^{2}+3\beta_{1}r\right),\label{ah} \\
\gamma_{\gc} & = & -\frac{1}{2}\left(\beta_{0}+\beta_{2}r^{2}+2\beta_{1}r\right),\\
\epsilon_{\gc} & = & \frac{1}{2}\left(\beta_{0}+\frac{(3c+2)r}{c+1}\beta_{1}+\beta_{2}\frac{(3c+1)r^{2}}{c+1}+\frac{\beta_{3}cr^{3}}{c+1}\right),\\
\eta_{\gc} & = & \frac{1}{2}\left(\beta_{0}+\beta_{2}cr^{2}+\beta_{1}(c+1)r\right),\\
\sigma_{\gc} & = & -\frac{1}{2}\left(2\beta_{0}+\beta_{3}cr^{3}+\beta_{2}(3c+1)r^{2}+\beta_{1}(2c+3)r\right),\\
\nonumber\\
\alpha_{\fc} & = & -\frac{1}{2c^{3}}\left(\beta_{4}+3\beta_{3}r+3\beta_{2}r^{-2}+\beta_{1}r^{-3}\right),\\
\gamma_{\fc} & = & -\frac{1}{2c}\left(\beta_{4}+2\beta_{3}r^{-1}+\beta_{2}r^{-2}\right),\\
\epsilon_{\fc} & = & \frac{1}{2c}\left(\beta_{4}+\frac{\beta_{3}(2c+3)r^{-1}}{(c+1)}+\frac{\beta_{2}(c+3)r^{-2}}{(c+1)}+\frac{\beta_{1}r^{-3}}{(c+1)}\right),\\
\eta_{\fc} & = & \frac{1}{2}\left(\beta_{4}c+\beta_{3}(c+1)r^{-1}+\beta_{2}r^{-2}\right),\\
\sigma_{\fc} & = & -\frac{1}{2}\left(2\beta_{4}c+\beta_{3}(3c+2)r^{-1}+\beta_{2}(c+3)r^{-2}+\beta_{1}r^{-3}\right),\\
\nonumber \\
\alpha_{\gc\fc} & = & 0,\\
\gamma_{\gc\fc} & = & -\frac{1}{r}\left(\beta_{1}+2\beta_{2}r+\beta_{3}r^{2}\right),\\
\gamma_{\fc\gc} & = & -\frac{1}{rc}\left(\beta_{1}+2\beta_{2}r+\beta_{3}r^{2}\right),\\
\epsilon_{\gc\fc} & = & \frac{1}{(1+c)r}\left(\beta_{1}+2\beta_{2}r+\beta_{3}r^{2}\right),\\
\eta_{\gc\fc} & = & \frac{1}{r}\left(\beta_{1}+\beta_{3}cr^{2}+\beta_{2}(c+1)r\right),\\
\sigma_{\gc\fc} & = & -\frac{1}{r}\left(\beta_{1}+\beta_{3}cr^{2}+\beta_{2}(c+1)r\right).\label{shl}
\end{eqnarray}

Note that for $c=1$ an arbitrary coefficient $\mu_f$ can be obtained from the corresponding $\mu_h$ by replacing $\beta_i\ra\beta_{4-i}$ and $r\ra r^{-1}$ as a consequence of the symmetry (\ref{e:sym}). As $c$ has been set to $1$ for the metric $\bar g$ the behaviour with $c$ is less evident. Setting $c=1$ gives the general mass term for conformally related metrics $g$ and $f$. In this case, in the algebraic branch all coupling terms $\mu_{gf}$ vanish and the mass matrix completely decouples. For conformally related metrics the algebraic branch reduces to two independent copies of Einstein gravity.

We can  conveniently express all parametrization functions in terms of $\rho_g$, $\rho_f$, $p_g$, $p_f$  introduced in eqs.~(\ref{eqF1}-\ref{ac2}) and two functions $\sigma_{1}\equiv\beta_{1}+2\beta_{2}r+\beta_{3}r^{2}$ and $\sigma_{2}\equiv\beta_{1}+\beta_{2}(c+1)r+\beta_{3}cr^{2}$ as follows:

\begin{equation}
\alpha_{\gc}=-\frac{8\pi G\rho_{\gc}}{2m^{2}}\,,\hspace{0.4cm}\epsilon_{\gc}=-\frac{r\sigma_{1}}{2(c+1)}+\frac{8\pi G \rho_{\gc}}{2m^{2}},\hspace{0.4cm}\eta_{\gc}=-\frac{r\sigma_{2}}{2}-\frac{8\pi G p_{\gc}}{2m^{2}}\,,
\end{equation}
\begin{equation}
\sigma_{\gc}=\frac{r\sigma_{2}}{2}+\frac{8\pi G p_{g}}{m^{2}}\,,\hspace{0.5cm}\gamma_{\gc}=\frac{r\sigma_{1}}{2}-\frac{\rho_{\gc}}{2m^{2}}\,,
\end{equation}

\begin{equation}
\alpha_{f}=-\frac{ 8\pi G\rho_{f}}{2m^{2}c^{3}}\,,\hspace{0.4cm}\epsilon_{\fc}=-\frac{\sigma_{1}}{2(c+1)r^{3}}+\frac{8\pi G \rho_{f}}{2m^{2}c},\hspace{0.4cm}\eta_{\fc}=-\frac{\sigma_{2}}{2r^{3}}-\frac{8\pi G p_{f}}{2m^{2}}\,,
\end{equation}
\begin{equation}
\sigma_{\fc}=\frac{\sigma_{2}}{2r^{3}}+\frac{8\pi G p_{f}}{m^{2}}\,,\hspace{0.5cm}\gamma_{\fc}=\frac{\sigma_{1}}{2cr^{3}}-\frac{8\pi G \rho_{f}}{2m^{2}c}\,,
\end{equation}

\begin{equation}
\alpha_{\gc\fc}=0\,,\hspace{1.0cm}\epsilon_{\gc\fc}=\frac{\sigma_{1}}{r(1+c)},\hspace{0.6cm}\eta_{\gc\fc}=\frac{\sigma_{2}}{r}\,,
\end{equation}
\begin{equation}
\sigma_{\gc\fc}=-\frac{\sigma_{2}}{r}\,,\hspace{0.6cm}\text{\ensuremath{\gamma_{\gc\fc}}}=c\text{\ensuremath{\gamma_{\fc\gc}=-\frac{\sigma_{1}}{r}}}\,.
\end{equation}
 Once the Friedmann equations (\ref{eqF1}-\ref{ac2}) are used to replace $\rho_\bullet$ and $p_\bullet$ by $\HH_\bullet$ and $\HH'_\bullet$, all these terms cancel with contributions from the in the kinetic part of the action and $\sigma_1$ and $\sigma_2$ are the only functions of the $\beta_i$ which remain\footnote{The functions $\sigma_1$ and $\sigma_2$  correspond respectively to $f_2$ and $f_1$ in \cite{Comelli:2012db}}. The final action for scalars, vectors and tensors after this simplification is given in \cite{DeFelice:2014nja} for a  specific gauge choice.

\section{Analytic results for the evolution of tensor perturbations}\label{analytic tens}

In this appendix we collect the results obtained by solving analytically the equations of tensor perturbations (\ref{peg}) and (\ref{pef}) in the early de\,Sitter, radiation- , matter and late de\,Sitter-dominated epochs.

\subsection{Analytic results during de\,Sitter inflation}\label{Hig-tach}
We start by embedding the model in inflation.  We consider a model of inflation with a single scalar filed coupled to the $g$-metric, with potential $V(\phi)=M_{\phi}^2 \phi^2/2$. Deep in the inflationary era the inflaton is slowly rolling. Since $p_{\phi}=-\rho_{\phi}$, it is legitimate to model this period as a de\,Sitter phase with constant Hubble parameter $H\simeq H_{I}=const$. It follows that
\be
c\simeq \sqrt{\frac{\rho_{\phi}}{\Lambda_c}}\approx \frac{H_I}{H_0}\,.
\ee
With this and $(a'/a)^2 =\frac{1}{2}a''/a =1/\tau^2$, the equations of motion for the canonically normalized variables become
\be
Q_h''-\frac{2}{\tau^2}Q_h+k^2 Q_h+\frac{1}{\tau^2}\left(\frac{H_0}{H_I}\right)\left(Q_h-\sqrt{\frac{H_I}{H_0}}Q_\lc\right)=0\,,
\ee
\be
Q_\lc''-\frac{2}{\tau^2}Q_\lc+\left(\frac{H_I}{H_0}\right)^2 k^2 Q_\lc-\frac{1}{\tau^2} \sqrt{\frac{H_0}{H_I}}\left(Q_h-\sqrt{\frac{H_I}{H_0}}Q_\lc\right)=0\,.
\ee
Taking into account that $H_I \gg H_0$, these equations decouple and they can be approximated as
\be\label{Q11}
Q_h''-\frac{2}{\tau^2}Q_h+k^2 Q_h=0\,,
\ee
\be\label{Q22}
Q_\lc''-\frac{1}{\tau^2}Q_\lc+\left(\frac{H_I}{H_0}\right)^2 k^2 Q_\lc=0\,.
\ee
Eq. (\ref{Q11}) is equal to the one for the canonically normalized tensor mode in GR. Requiring that for $|k\tau|\gg1$ we recover the quantum vacuum solution
\be
Q_h=\frac{1}{\sqrt{2 k}}e^{-i k\tau}\,,
\ee
we find the standard result
\be
Q_h=\frac{1}{\sqrt{2 k}}\left(1-\frac{i}{k\tau}\right)e^{-i k\tau}\,.
\ee
On super-Hubble scales, $|\tau k|<1$, the canonical variable grows, $Q_h\propto -1/\tau \propto a$ so that $h\propto a^{-1}Q_h$ remains constant.

In eq. (\ref{Q22}) the mass term $-1/\tau^2$ becomes relevant only for 
\be
\frac{1}{\tau^2}= H_I^2 a^2 = \frac{H_I^2}{(1+z)^2} > k^2\frac{H_I^2}{H_0^2}\, ,\, \quad \mbox{ hence for} \quad
k\, \lsim\,\frac{H_0}{(1+z)}  < \frac{H_0}{(1+z_e)} \,, 
\ee
where $z_e$ is the redshift at the end of inflation. This corresponds to a huge scale irrelevant for cosmological observations. On presently measurable scales, $k\gsim H_0$,  eq. (\ref{Q22}) can be approximated as an harmonic oscillator equation, with solution
\be
Q_\lc=\frac{1}{\sqrt{2 c k}}\,e^{-i\,k c \tau}\,,\hspace{1 em} c\approx \frac{H_I}{H_0}\,.
\ee
As $c$ is very large, this solution  oscillates rapidly with small amplitude.
 
To conclude this analysis, we determine the power spectra for super horizon modes $|k\tau|\ll1$ at the end of inflation. We find
\be
P_h(k)=2\cdot \frac{k^3|Q_h|^2}{a^2 M_P^2}\simeq\left(\frac{H_I}{M_P}\right)^2\,,
\ee
\be
P_\lc(k)=2\cdot c\cdot \frac{k^3|Q_\lc|^2}{a^2 M_P^2}\simeq \left(\frac{H_I}{M_P}\right)^2 \left(\frac{k}{H_0}\right)^2\left(\frac{H_0}{H_I}\right)^2(1+z_e)^2\simeq P_h(k) \left(\frac{k}{H_0}\right)^2\left(\frac{H_0}{H_I}\right)\,,
\ee
where we have approximated at the end of inflation $(1+z_e)^2\simeq H_I/H_0$. 

The power spectrum of the $\lc$-mode is highly suppressed for all modes of cosmological interest. Therefore, we expect that at very high redshift, the evolution of the physical tensor mode $h$ will not be affected by the coupling with $\lc$.  

However, typical inflation scales are expected to lie above the strong coupling scale and so the results for the inflationary power spectra may not be relevant.

\subsection{Analytic results in radiation-dominated epoch}

In the  radiation dominated era $a\propto \tau$, $\mathcal{H}=1/\tau$ and 
\be
c\simeq \sqrt{\frac{\rho_r}{\Lambda_c}}\approx \sqrt{\frac{3\mathcal{H}_0^2\,\Omega_{r 0}}{m^2\left(\beta_4 \bar r^2+2 \bar r\beta_3+\beta_2\right)}}\,\,\frac{1}{a^2}\equiv \frac{\bar{c}_r}{a^2}\,,\hspace{2.5 em}
\frac{c'}{c}=-2\mathcal{H}\,,
\ee
where $\bar c_r\sim \sqrt{3H_0^2\Omega_{r0}/\La_c}\simeq 0.12$ for our parameter choice. The two equations for the tensor modes (\ref{peg}) and (\ref{pef}) can now be written as 
\be\label{eqq1}
h^{''}+\frac{2}{\tau}\,h^{'}+k^2 h+\mathcal{R}_r\left(h-\lc\right)=0\,,
\ee
\be\label{eqq2}
\lc^{''}+\frac{4}{\tau}\,\lc^{'}+\frac{\mathcal{K}^2_r}{a^4}\,\lc-\mathcal{S}_r \left(h-\lc\right)=0\,,
\ee
where 
\be
\mathcal{K}^2_r=\bar{c}^2_r\,k^2\,,
\ee
\be
\mathcal{R}_r=m^2 a^2 C+ m^2 D\, \bar{c}_r \approx m^2 D\, \bar{c}_r \simeq -0.2H_0^2\,,
\ee
\be
\mathcal{S}_r=\frac{m^2 C}{\bar r^2}\,\bar{c}_r+\frac{m^2 D}{\bar r^2} \,\frac{\bar{c}^2_r}{a^2}\approx \frac{m^2 D}{\bar r^2} \,\frac{\bar{c}^2_r}{a^2}\equiv\frac{\bar{\mathcal{S}}_r}{a^2} \simeq -0.014H_0^2/a^2\,.
\ee
Here we have used that $a=1/(1+z)\ll 1$ with our normaization of $a_0=1$.

 If we consider super-Hubble modes and we neglect the $k$-term in eq. (\ref{eqq1}), also the term proportional to $\mathcal{R}_r\, h$ has to be neglected since $k^2\,h\geq |\mathcal{R}_rh| \approx 0.2\mathcal{H}_0^2\,h$. Therefore, recalling that at the end of inflation $h>\lc$, eq. (\ref{eqq1})  is decoupled and given by 
\be\label{1def}
h^{''}+\frac{2}{\tau}\,h^{'}=0\,,
\ee
with solution
\be
h=d_1+d_2\,\frac{\tau_0}{\tau}\,.
\ee
In eq. (\ref{eqq2}), the coupling is negligible if  % given that in the early radiation era 
\be\label{e:limdec}
\left(\frac{\mathcal{\bar{S}}_r}{a^2}\,h\right)/\left(\frac{\mathcal{K}_r^2}{a^4}\,\lc\right)\simeq \frac{10}{(1+z)^2}\left(\frac{H_0}{k}\right)^2\left(\frac{h}{\lc}\right)\approx \frac{10}{(1+z)^2}\frac{H^2_0}{kH_I}\ll1\,.
\ee
In the last inequality we have used the inflationary result for $h/\lc$. This is certainly satisfied early on but may be violated  for large scales at late time, when $z\ra 3000$.  
If we set $h/\lc \sim 1$ it is even satisfied until equality. 
As long as eq.~(\ref{e:limdec}) is satisfied also eq. (\ref{eqq2}) is decoupled and for  $\lc\ll h$, it can be approximated as
\be\label{2def}
\lc^{''}+\frac{4}{\tau}\,\lc^{'}+\frac{T^2}{\tau^4}\lc=0\,,
\ee
where $T\equiv \bar{c}_r\,\left(k/\mathcal{H}_0\right)\left(\Omega_{r0}\mathcal{H}_0\right)^{-1} \simeq 3\times 10^3k/H_0^2$. The analytic solution is given by
\begin{align}\label{sol2}
\lc &= c_1\left(\frac{T}{\tau}+i\right) e^{i \frac{T}{\tau}}+c_2\left(\frac{T}{\tau}-i\right) e^{-i \frac{T}{\tau}}
\simeq c_1 \frac{T}{\tau} e^{i \frac{T}{\tau}}+c_2 \frac{T}{\tau} e^{-i \frac{T}{\tau}}\,.
\end{align}

Summarizing, we found that in radiation for super-horizon scales, $h$ evolves as in GR while $\lc$ has a constant and a decaying mode and it oscillates with the phase $3\times 10^3k/(\tau H_0^2)$.

\subsection{Analytic results in matter-dominated epoch}
In the matter dominated era $a\propto \tau^2$, $\mathcal{H}=2/\tau$ and 
\be
c\simeq \sqrt{\frac{\rho_m}{\Lambda_c}}\approx \sqrt{\frac{3\mathcal{H}_0^2\,\Omega_{m0}}{m^2\left(\beta_4 \bar r^2+2 \bar r\beta_3+\beta_2\right)}}\,\,\frac{1}{a^{3/2}}\equiv \frac{\bar{c}_m}{a^{3/2}}\,,\hspace{2.5 em}
\frac{c'}{c}=-\frac{3}{2}\mathcal{H}\,.
\ee
With our choice of parameters $\bar c_m\simeq 10$.
The two equations for the tensor modes (\ref{peg}) and (\ref{pef}) can be written as 
\be\label{eqqq1}
h^{''}+\frac{4}{\tau}\,h^{'}+k^2 h+\mathcal{R}_m\left(h-\lc\right)=0\,,
\ee
\be\label{eqqq2}
\lc^{''}+\frac{7}{\tau}\,\lc^{'}+\frac{\mathcal{K}^2_m}{a^3}\,\lc-\mathcal{S}_m \left(h-\lc\right)=0\,,
\ee
where 
\be
\mathcal{K}^2_m=\bar{c}^2_m\,k^2\,,
\ee
\be
\mathcal{R}_m=m^2 a^2 C+ m^2 D\, \bar{c}_m\,a^{1/2} \approx m^2 D\, \bar{c}_m\,a^{1/2} \simeq -19H_0^2 a^{1/2}\,,
\ee
\be
\mathcal{S}_m=\frac{m^2 C}{\bar r^2}\,\bar{c}_m\,a^{1/2}+\frac{m^2 D}{\bar r^2} \,\frac{\bar{c}^2_m}{a}\approx \frac{m^2 D}{\bar r^2} \,\frac{\bar{c}^2_m}{a} \equiv \frac{\bar{\mathcal{S}}_m}{a}  \simeq -\frac{95H_0^2}{a} \,.
\ee
In eq. (\ref{eqqq1}),  $\mathcal{R}_m\, h\simeq 19H_0^2\,a^{1/2}\,h\simeq \left(\mathcal{H}_0/\mathcal{H}\right)\,\mathcal{H}_0^2\,h$. As long as this is smaller than $k^2\,h$, and the term $\mathcal{R}_m\,h$ is subdominant, since we expect $\lc\leq h$, eq. (\ref{eqqq1}) reduces to
 \be\label{eqqqq1}
h^{''}+\frac{4}{\tau}\,h^{'}+k^2 h=0\,.
\ee
This is the same equation that we find in GR during matter domination. On super-horizon scales, it  has the following solution
\be
h=d_1+\frac{d_2}{\tau^3}\,.
\ee
At late time, when $a^{1/2}\sim 1$, the coupling becomes significant and this approximation no longer holds on large scales.
In eq. (\ref{eqqq2}), the coupling is negligible as long as 
 \be
\left(\frac{\mathcal{\bar{S}}_m}{a}\,h\right)/\left(\frac{\mathcal{K}_m^2}{a^3}\,\lc\right)\simeq \frac{10H^2_0}{k^2} \frac{1}{\left(1+z\right)^2}\left(\frac{h}{\lc}\right)\ll 1\,.
\ee
This condition is satisfied for $h/\lc\ll (1+z)^2$. If we neglect the coupling, eq. (\ref{eqqq2}) becomes 
\be
\lc^{''}+\frac{7}{\tau}\,\lc^{'}+\frac{T^4}{\tau^6}\,\lc=0\,,
\ee
where $T^2=8\,\bar{c}_m\, \frac{k}{\mathcal{H}_0}\,\mathcal{H}_0^{-2}\,\Omega_{m0}^{-3/2} \simeq 490 k/H_0^3$. This equation is decoupled and can be solved analytically. The solution is given by 
\begin{align*}
\lc&=\sqrt{4+\frac{T^4}{\tau^4}}\left[c_1\,e^{-i \arctan\left(\frac{T^2}{2\,\tau^2}\right)}e^{\frac{i}{2}\left(\frac{T}{\tau}\right)^2}+c_2\,e^{i \arctan \left(\frac{T^2}{2\,\tau^2}\right)}\,e^{-\frac{i}{2}\left(\frac{T}{\tau}\right)^2}\right]\\
&\simeq \sqrt{4+\frac{T^4}{\tau^4}}\left[c_1\,e^{\frac{i}{2}\left(\frac{T}{\tau}\right)^2}+c_2\,e^{-\frac{i}{2}\left(\frac{T}{\tau}\right)^2}\right]\,.
\end{align*}

Summarizing, we find that also during matter domination, as long as couplings can be neglected, $h$ evolves as in GR while $\lc$ has a constant and a decaying mode and it oscillates rapidly. At late times, especially for the large modes with $k\sim H_0$ coupling can in general not be neglected and the system has to be solved numerically.

 \subsection{Analytic results in the late de\,Sitter phase}
Let us finally 
find approximate solutions for eqs. (\ref{peg}) and (\ref{pef}) valid in the late de\,Sitter phase, where $\Lambda_{\rm eff}\gg 8\pi G \rho\rightarrow 0$. In this phase $a\simeq -\left(\mathcal{H}_0\,\tau\right)^{-1}$, $\mathcal{H}\simeq a\,\mathcal{H}_0\simeq -1/\tau$ and 
%\be
%c\simeq \sqrt{\frac{\Lambda_{\rm eff}}{\Lambda_c}}\approx 15\,,\hspace{1 cm} \eta(\beta_i, c)\simeq \text{const.}\equiv \eta  \simeq -30\,.%\,,\hspace{1 cm}\frac{c}{r^2}\simeq 1\,.
%\ee
\be
c\simeq \sqrt{\frac{\Lambda_{\rm eff}}{\Lambda_c}}\approx 15\,,\hspace{1 cm} \bar r \sigma_2(\beta_i, c)\simeq \text{const.} \simeq -30\,.%\,,\hspace{1 cm}\frac{c}{r^2}\simeq 1\,.
\ee
Therefore, the two equations for the tensor modes can be written as 
\be\label{ds1}
h^{''}-\frac{2}{\tau}\,h^{'}+k^2 h+\frac{\mathcal{R}_{\Lambda}}{\tau^2}\left(h-\lc\right)=0\,,
\ee
\be\label{ds2}
\lc^{''}-\frac{2}{\tau}\,\lc^{'}+c^2\,k^2\,\lc-\frac{\mathcal{R}_{\Lambda}}{\tau^2} \,\frac{c}{\bar r^2}\left(h-\lc\right)=0\,,
\ee
where 
\be
\mathcal{R}_{\Lambda}=\frac{m^2}{\mathcal{H}_0^2}\,\bar r \sigma_2 \simeq -30 \,.%\frac{m^2}{\mathcal{H}_0^2}\,\frac{c}{r^2}\,\eta\,.
\ee

 For super-Hubble scales, $k^2\ll 1/\tau^2$,  eqs. (\ref{ds1}) and (\ref{ds2}) can be diagonalized  in terms of  $H_+\equiv h+\lc$ and $H_-\equiv h-\lc$.
 \be
H_+''-\frac{2}{\tau}H_+'=0\,,
\ee
\be
H_-''-\frac{2}{\tau}H_-'+\frac{\mathcal{R}_-}{\tau^2}\,H_-=0\,,
\ee
where $\mathcal{R}_-\equiv\left(1+\frac{c}{r^2}\right)\mathcal{R}_{\Lambda}\simeq -255$. These equations are solved by
\be
H_+=c_1+c_2\,\tau^3\,,
\ee
\be
H_-=d_1\,\tau^{\frac{1}{2}\left(3-\sqrt{9-8\,\mathcal{R}_-}\right)}+d_2\,\tau^{\frac{1}{2}\left(3+\sqrt{9-8\,\mathcal{R_-}}\right)}\,.
\ee
The physical tensor modes $h$ and $\lc$ can be expressed as linear combinations of $H_+$ and $H_-$
\be
h=\frac{1}{2}\left(H_++H_-\right)=a_1+a_2\,\tau^3+a_3\,\tau^{\frac{1}{2}\left(3-\sqrt{9-8\,\mathcal{R}_-}\right)}+a_4\,\tau^{\frac{1}{2}\left(3+\sqrt{9-8\,\mathcal{R}_-}\right)}\,,
\ee
\be
\lc=\frac{1}{2}\left(H_+-H_-\right)=a_1+a_2\,\tau^3-a_3\,\tau^{\frac{1}{2}\left(3-\sqrt{9-8\,\mathcal{R}_-}\right)}-a_4\,\tau^{\frac{1}{2}\left(3+\sqrt{9-8\,\mathcal{R}_-}\right)}\,.
\ee
Since $\mathcal{R}_- \ll 0$ the $a_3$-mode is growing while all other modes are decaying in time (we recall that in the parametrization chosen, conformal time $\tau\in [-\infty, 0]$).

For sub-Hubble scales, we expect the couplings in eqs. (\ref{ds1}) and (\ref{ds2}) to be suppressed with respect to the $k$-terms. If this is the case,  eqs. (\ref{ds1}) and (\ref{ds2}) decouple and are the same as in GR% during the late-time de\,Sitter phase. 
\be\label{dss1}
h^{''}-\frac{2}{\tau}\,h^{'}+k^2 h=0\,,
\ee
\be\label{dss2}
\lc^{''}-\frac{2}{\tau}\,\lc^{'}+c^2\,k^2\,\lc=0\,,
\ee
with solutions, $|k\tau|\gg 1$,
\be
h=c_1\,\tau\left(\sin\left(k\tau\right)+\frac{\cos\left(k\tau\right)}{k\,\tau}\right)+c_2\,\tau\left(\frac{\sin\left(k\tau\right)}{k\,\tau}-\cos\left(k\tau\right)\right)\,,
\ee
\be
\lc=d_1\,\tau\left(\sin\left(kc\tau\right)+\frac{\cos\left(kc\tau\right)}{kc\,\tau}\right)+d_2\,\tau\left(\frac{\sin\left(kc\tau\right)}{kc\,\tau}-\cos\left(kc\tau\right)\right)\,.
\ee
As $|\tau|$ decreases the coupling terms in eqs. (\ref{ds1}) and (\ref{ds2}) can become relevant, when  $\mathcal{R}_{\Lambda}/\tau^2\simeq m^2\,\sigma_2 \bar r\,\left(\frac{\mathcal{H}}{\mathcal{H}_0}\right)^2\gg\mathcal{H}_0^2\simeq k^2$. More precisely 
\be
\left(\frac{|\mathcal{R}_{\Lambda}|}{\tau^2}\,\lc\right)/\left(k^2\,h\right)\simeq %\frac{m^2\,f_1}{\mathcal{H}_0^2}\left(\frac{\mathcal{H}}{\mathcal{H}_0}\right)^2\left(\frac{\mathcal{H}_0}{k}\right)^2\left(\frac{l}{h}\right)\simeq 
30\left(\frac{\mathcal{H}}{k}\right)^2\left(\frac{\lc}{h}\right)\,.
\ee
Hence, once the initial conditions for the two tensor perturbations are fixed, modes with smaller $k$ will experience the effects of the coupling earlier. On the other hand, for a given mode $k$, the effects of the coupling at late times will be proportional to the ratio $(\lc/h)$. This result is in line with what we have found in section \ref{Hig-tach} by examining the positivity of the mass matrix for the canonically normalized tensor modes. 
At the end of inflation, the amplitude of the tensor mode $\lc$ is suppressed by a factor $\left(H_0/H_I\right)^{1/2}$ with respect to the one of $h$. In the following epochs, $h$ evolves like in $\Lambda$CDM while $\lc$ has a constant and a decaying mode during the radiation and matter era. We therefore expect that the ratio $\lc/h$ during radiation and matter will stay almost constant and, if the inflationary result is to be trusted,  of the order of $\left(H_0/H_I\right)^{1/2}$.

\bibliographystyle{utphys}
\bibliography{bi-grefs}

\end{document}